\documentclass{statsoc}
\usepackage{amsmath,amsfonts,amssymb}
\usepackage[authoryear]{natbib}
\usepackage[colorlinks,citecolor=blue,urlcolor=blue]{hyperref}
\usepackage{graphicx}
\usepackage[a4paper, textwidth=426pt]{geometry}

\title[LCCA]{Longitudinal Canonical Correlation Analysis}

\author[Lee {\it et al.}]{Seonjoo Lee, Jongwoo Choi, Zhiqian Fang and F. DuBois Bowman\\
For the Alzheimer’s Disease Neuroimaging Initiative$^*$}

\address{Columbia University and New York State Psychiatric Institute, New York, U.S.A.\\
Department of Biostatistics, University of Michigan, New York, U.S.A.}
\email{seonjoo.lee@nyspi.columbia.edu}

\begin{document}

\begin{abstract}
This paper considers canonical correlation analysis for two longitudinal variables that are possibly sampled at different time resolutions with irregular grids. We modeled trajectories of the multivariate variables using random effects and found the most correlated sets of linear combinations in the latent space. Our numerical simulations showed that the longitudinal canonical correlation analysis (LCCA) effectively recovers underlying correlation patterns between two high-dimensional longitudinal data sets. We applied the proposed LCCA to data from the Alzheimer’s Disease Neuroimaging Initiative and identified the longitudinal profiles of morphological brain changes and amyloid cumulation.
\end{abstract}


\footnotetext{$^*$Data used in preparation of this article were obtained from the Alzheimer’s Disease Neuroimaging Initiative (ADNI) database (adni.loni.usc.edu). As such, the investigators within the ADNI contributed to the design and implementation of ADNI and/or provided data but did not participate in analysis or writing of this report. A complete listing of ADNI investigators can be found at: \url{http://adni.loni.usc.edu/wp-content/uploads/how_to_apply/ADNI_Acknowledgement_List.pdf.}}


\section{Introduction}\label{introduction}

Canonical correlation analysis (CCA) aims to find the correlation structures between two sets of multivariate variables. CCA seeks linear combinations within each set, such that the resulting linear combinations from variables are maximally correlated, but orthogonal with all other linear combinations in either set. Recently, CCA has been applied to high-dimensional multivariate variables via dimension reduction \citep{song2016canonical}, penalization \citep{avantsDementiaInducesCorrelated2010,baoCanonicalCorrelationCoefficients2019,fangJointSparseCanonical2016,gossmannFDRCorrectedSparseCanonical2018,linCorrespondenceFMRISNP2014a,wilmsRobustSparseCanonical2016,wittenPenalizedMatrixDecomposition2009} and combining multiple datasets \citep{deleusFunctionalConnectivityAnalysis2011,zhangFrequencyRecognitionSSVEPbased2014,kimJointconnectivitybasedSparseCanonical2019}.

CCA has been used frequently in medical applications to identify associations between clinical/behavior/imaging/genetic variables. For example, CCA has been applied to explore the associations between clinical symptoms and behavioral measures \citep{mihalikBrainbehaviourModesCovariation2019}, functional connections in the brain and cognitive deficits \citep{adhikariFunctionalNetworkConnectivity2019} or clinical symptoms \citep{kang2016depression,grosenickFunctionalOptogeneticApproaches2019}, two sets of imaging data \citep{avantsDementiaInducesCorrelated2010}, and gene-imaging associations \citep{linCorrespondenceFMRISNP2014a,kimJointconnectivitybasedSparseCanonical2019}.

When one of the multivariate variables is measured over time, CCA can be extended using a multi-set sparse canonical correlation approach via group lasso penalization \citep{haoIdentificationAssociationsGenotypes2017} and temporal multi-task SCCA (T-MTSCCA) \citep{duIdentifyingProgressiveImaging2019}. However, existing methods cannot handle missing values, irregular temporal sampling, and temporal misalignment between two variables.

To address this gap, we propose a new LCCA method that identifies the patterns of canonical variates that maximize the association between longitudinal trajectories. We model the longitudinal trajectory of each variable using random effects (e.g. random intercept and random slope). Then we find the linear combinations of the random effects that maximize the correlations. Since the dimensions of the two multivariate variables can be greater than the sample size, we employ dimension reduction via eigen decomposition. We implement estimation in the longitudinal principal component analysis framework. We conduct extensive simulation experiments to evaluate the performance of LCCA. We also apply LCCA to data from the Alzheimer’s Disease Neuroimaging Initiative (ADNI) cohort \citep{muellerAlzheimerDiseaseNeuroimaging2005,weinerAlzheimerDiseaseNeuroimaging2012}. The ADNI data include longitudinal $^{18}$F-AV-45 (florbetapir) Positron emission tomography (PET) images to quantify brain amyloid loads over eight years and structural MRI data during the same follow-up. Our LCCA method shows stable performance in terms of dimension selection and yields accurate signal identification, and the method identifies distinct AD- related brain patterns in the ADNI data.

\section{Method}\label{method}

\subsection{Longitudinal Canonical Correlation
Analysis}\label{longitudinal-canonical-correlation-analysis}

For each subject \(i=1,...,n\) and visits \(j=1, \ldots, J_i\), we
observe the \(p\)-dimensional vector \(\mathbf{X}_{ij}\) at time \(t_{ij}\).
Similarly, for the visits \(k=1, \ldots, K_i\), we observe a
\(q\)-dimensional vector \(\mathbf{Y}_{ik}\) at time \(s_{ik}\). For
illustration, we consider the linear trajectories for each variable and
later we generalize to the nonlinear trajectories via spline
regression. We model each observation using a random intercept and slope model: 
\begin{eqnarray}\label{intro:lmm}
  \mathbf{X}_{ij} &= \mathbf{\mu}_{ij}^{X} + \mathbf{Z}_{i,0}^{X} + t_{ij} \mathbf{Z}_{i,1}^{X}  + {\boldsymbol\epsilon}_{ij}^{X},\\
  \mathbf{Y}_{ik} &= \mathbf{\mu}_{ik}^{Y} + \mathbf{Z}_{i,0}^{Y} + s_{ik} \mathbf{Z}_{i,1}^{Y}  + {\boldsymbol\epsilon}_{ik}^{Y},
\end{eqnarray} where \(\mathbf{\mu}_{ij}^{X}\) and
\(\mathbf{\mu}_{ik}^{Y}\) are fixed effects;
\(\mathbf{Z}_{i,0}^{X}\) and \(\mathbf{Z}_{i,0}^{Y}\) are random intercepts;
\(\mathbf{Z}_{i,1}^{X}\) and \(\mathbf{Z}_{i,1}^{Y}\) are random slopes;
\({\boldsymbol\epsilon}_{ij}^{X}\) and
\({\boldsymbol\epsilon}_{ik}^{Y}\) are errors. The random effects are
stacked to a \(2p\)-dimensional vector
\(\mathbf{Z}_i^X := \big( { {\mathbf Z}_{i,0}^{X}}^{\prime}, {{\mathbf Z}_{i,1}^{X}}^{\prime} \big)^{\prime} \in \mathbb{R}^{2p}\)
, and a \(2q\)-dimensional vector
\({\mathbf Z}_i^Y := \big( {{\mathbf Z}_{i,0}^{Y}}^{\prime}, {{\mathbf Z}_{i,1}^{Y}}^{\prime} \big)^{\prime} \in \mathbb{R}^{2q}\), 
and are assumed to be distributed with zero mean and
a covariance
\({\boldsymbol\Sigma}_{c} = \left[\begin{array} {rr} {\boldsymbol\Sigma}_{c}^{(0,0)} & {\boldsymbol\Sigma}_{c}^{(0,1)} \\ {\boldsymbol\Sigma}_{c}^{(1,0)} & {\boldsymbol\Sigma}_{c}^{(1,1)} \\ \end{array}\right]\),
where
\({\boldsymbol\Sigma}_{c}^{(a, b)}=\mathbb{E}\left( {\mathbf Z}_{i,a}^{c} {{\mathbf Z}_{i,b}^c}^{\prime} \right)\),
\(a, b \in \{ 0,1\}, c \in \{X,Y\}\) and uncorrelated with
\({\boldsymbol\epsilon}_{ij}^{X}\) and
\({\boldsymbol\epsilon}_{ik}^{Y}\). The random intercepts, slopes and errors are assume to follow a Gaussian distribution. 

Then, the objective function defining the canonical correlation is
\begin{equation}\label{met:cca}
r_m^{XY} = max_{\mathbf{u}_m \in \mathcal{R}^{2p},\mathbf{v}_m \in \mathcal{R}^{2q}} <\mathbf{u}_m^{'} {\mathbf Z}_i^{X},\mathbf{v}_m^{'} {\mathbf Z}_i^{Y}>.
\end{equation}

When $p$ is greater than $n/2$, we can employ principal component analysis to represent
\(\mathbf{Z}_i^X \approx {\boldsymbol\Phi}_{X} {\boldsymbol\xi}_i^X\),
where
\({\boldsymbol\Phi}_{X} =\left({{\boldsymbol\Phi}_0^X}^{\prime}, {{\boldsymbol\Phi}_1^X}^{\prime} \right)^{\prime}\)
is a \(2p \times N_{X}\) matrix of the leading \(N_X\) eigenvectors of
\({\boldsymbol\Sigma}_{X} = Var (\mathbf{Z}_i^X) \) and
\({\boldsymbol \xi}_i^X = ( \xi_{i1}^X,\ldots,\xi_{i N_X}^X)^{\prime} \in \mathbb{R}^{N_X}\)
are the associated subject \(i\)-specific eigenscores. Using longitudinal principal component analysis (LPCA) \citep{grevenLongitudinalFunctionalPrincipal2011,zipunnikovLongitudinalHighdimensionalPrincipal2014,leeStatisticalImageAnalysis2015}, each term is expressed as
\begin{equation}\label{eqn:lpcax}
\mathbf{X}_{ij} \approx \mathbf{\mu}_{ij}^{X} + \sum_{l=1}^{N_X} \xi_{il}^{X} \left(\Phi_l^{X,0} + t_{ij}\Phi_l^{X,1}\right) + \epsilon_{ij}^{X},
\end{equation} where
\((\xi_{il_1}^{X}, \xi_{il_2}^{X}) \sim (0, 0; \lambda_{X}^{l_1}, \lambda_{X}^{l_2}; 0 )\),
in which
``\(\cdot \sim (\mu_1, \mu_2; \sigma_1^2, \sigma_2^2; \rho )\)''
represents that a pair of gaussian variables has a distribution with mean
\((\mu_1, \mu_2)\), variance \((\sigma_1^2, \sigma_2^2)\), and
correlation \(\rho\). Similarly, \(\mathbf{Y}_{ik}\) are expressed as 

\begin{equation}\label{eqn:lpcay}
\mathbf{Y}_{ik} \approx \mathbf{\mu}_{ik}^{Y} + \sum_{l=1}^{N_Y} \xi_{il}^{Y} \left(\Phi_l^{Y,0} + s_{ik}\Phi_l^{Y,1}\right) + \epsilon_{ik}^{Y},
\end{equation} where
\((\xi_{il_1}^{Y}, \xi_{il_2}^{Y} ) \sim (0, 0; \lambda_{Y}^{l_1}, \lambda_{Y}^{l_2}; 0 )\).
The eigenvectors of \eqref{eqn:lpcax} and \eqref{eqn:lpcay} can be obtained
by the least-squares estimation of the covariance matrices. Without loss of generality, we assume $\mathbf{X}_{ij}$ is demeaned to have mean zero. The p × p-covariance of $\mathbf{X}_{ij_1}$  and $\mathbf{X}_{ij_2}$ is given by
\begin{equation}
	E\mathbf{X}_{ij_1}\mathbf{X}_{ij_2}^{\top} = {\boldsymbol\Sigma}_X^{(0,0)} +  t_{ij_2}{\boldsymbol\Sigma}_X^{(0,1)}  +  t_{ij_1}{\boldsymbol\Sigma}_X^{(1,0)}  + 
t_{ij_1} t_{ij_2}{\boldsymbol\Sigma}_X^{(1,1)} + \delta_{j_1,j_2} {\boldsymbol\Sigma}_X^{\epsilon},  
\end{equation}
where \({\boldsymbol\Sigma}_X^{\epsilon}\) is the covariance matrix of $ \epsilon_{ij}^{X}$ and $\delta_{j_1,j_2}=1$ if $j_1=j_2$ and $\delta_{j_1,j_2}=0$ otherwise. Denote $vec(\cdot)$ as the vectorization of a matrix by stacking columns of the matrix on top of one another. 
Then, we can form a matrix 
$\mathbf{K}_X = \{ vec({\boldsymbol\Sigma}_X^{(0,0)}), vec({\boldsymbol\Sigma}_X^{(0,1)}), vec({\boldsymbol\Sigma}_X^{(1,0)}), vec({\boldsymbol\Sigma}_X^{(1,1)}),$ $vec({\boldsymbol\Sigma}_X^{\epsilon})\} $ and $\mathbf{f}_{i j_1 j_2}=(1, t_{ij_2}, t_{ij_1}, t_{ij_1}t_{ij_2}, \delta_{j_1,j_2} )^{\top}$ such that 
$E vec (\mathbf{X}_{ij_1}\mathbf{X}_{ij_2}^{\top}) = \mathbf{K}_X \mathbf{f}_{i j_1 j_2}$. By concatenating all vectors across all subjects and visits, we obtain a moment matrix identity for the $p^2 \times J$ matrix $\mathbf{X}$: $\mathbf{X} = \mathbf{K}_X \mathbf{F}$, where $J = \sum_{i=1}^N J_i^2$. Then covariance parameters $ \mathbf{K}_X$ can be unbiasedly estimated by using ordinary least squares (OLS): $\widehat{\mathbf{K}}_X=\mathbf{XF}^{\top} (\mathbf{FF}^{\top})^{-1}$. Given the estimated covariance matrix, the eigenvectors are computed via eigen decomposition. Given the eigenvectors, the eigenscores are estimated by the best linear unbiased predictors (BLUP) from the equations \eqref{eqn:lpcax} and \eqref{eqn:lpcay} as discussed in \cite{grevenLongitudinalFunctionalPrincipal2011,zipunnikovLongitudinalHighdimensionalPrincipal2014}.

Denote the $N \times N_X$ matrix $ {\boldsymbol \xi}^{X} = \{ \xi_{il}^X\}_{i=1,\ldots,N; l=1,\ldots, N_X}$ and $N \times N_Y$ matrix $ {\boldsymbol \xi}^{Y} = \{ \xi_{il}^Y\}_{i=1,\ldots,N; l=1,\ldots, N_Y}$. 
Then, the objective function defining the canonical correlation is
\begin{equation}\label{met:ccalow}
r_m^{XY} = max_{\mathbf{u}_m \in \mathcal{R}^{N_X},\mathbf{v}_m \in \mathcal{R}^{N_Y}}
<{\boldsymbol \xi}^{X}\mathbf{u}_m , {\boldsymbol \xi}^{Y}\mathbf{v}_m>
\end{equation}

The objective function in \eqref{met:ccalow} is maximized under the
restriction that each \(\mathbf{u}_m\) is orthogonal
to the lower order \(\mathbf{u}_{m^{\prime}}\), with
\(1 \leq m < m^{\prime} \leq min(N_X,N_Y)\). Same restriction is applied for \(\mathbf{v}_m\) of the second set of variates. For identification
purpose, we require a normalization condition for the canonical variates, \( \mathbf{u}_m^{\top} {\boldsymbol \xi}_i^{X}, \mathbf{v}_m^{\top} {\boldsymbol \xi}_i^{Y}, m=1,.\dots, min(N_X, N_Y)\), 
to have unit variance. The canonical vectors are estimated in the
lower-dimensional space, and the $m$-th longitudinal canonical vectors are
computed as $ \left(\mathbf{\Phi}_0^{X} + t_{ij}\mathbf{\Phi}_1^{X}\right) \mathbf{u}_{m} $ and 
$ \left(\mathbf{\Phi}_0^{Y} + s_{ik}\mathbf{\Phi}_1^{Y}\right) \mathbf{v}_{m} $ for
\(m=1,...,min(N_X,N_Y)\).

Depending on the availability of time points, this formation can be
naturally extended to higher-order trends or modeled via B-spline basis
expansion. Assume that the trajectory of the imaging measure at location $v$ is represented by a spline function with fixed knot sequence 
$\tau_1<\cdots< \tau_D$ and fixed degree $d$. 

\begin{equation}\label{eq:splinex}
\mathbf{X}_{ij}(v)= 
\mathbf{\mu}^{X}(v) + \sum_{r=1}^{D+d+1} Z_{i,r}(v) b_r(t_{ij}) + \epsilon_{ij}^{X},
\end{equation}

where the $b_r$ are a set of basis functions and $Z_{i,r}(v)$ are the associated spline coefficients. The $(D+d+1)V$-dimensional stacked vector of $\mathbf{Z_{i}} = (\mathbf{Z}_{i,1}^{\top},\ldots, \mathbf{Z}_{i,D+d+1}^{\top})^{\top}$ can be approximated by the principal component analysis: 
$\mathbf{Z_{i}} \approx  \sum_{l=1}^{N_X} \xi_{il}^{X}  \mathbf{\Phi}_l^{X}$ with each element expressed as $Z_{i,r}(v) \approx  \sum_{l=1}^{N_X} \xi_{il}^{X}  \Phi_l^{X,r}(v)$, for $r=1,\ldots, D+d+1$. Then, (\ref{eq:splinex}) is reorganized as

\begin{equation}\label{eq:splinex1}
\mathbf{X}_{ij}(v)= 
\mathbf{\mu}^{X}(v) + \sum_{l=1}^{N_X}  \xi_{il}^{X}   \sum_{r=1}^{D+d+1}  \Phi_l^{X,r}(v) b_r(t_{ij}) + \epsilon_{ij}^{X}.
\end{equation}
Similarly, $Y$ is modeled as
\begin{equation}\label{eq:splinex2}
\mathbf{Y}_{ik}(v)= 
\mathbf{\mu}^{Y}(v) + \sum_{l=1}^{N_Y}  \xi_{il}^{Y}   \sum_{r=1}^{D+d+1}  \Phi_l^{Y,r}(v) b_r(s_{ik}) + \epsilon_{ik}^{Y}.
\end{equation}

This method does not require that \(X\)'s and \(Y\)'s are observed at
the same time. As long as they are observed within a time frame that the
research question asks, the proposed method can extract association
patterns.

\subsection{The number of canonical
covariates}\label{number-of-canonical-covariates}

To determine the dimension of the CCA, we employ a traditional likelihood ratio test-based approach in the latent space.
Starting with \(m= 0\), we test the null hypothesis \(H_0\): \(d=m\) versus the alternative hypothesis \(H_1\):
\(d > m\). If \(H_0\) is rejected, \(m\) is incremented and a new test is
conducted. This proceeds until \(H_0\) is not rejected or
\(m\) reaches $M=min(N_X, N_Y)$. For a given number of canonical variates, $m$, the Wilk's test statistics \citep{wilks1935independence, bartlett1947general, friederichs2003statistical} is given by 

\begin{equation}
\Lambda_m = \Pi_{j=m+1}^{min(N_X,N_Y)}(1-\hat{r}_j^2).
\end{equation}

Based on Rao's F-approximation \citep{rao1973linear},  $F = df_2/df_1  (1-\Lambda_m/(\Lambda_m))^{1/\nu}$ follows asymptotically 
$F_{df_1,  df_2}$, where $\nu = \sqrt{(df_1^2 - 4)/((N_X - m)^2 + (N_Y - m)^2 - 5)}$, $df_1=(N_X -m)*(N_Y -m)$, and 
$df_2 =(n- 1.5 - (N_X + N_Y)/2)\nu -  df_1/2 + 1$.

The performance of LCCA depends on the selection of \(N_X\) and \(N_Y\). Previous works suggested using 80\% threshold of the variance explained by the number of individual components \citep{grevenLongitudinalFunctionalPrincipal2011,leeStatisticalImageAnalysis2015} or pre-specified numbers \citep{zipunnikovLongitudinalHighdimensionalPrincipal2014}. In practice, the performance depends on the signal-to-noise ratio, which is particularly relevant for imaging applications. Our numerical experiments show good performance using a threshold of 80-90\% as a rule of thumb. The algorithms are implemented as an R package LCCA and available in github (\url{https://seonjoo.github.io/lcca/}).

\section{Alzheimer's Disease Neuroimaging Data
Analysis}\label{alzheimers-disease-neuroimaging-data-analysis}

\subsection{Participants}\label{participants}

We apply our LCCA method to the ADNI data to identify
longitudinal associations between brain morphometry and amyloid
deposition. The data were downloaded from the ADNI database
(\url{http://adni.loni.usc.edu}). The initial phase (ADNI-1) recruited
800 participants, including approximately 200 healthy controls, 400
patients with late mild cognitive impairment (MCI), and 200 patients clinically diagnosed with
probable AD over 50 sites across the United States and Canada and
followed up at 6- to 12-month intervals for 2--3 years. ADNI has been
followed by ADNI-GO and ADNI-2 for existing participants and enrolled
additional individuals, including early MCI. To be classified as MCI in
ADNI, a subject needed an inclusive Mini-Mental State Examination score
of between 24 and 30, subjective memory complaint, objective evidence of
impaired memory calculated by scores of the Wechsler Memory Scale
Logical Memory II adjusted for education, a score of 0.5 on the Global
Clinical Dementia Rating, absence of significant confounding conditions
such as current major depression, normal or near-normal daily
activities, and absence of clinical dementia.

\begin{table}
\caption{\label{tab1} Demographic characteristics at baseline and 5 year conversion rate. }
\begin{tabular}{cccccc} \hline \hline
	&	AD (N=24)	&	MCI (N=365)	&	CN (N=291)	&	Total (N=680)	&	p value	\\ \hline
Sex,	Female, n (\%)	&	10 (41.7)	&	162 (44.4)	&	147 (50.5)	&	319 (46.9)	&	0.257	\\
Age, years, Mean (SD)	&	76.75 (7.15)	&	71.88 (7.74)	&	74.62 (6.47)	&	73.23 (7.35)	& $<$0.001		\\
Non-hispanic, n (\%)	&	22 (91.7)	&	351 (96.2)	&	279 (95.9)	&	652 (95.9)	&	0.245	\\
Race	, White, n (\%)	&	   23 (95.8)    	&	   342 (93.7)   	&	   267 (91.8)   	&	   632 (92.9)   	&	0.713	\\
MMSE$^1$, Mean (SD)	&	18.83 (4.90)	&	26.89 (3.34)	&	28.81 (1.63)	&	27.43 (3.39)	&	$<$ 0.001	\\
ADAS$^2$, Mean (SD)	&	37.35 (13.75)	&	16.64 (11.06)	&	9.05 (5.71)	&	14.10 (10.93)	&	$<$ 0.001	\\
5 Yr AD Transition$^3$,  n (\%)	&	N/A	&	104 (28.5)	&	16 (5.5)	&	120 (18.3)	&	$<$ 0.001	\\ \hline\hline
\multicolumn{6}{l}{\tiny $^1$ Eight participants has missing values. }\\
\multicolumn{6}{l}{\tiny$^2$ Nine participants had missing values.  }\\
\multicolumn{6}{l}{\tiny$^3$ Only non-demented participants at baseline were analyzed. }\\
\end{tabular}
\end{table}


All studies were approved by their respective institutional review
boards and all subjects or their surrogates provided informed consent
compliant with HIPAA regulations. In total, the analysis included PET-MRI 
scan pairs of 680 subjects on average  2.7 ($\pm$ 0.79) visits over 3.7 ($\pm$1.66) years on average. 
There are 291 cognitively normal (CN), 365 mild cognitive
impairment (MCI), and 24 AD participants at baseline. Detailed characteristics
of these individuals are given in Table \ref{tab1}.

\subsection{Structural imaging
processing}\label{structural-imaging-processing}

 For internal consistency, 3.0 MPRAGE T1-weighted MR images were used. 
 Cross-sectional image processing was performed using FreeSurfer Version 6. Region of
interest (ROI)-specific cortical thickness and volume measures
were extracted from the automated FreeSurfer v6 anatomical parcellation
using the Desikan-Killiany Atlas \citep{desikan2006automated} for cortical regions;
there were 68 ROIs (34 each on the left and right hemispheres), in which
the longitudinal cortical thickness and volume measures were collected. The volume
measures from 28 subcortical regions \citep{fischlWholeBrainSegmentation2002} were computed, including Lateral Ventricle, Inferior Lateral Ventricle, Thalamus,
Caudate, Putamen, Pallidum, Hippocampus, Amygdala, Accumbens area, Third
Ventricle, Fourth Ventricle, and five corpus callosum subregions. In addition, the total intracranial volume was included. 

We computed amyloid SUVR levels using the PetSurfer pipeline \citep{greveCorticalSurfacebasedAnalysis2014,greveDifferentPartialVolume2016}, 
which is available with Freesurfer version 6. The PetSurferpipeline first registers the PET image with the corresponding MRI scan, then applies Partial Volume Correction, and finally resamples the
voxel-wise SUVR values onto the cortical surface. The 81 ROI-level summary was computed based on Desikan Atlas for the cortical and subcortical
regions.

\subsection{Naive approach}
There are no existing CCA methods to handle the longitudinal data's missing and irregular temporal sampling. Thus, we consider the following approach. First, we computed the intercept and slope of each variable for each subject. Then, within each modality, the vectors of intercepts and slopes were stacked, and we performed a canonical correlation analysis. Finally, the number of canonical variates was estimated in the same principles described in Section \ref{number-of-canonical-covariates}. We named this approach the naive approach.

\subsection{Results}\label{results-1}

\begin{figure}
\begin{center}
\includegraphics[width=0.7\textwidth]{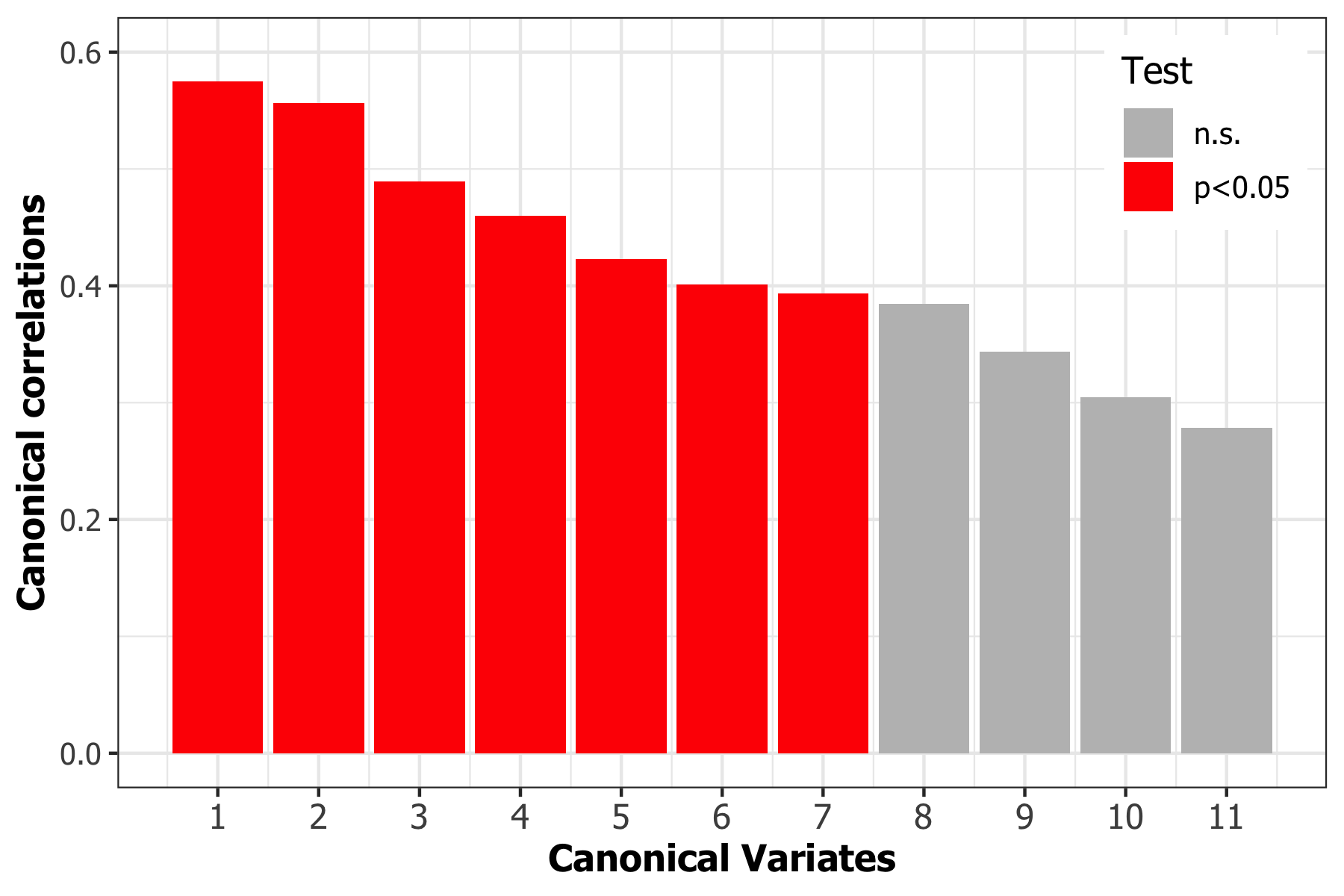}
\end{center}
\caption{Canonical correlation coefficients. 
Seven canonical correlation coefficients were significant ($p<0.05$) using 
F-test based on Rao's F-approximation of Wilk's Lambda.}\label{fig:cc}
\end{figure}

\begin{figure}
\includegraphics[width=0.32\textwidth]{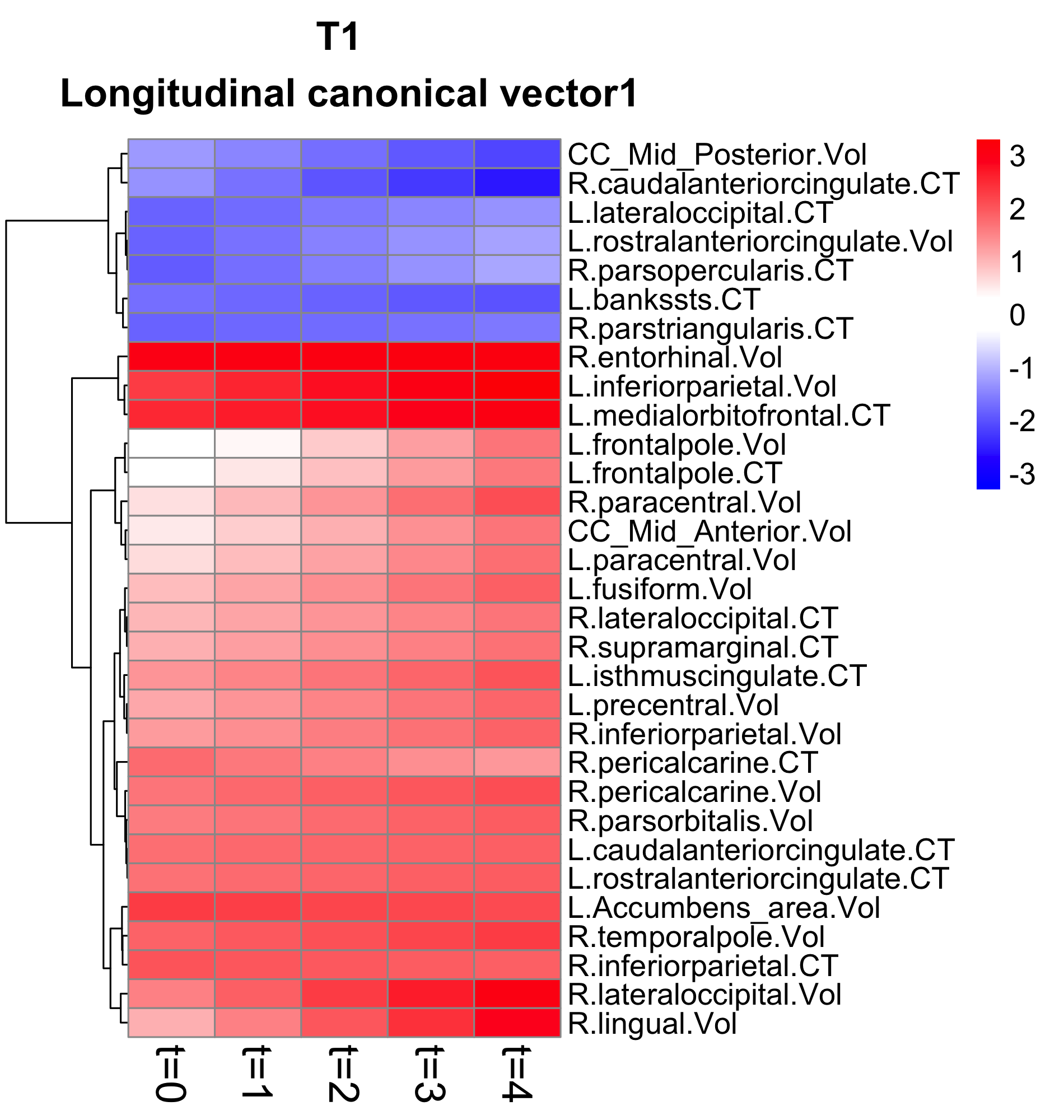}
\includegraphics[width=0.32\textwidth]{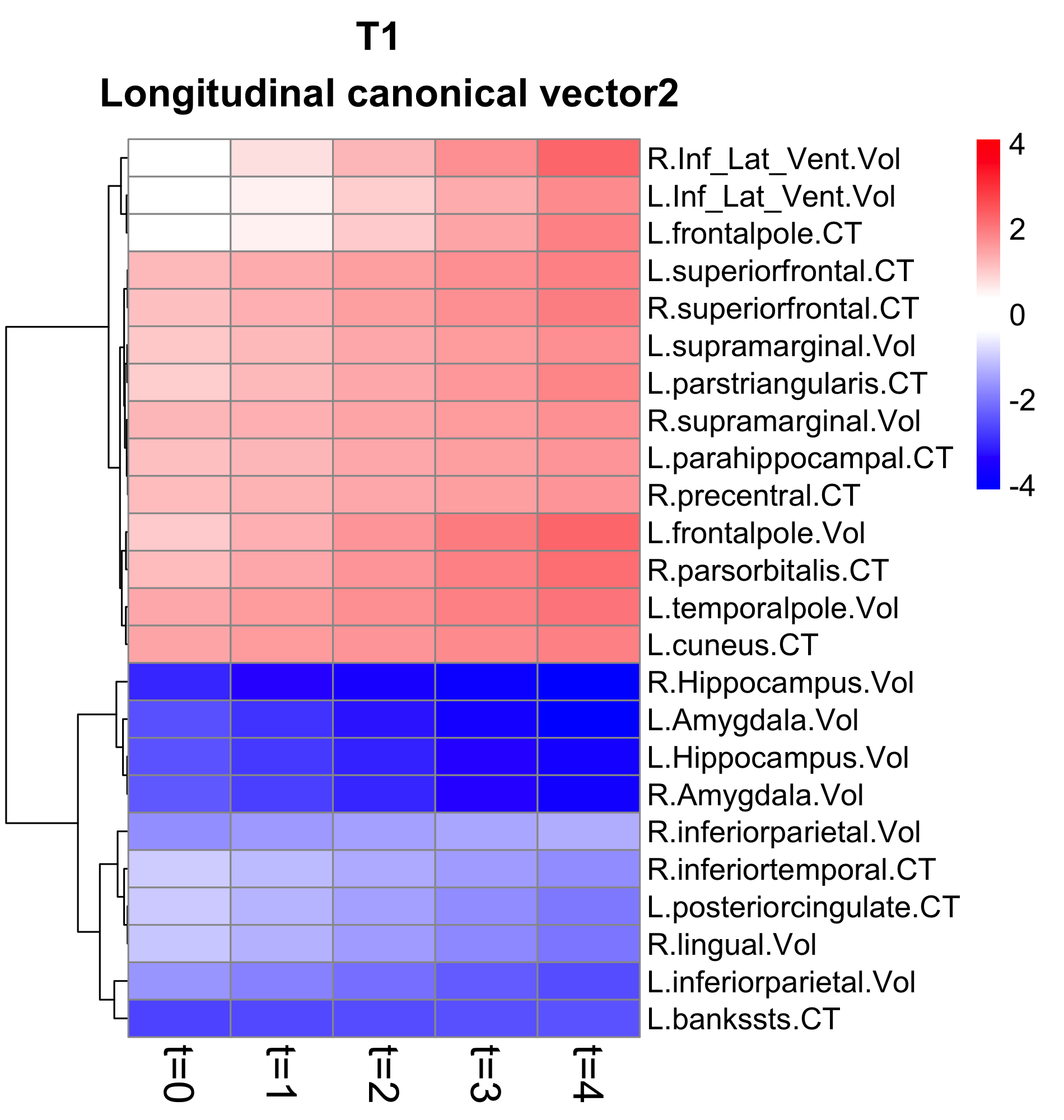}
\includegraphics[width=0.32\textwidth]{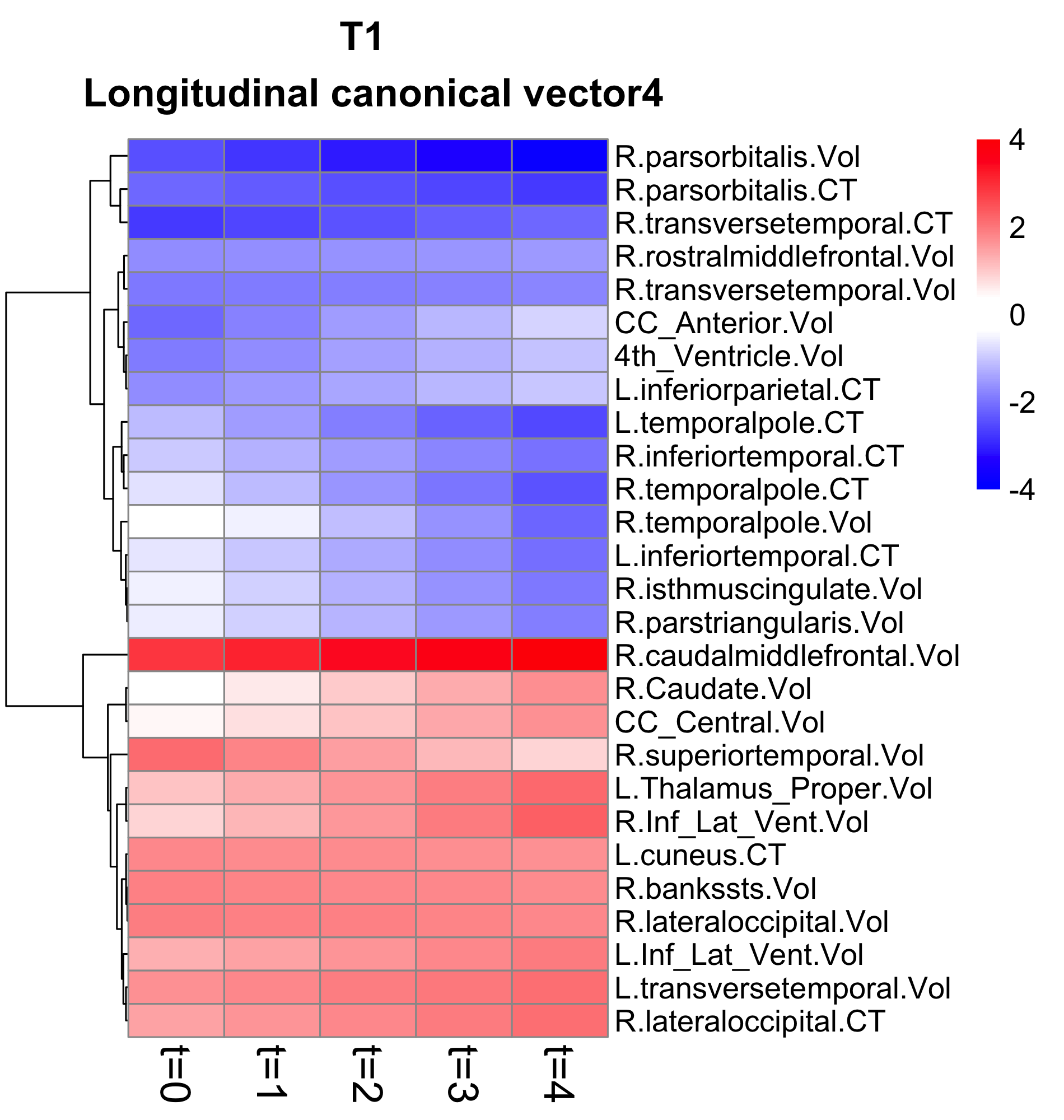}\\
\includegraphics[width=0.32\textwidth]{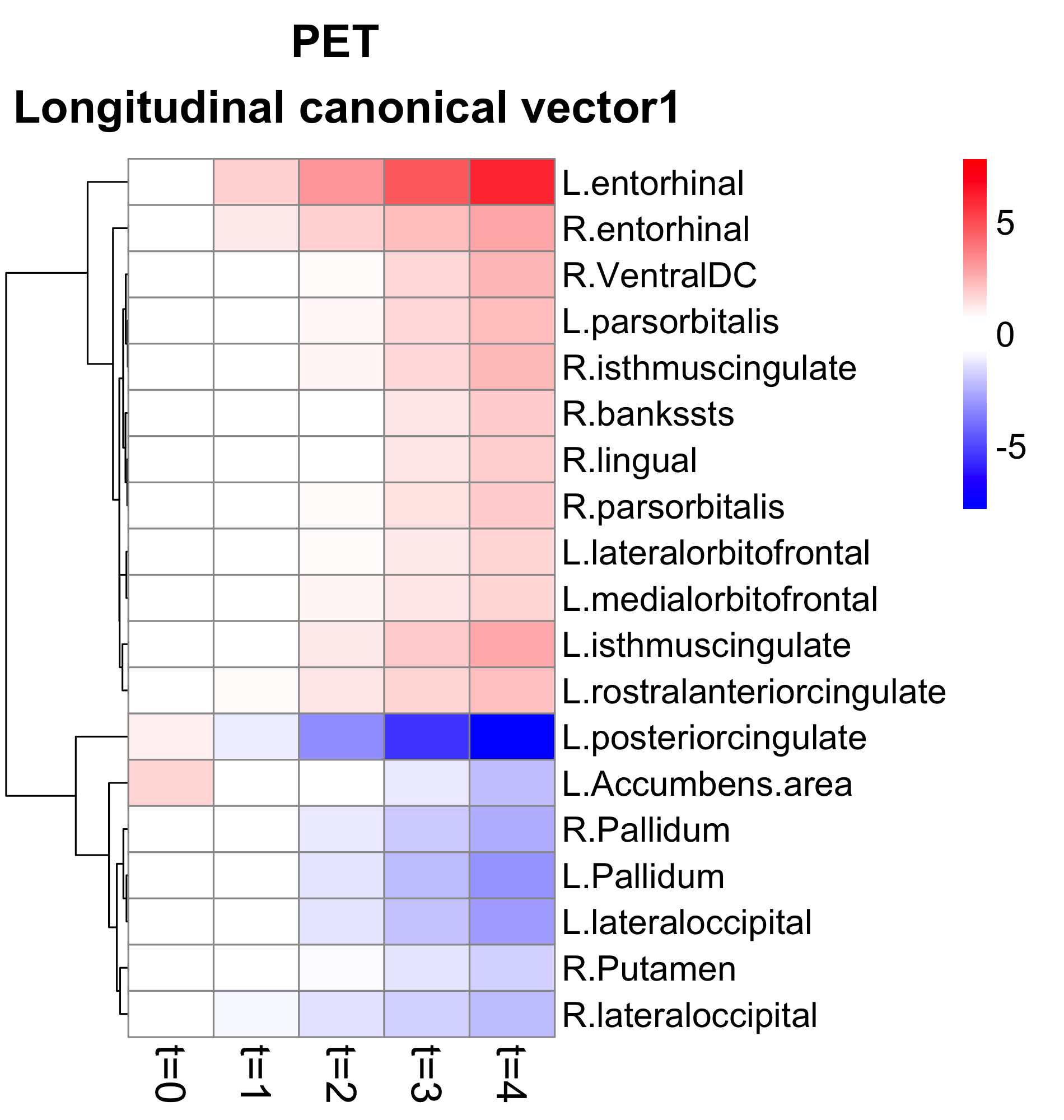}
\includegraphics[width=0.32\textwidth]{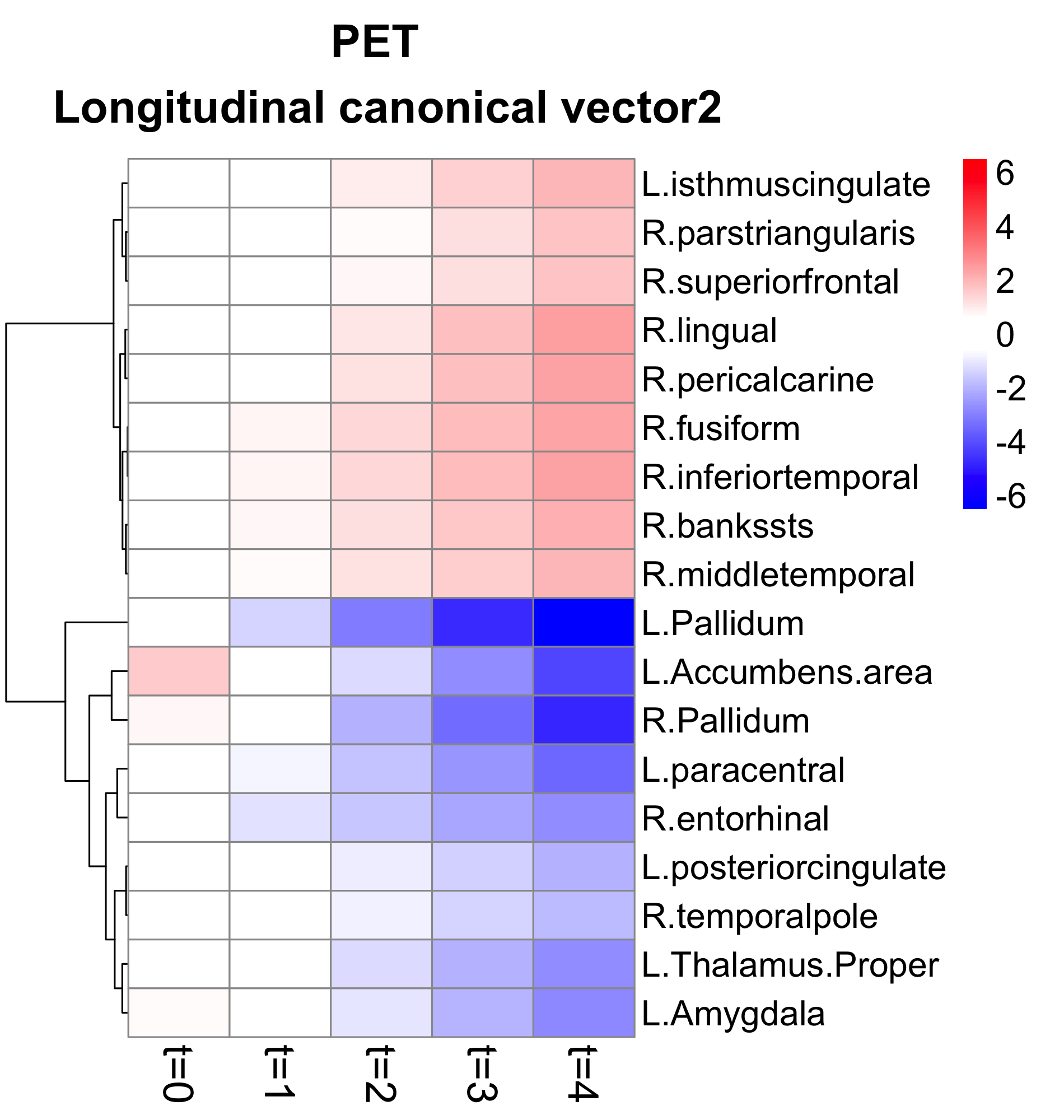}
\includegraphics[width=0.32\textwidth]{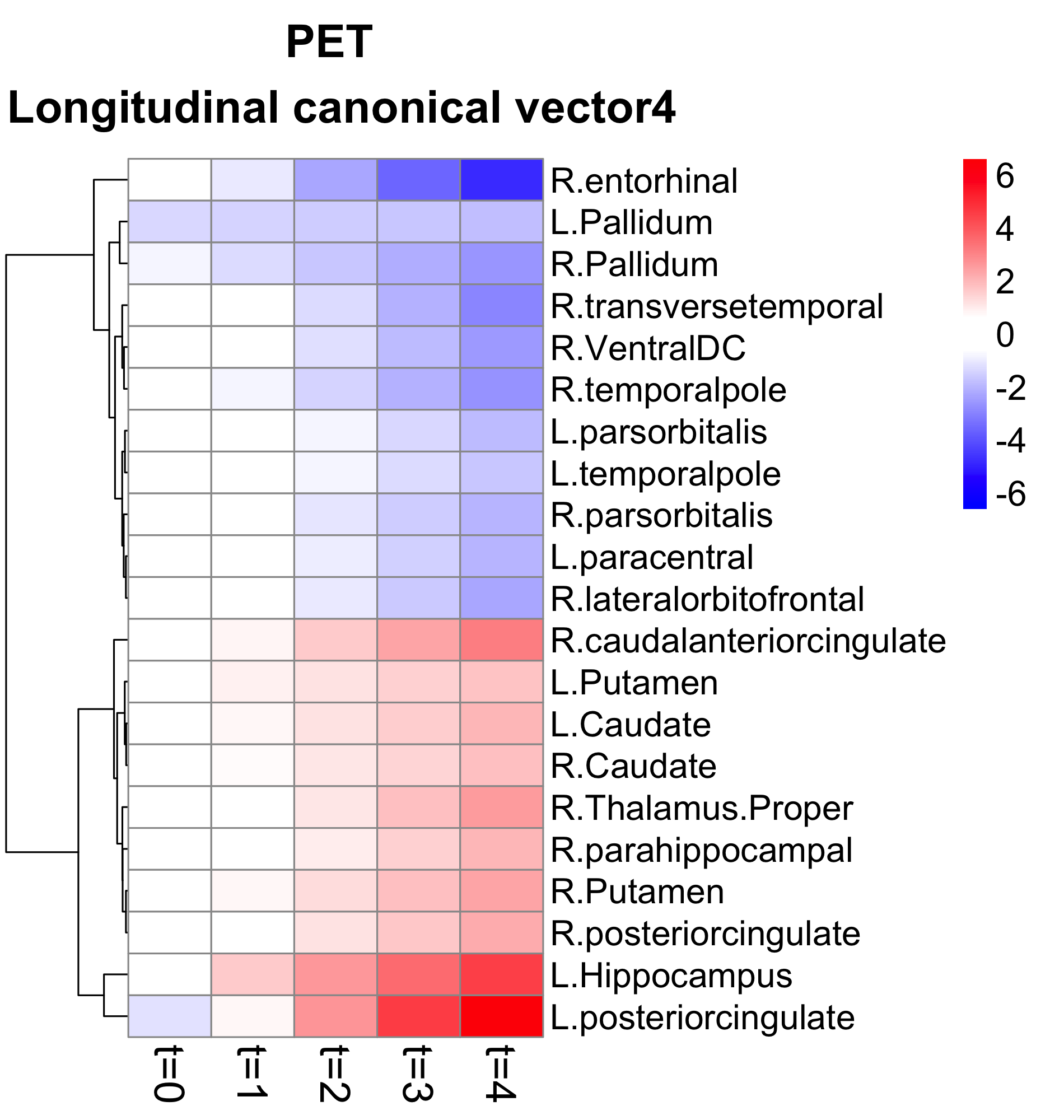}
\caption{LCCA: Pairs of the longitudinal canonical vectors are displayed at time=0,1,2,3,4.}\label{fig:cw1}
\end{figure}

\begin{figure}
\includegraphics[width=0.32\textwidth]{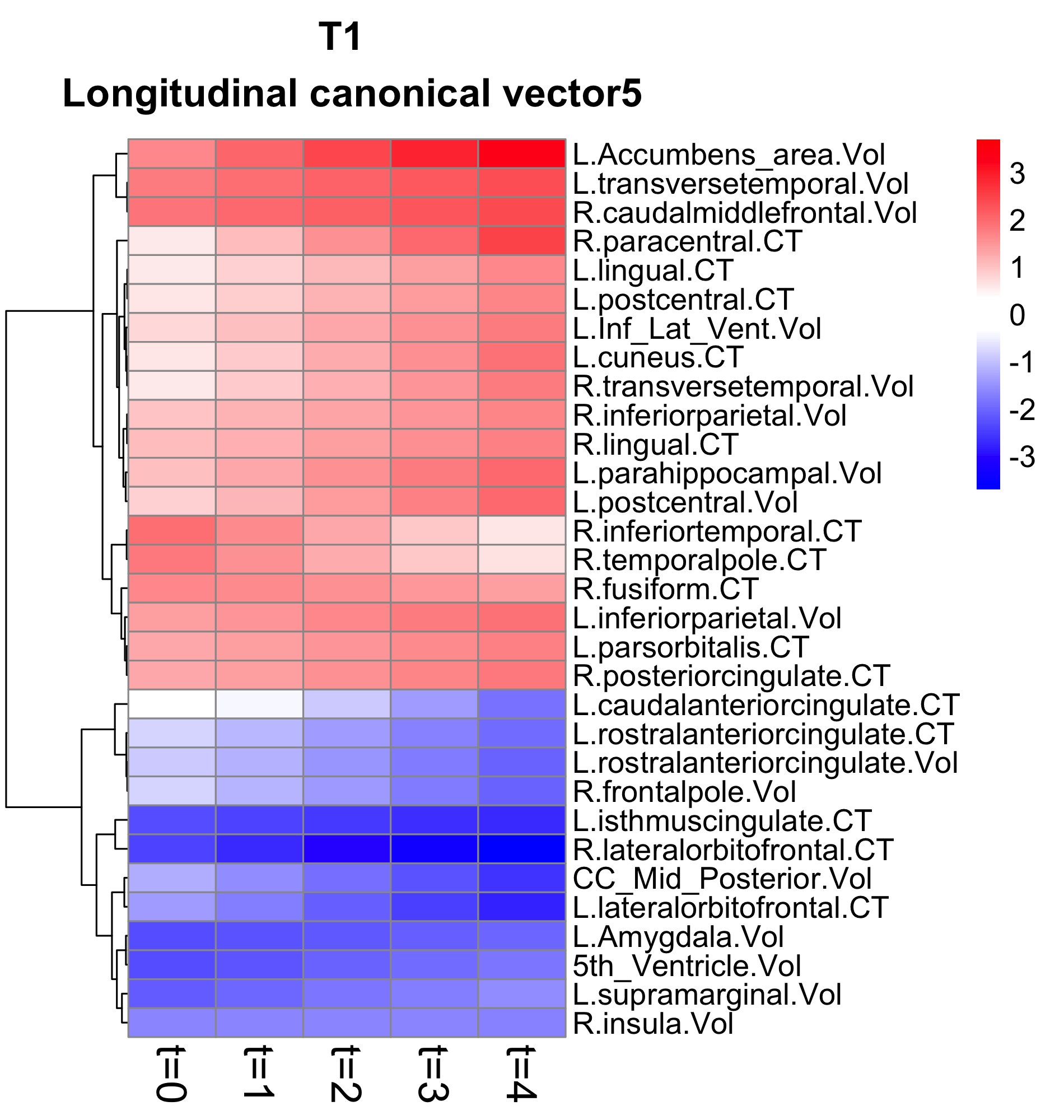}
\includegraphics[width=0.32\textwidth]{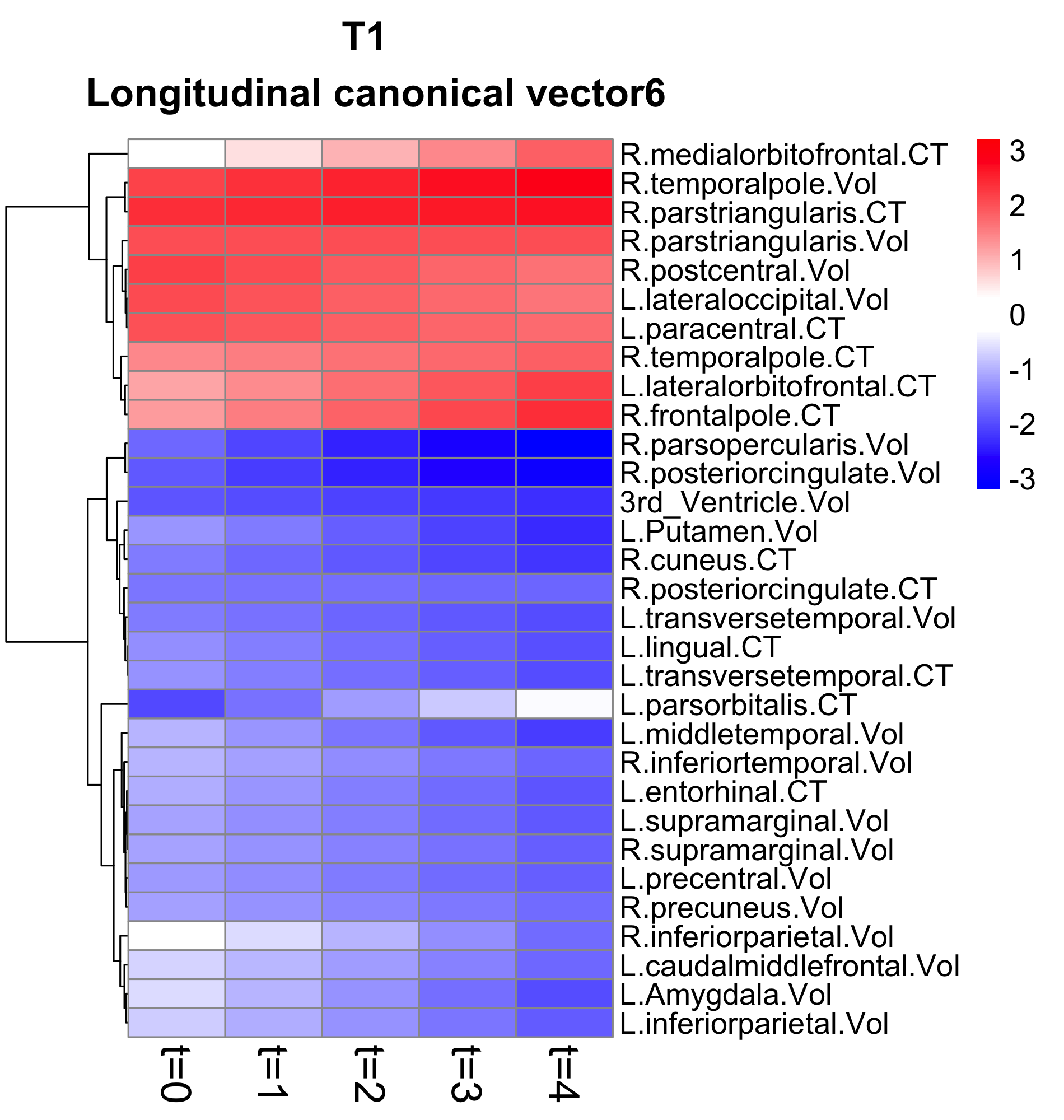}
\includegraphics[width=0.32\textwidth]{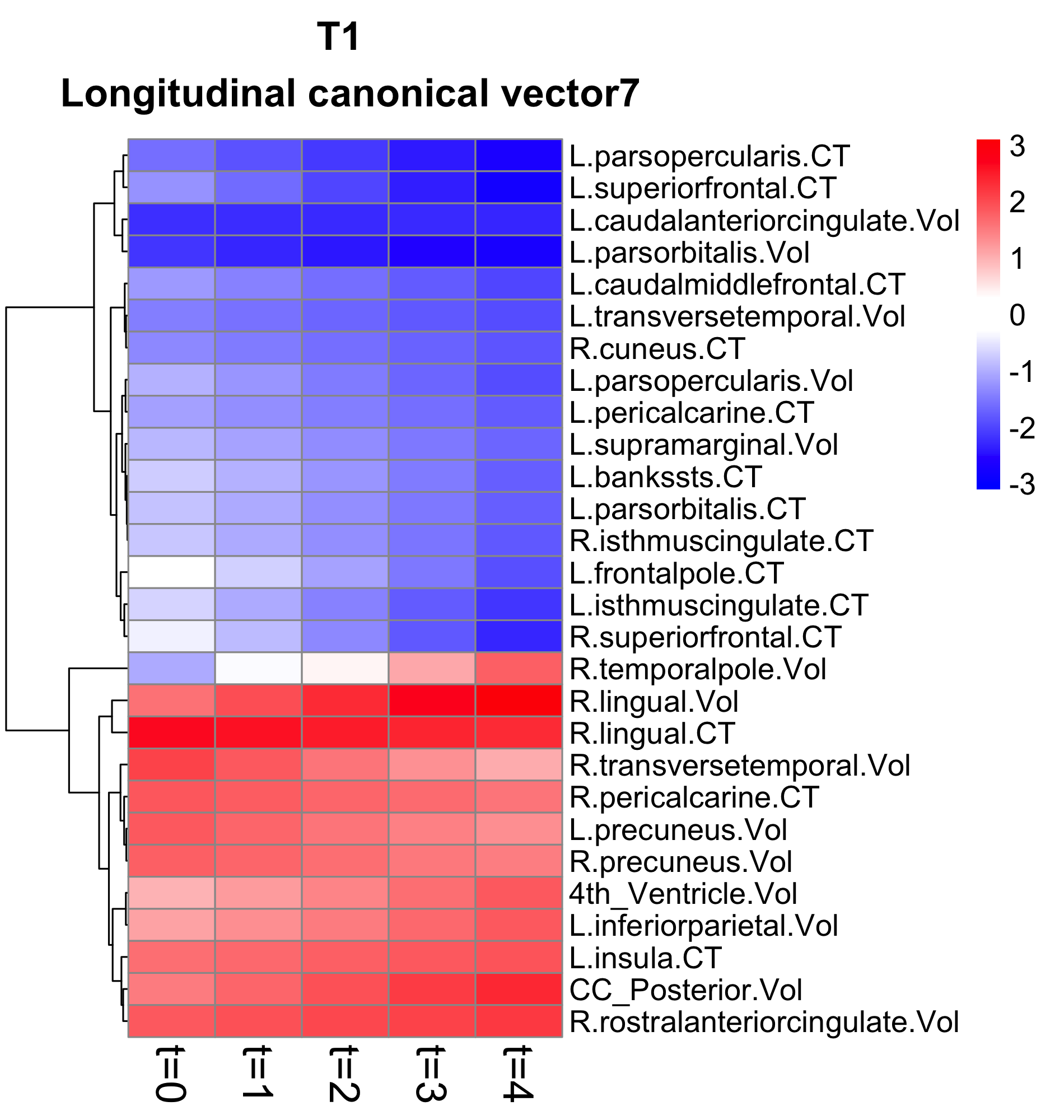}\\
\includegraphics[width=0.32\textwidth]{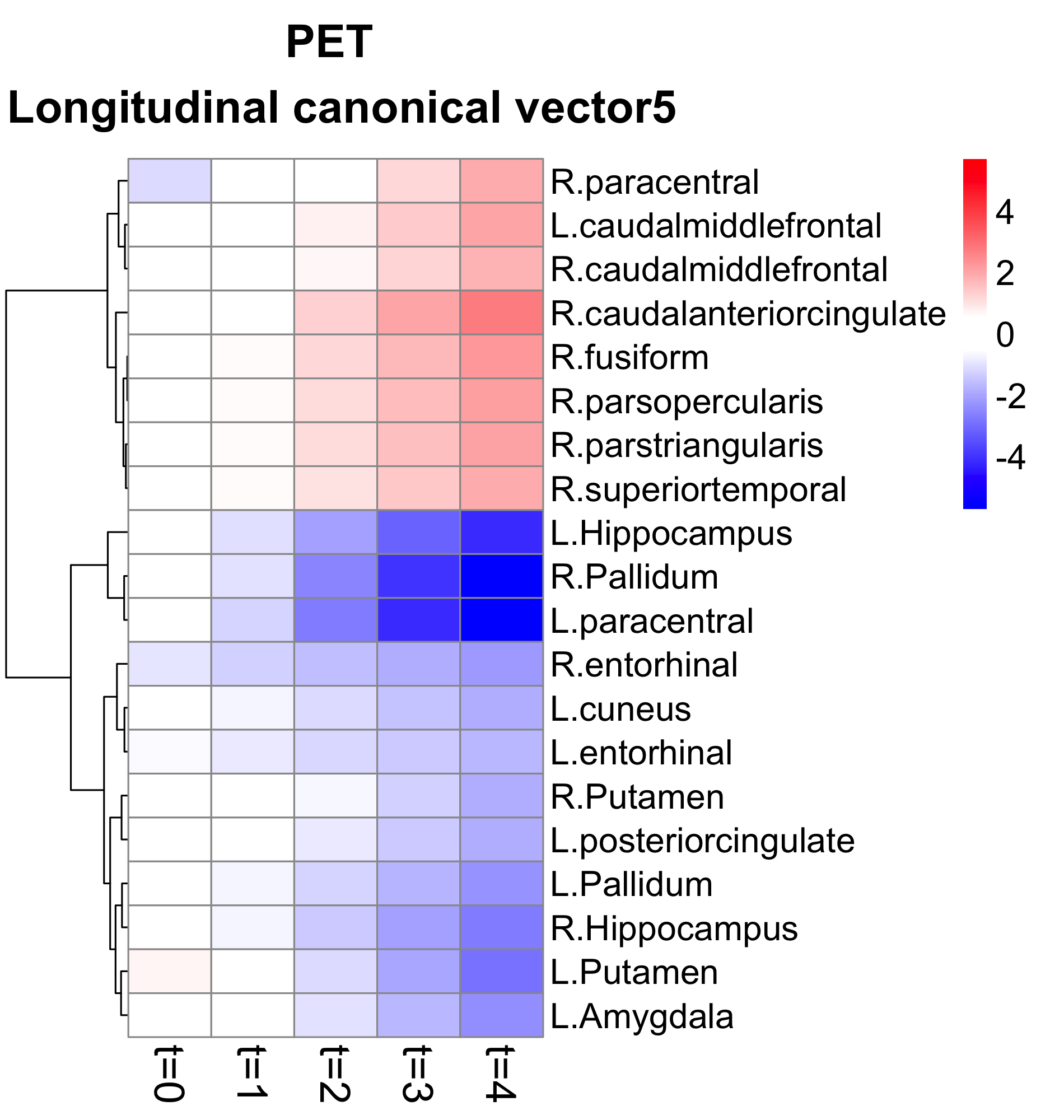}
\includegraphics[width=0.32\textwidth]{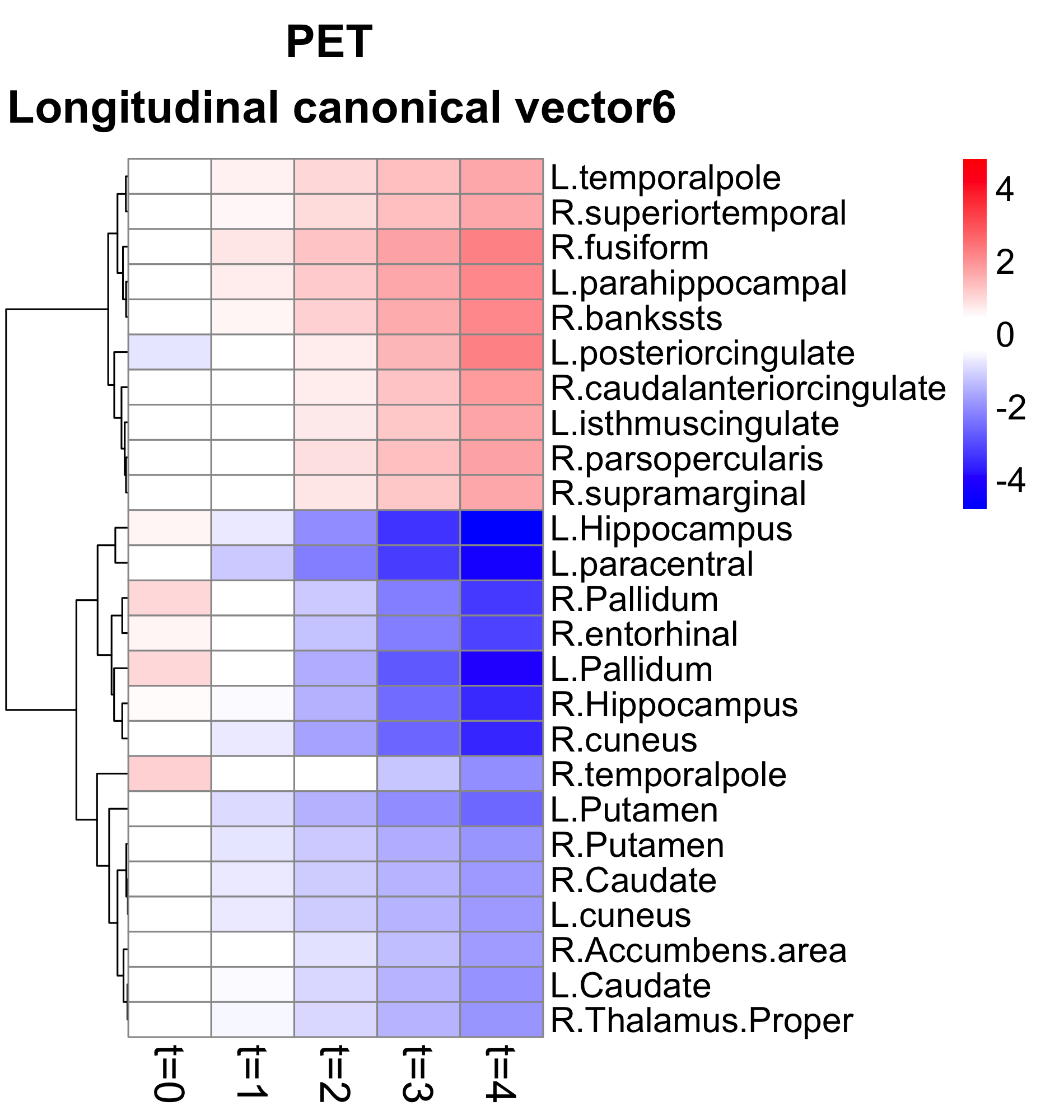}
\includegraphics[width=0.32\textwidth]{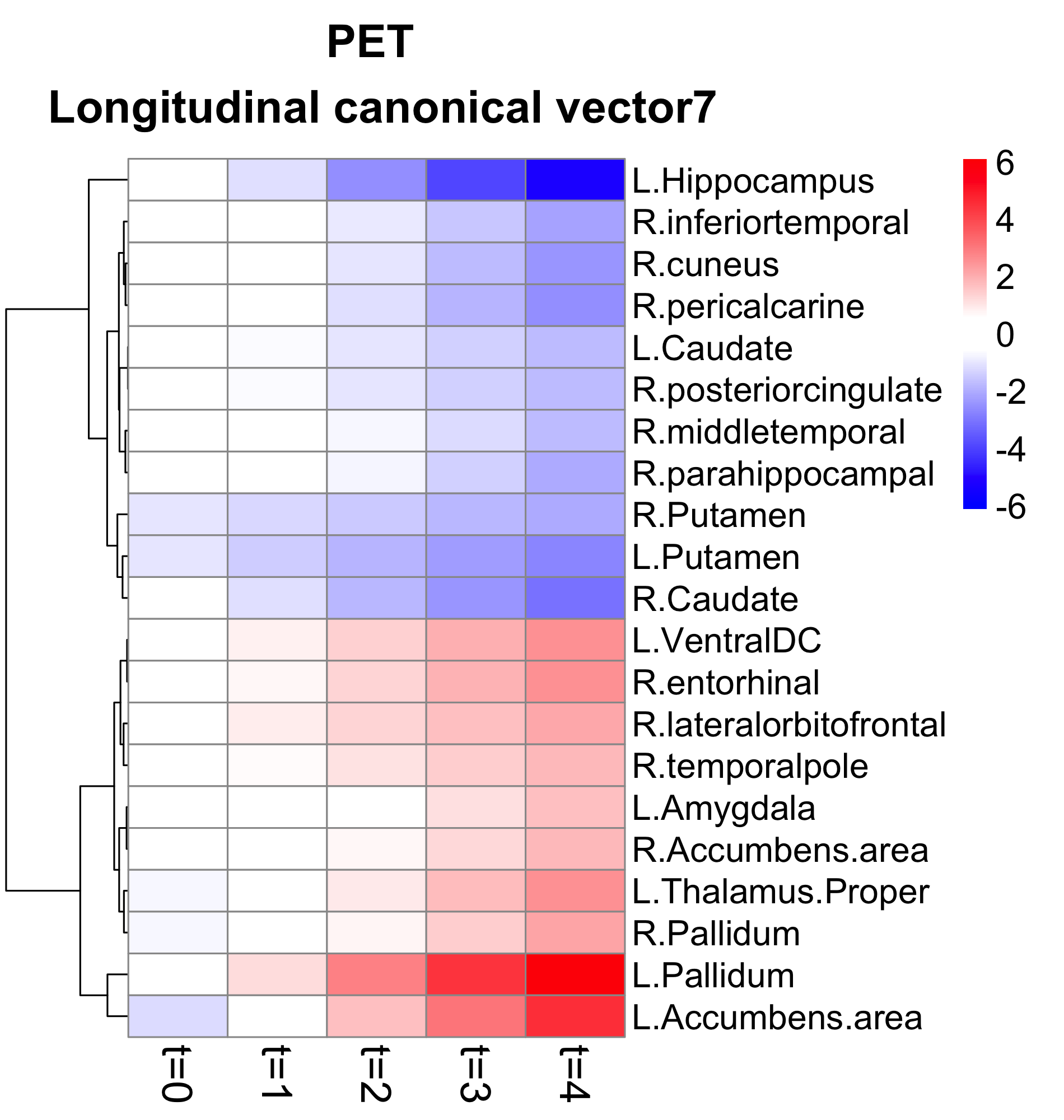}
\caption{LCCA: Pairs of the longitudinal canonical vectors are displayed at time=0,1,2,3,4.}\label{fig:cw2}
\end{figure}

The longitudinal CCA identified 7 canonical variates. The significant canonical correlation coefficients are displayed in Figure \ref{fig:cc}. 
The longitudinal canonical vectors were reorganized as a function of time $ \left(\mathbf{\Phi}_0^{X} + t\mathbf{\Phi}_1^{X}\right) \mathbf{u}_{m} $ and 
$ \left(\mathbf{\Phi}_0^{Y} + s\mathbf{\Phi}_1^{Y}\right) \mathbf{v}_{m} $ for
\(m=1,...,min(N_X,N_Y)\) are displayed in Figures \ref{fig:cw1} and \ref{fig:cw2} at time=0,1,2,3,4. For visualization, the vectors are standardized with the total variance of the vectors, and the ROIs with at least one element of the vector over 1.64 are included in the heatmaps. The rows are clustered using hierarchical cluster analysis with complete linkage to visualize which variables behave similarly over time. Figures \ref{fig:cw1} and \ref{fig:cw2} include the longitudinal canonical weights of the six longitudinal canonical variates that are associated with either baseline diagnosis status or AD transition. We also performed naive approach and compared its performance. 

\begin{figure}
\includegraphics[width=0.32\textwidth]{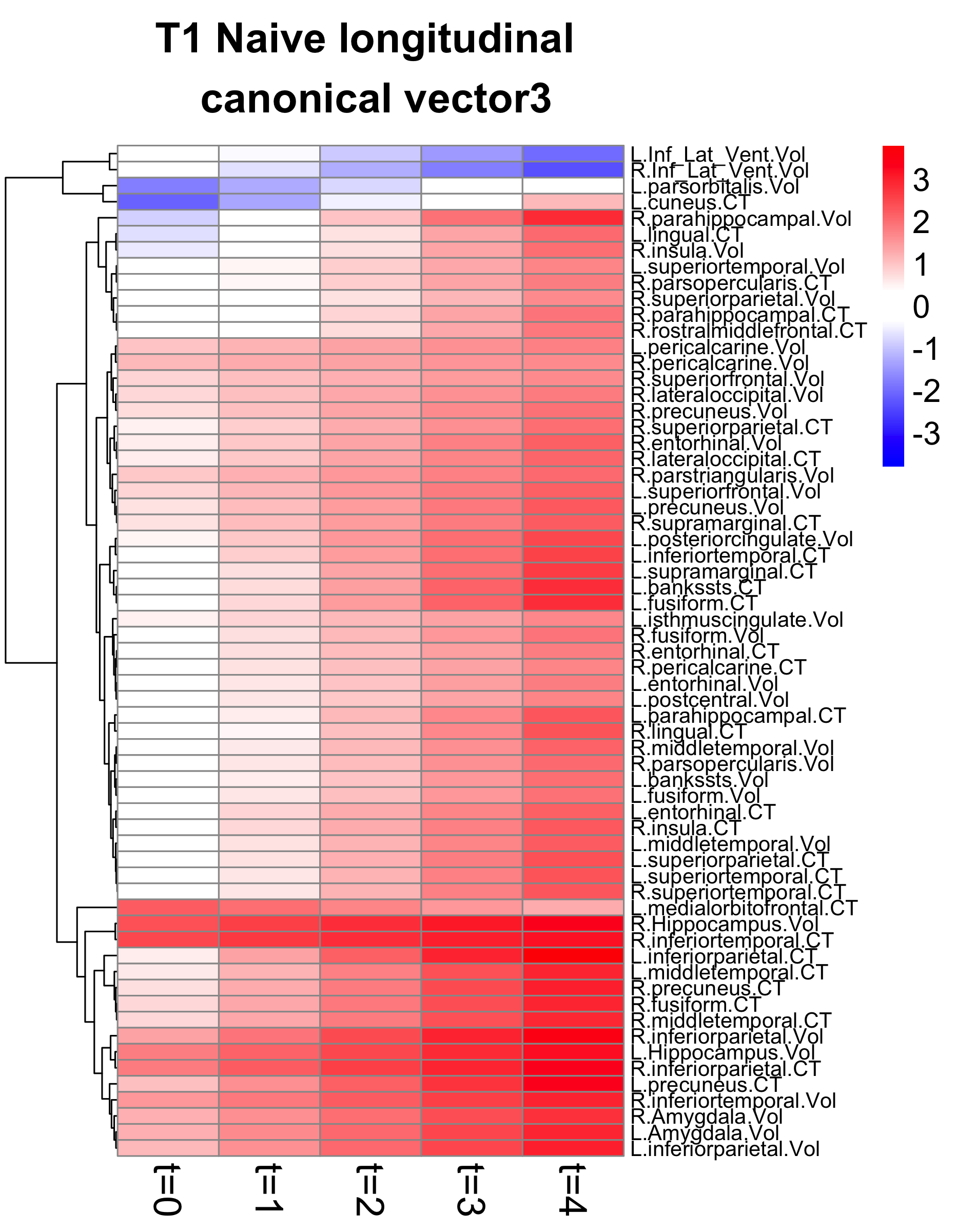}
\includegraphics[width=0.32\textwidth]{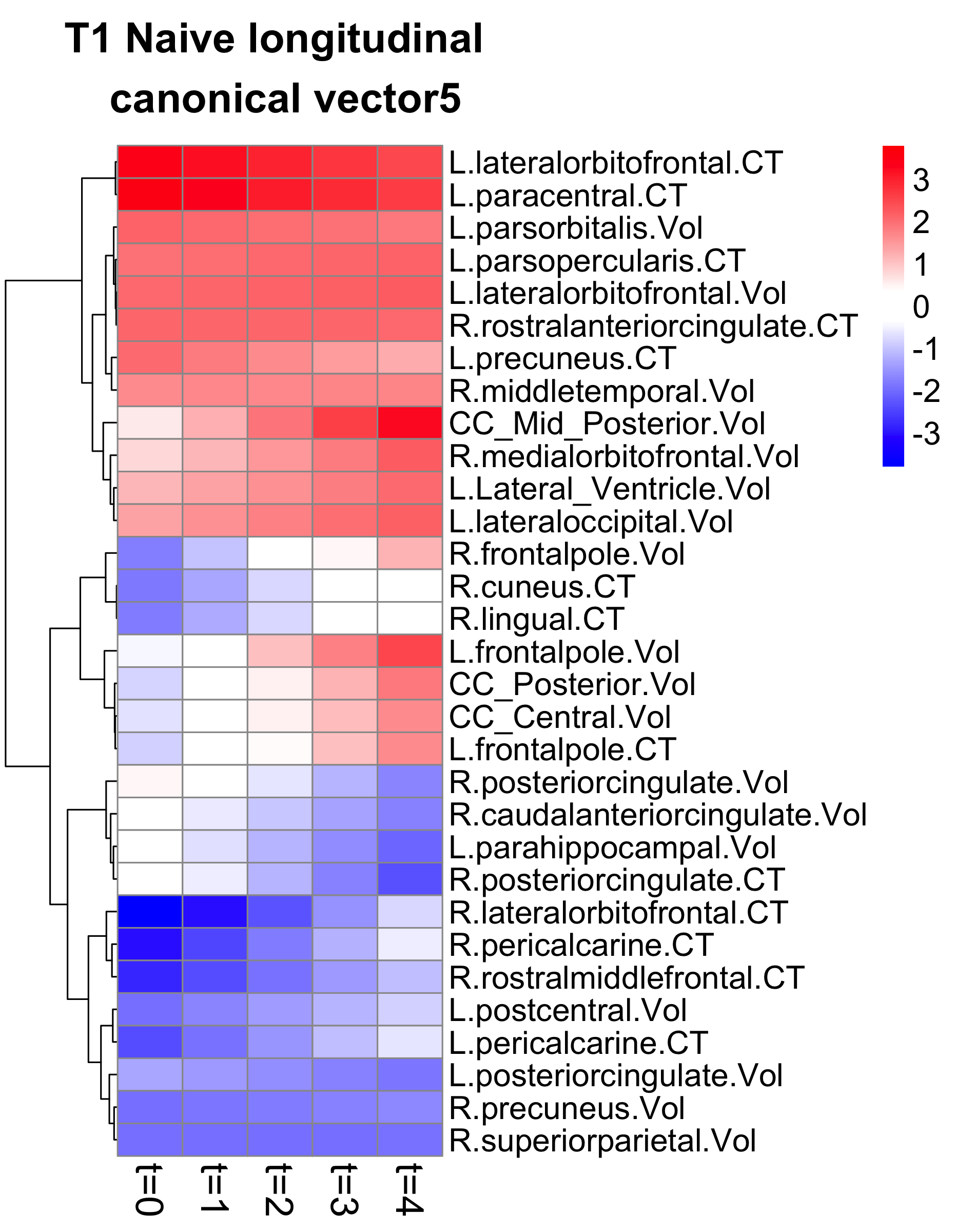}
\includegraphics[width=0.32\textwidth]{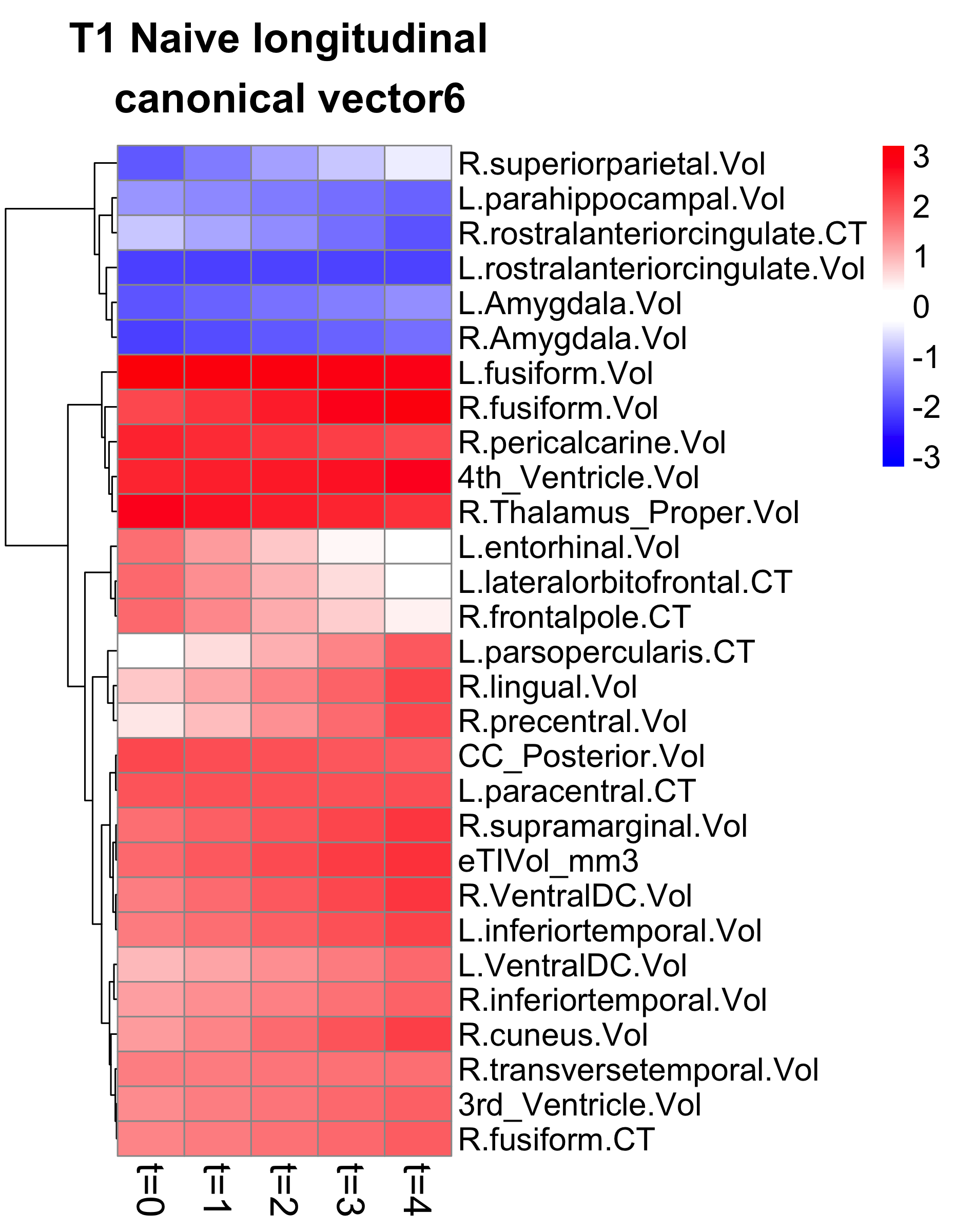}\\
\includegraphics[width=0.32\textwidth]{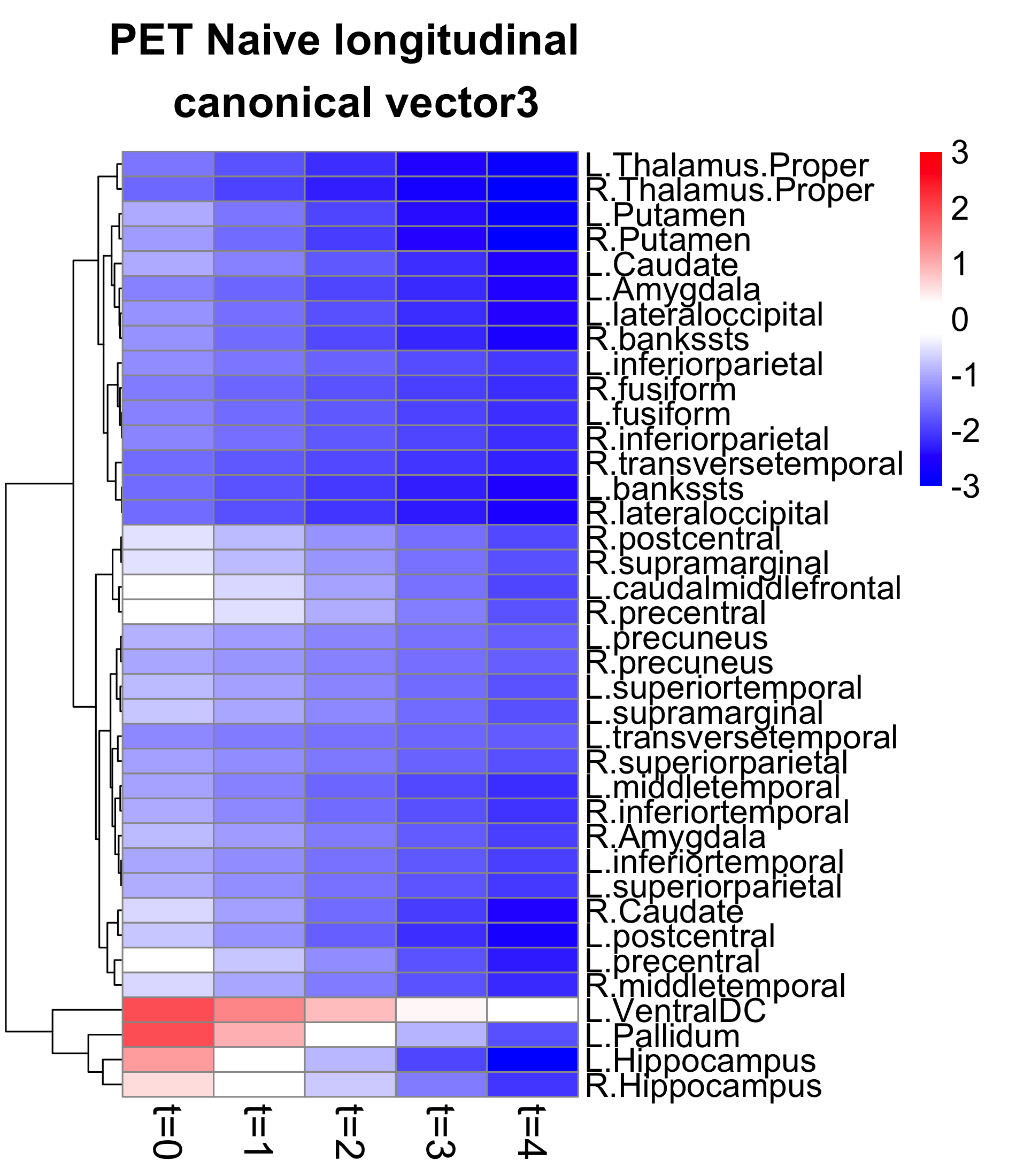}
\includegraphics[width=0.32\textwidth]{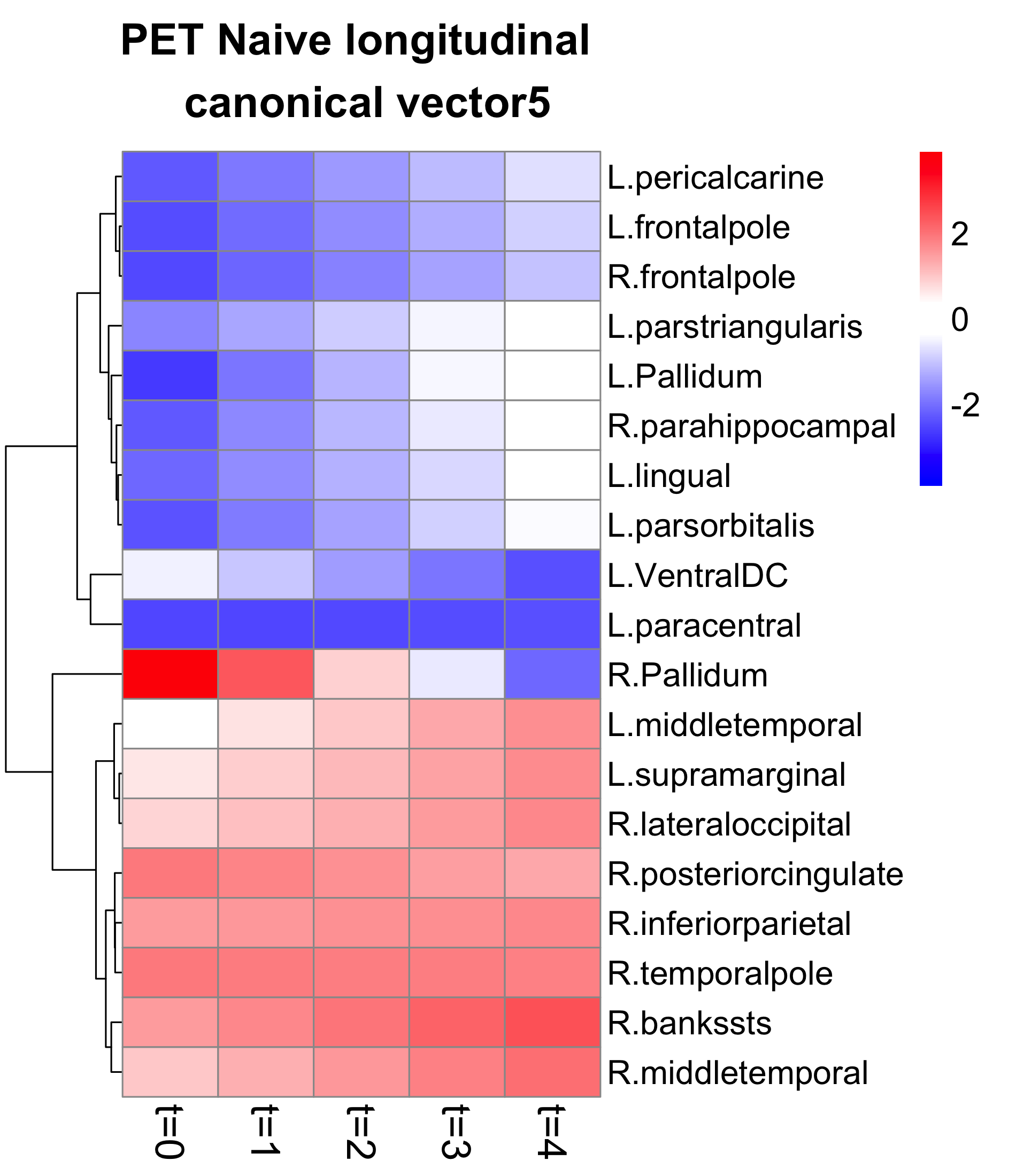}
\includegraphics[width=0.32\textwidth]{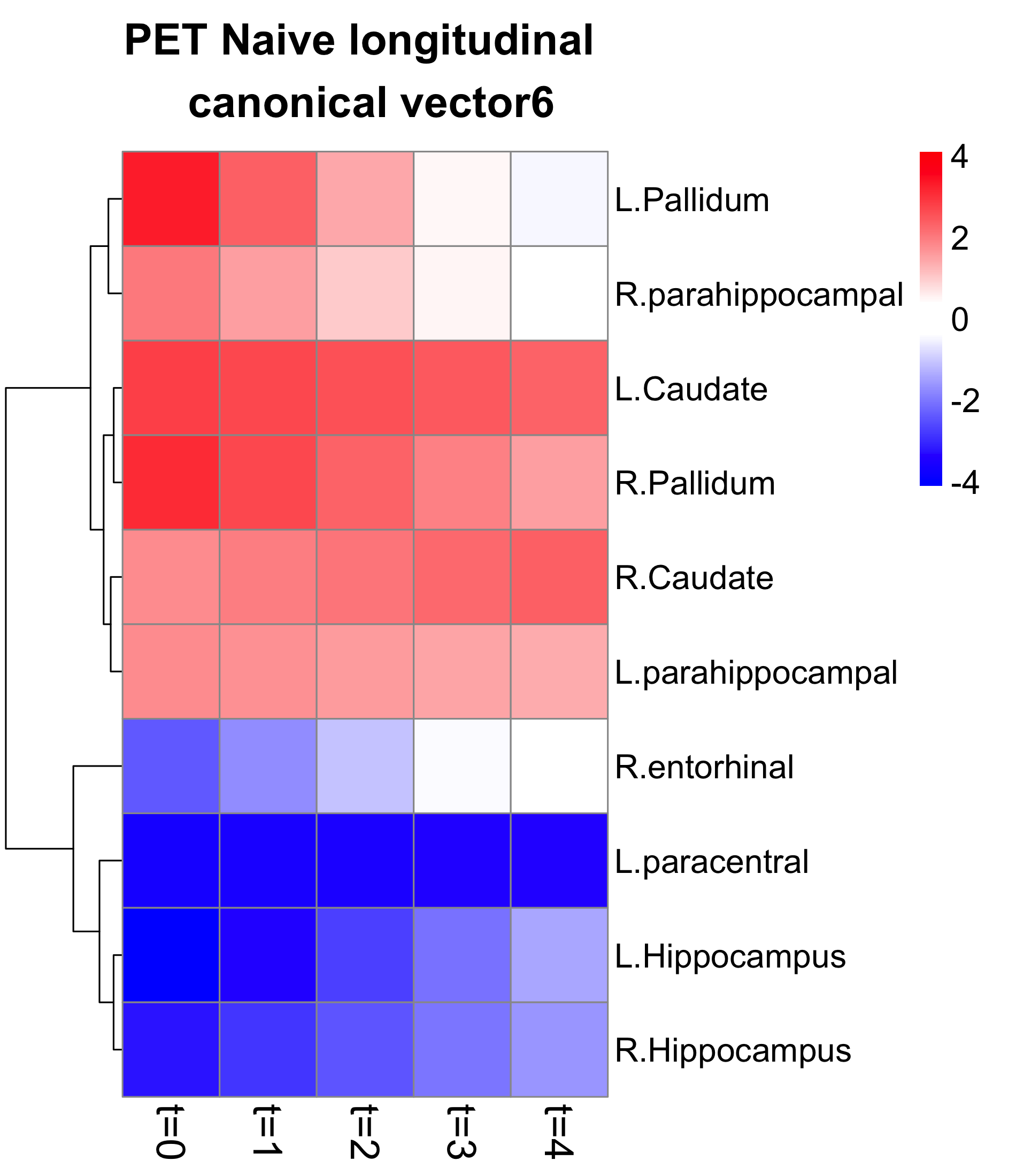}
\caption{Naive: Pairs of the longitudinal canonical vectors are displayed at time=0,1,2,3,4.}\label{fig:cw3}
\end{figure}

\begin{figure}
\includegraphics[width=0.32\textwidth]{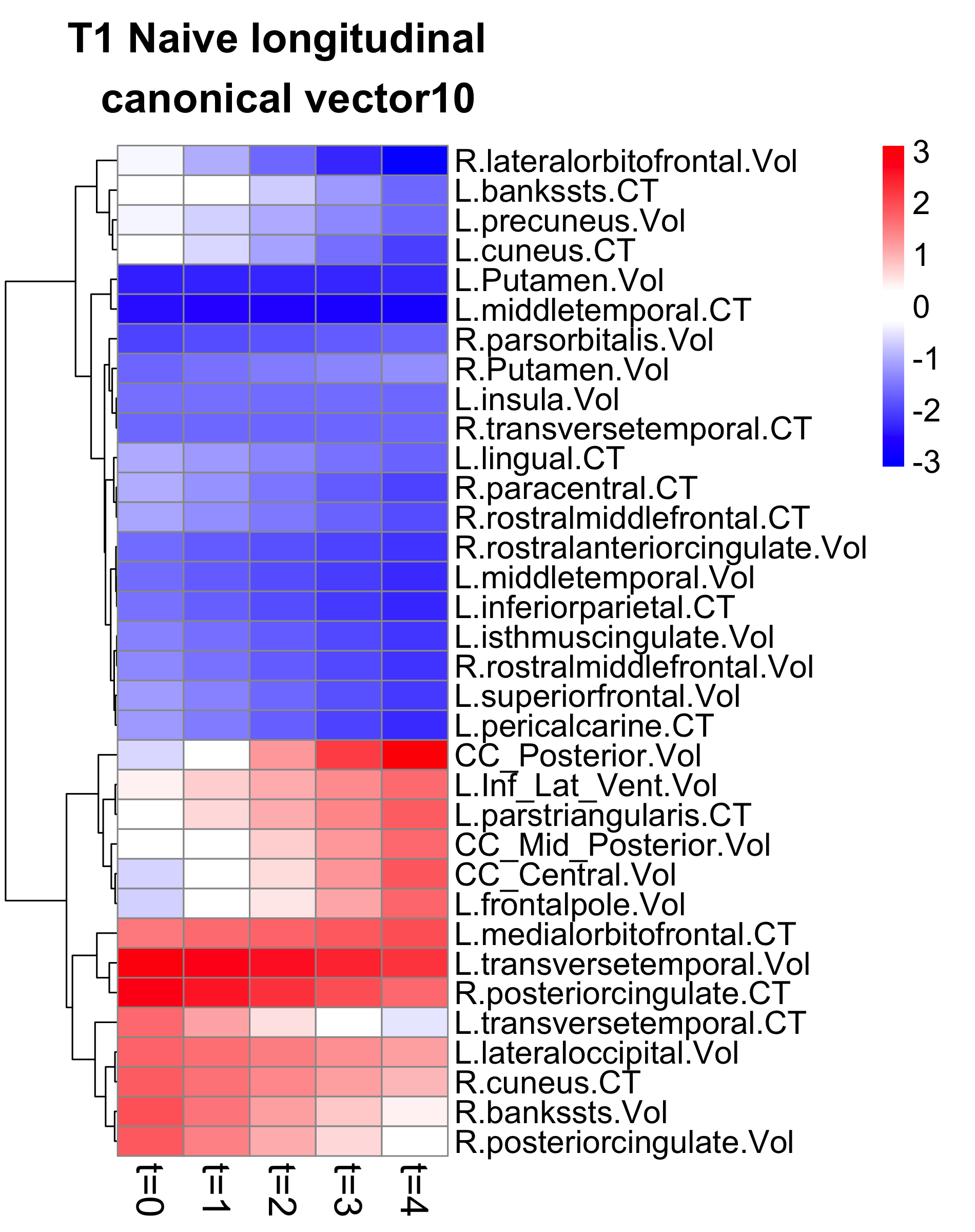}
\includegraphics[width=0.32\textwidth]{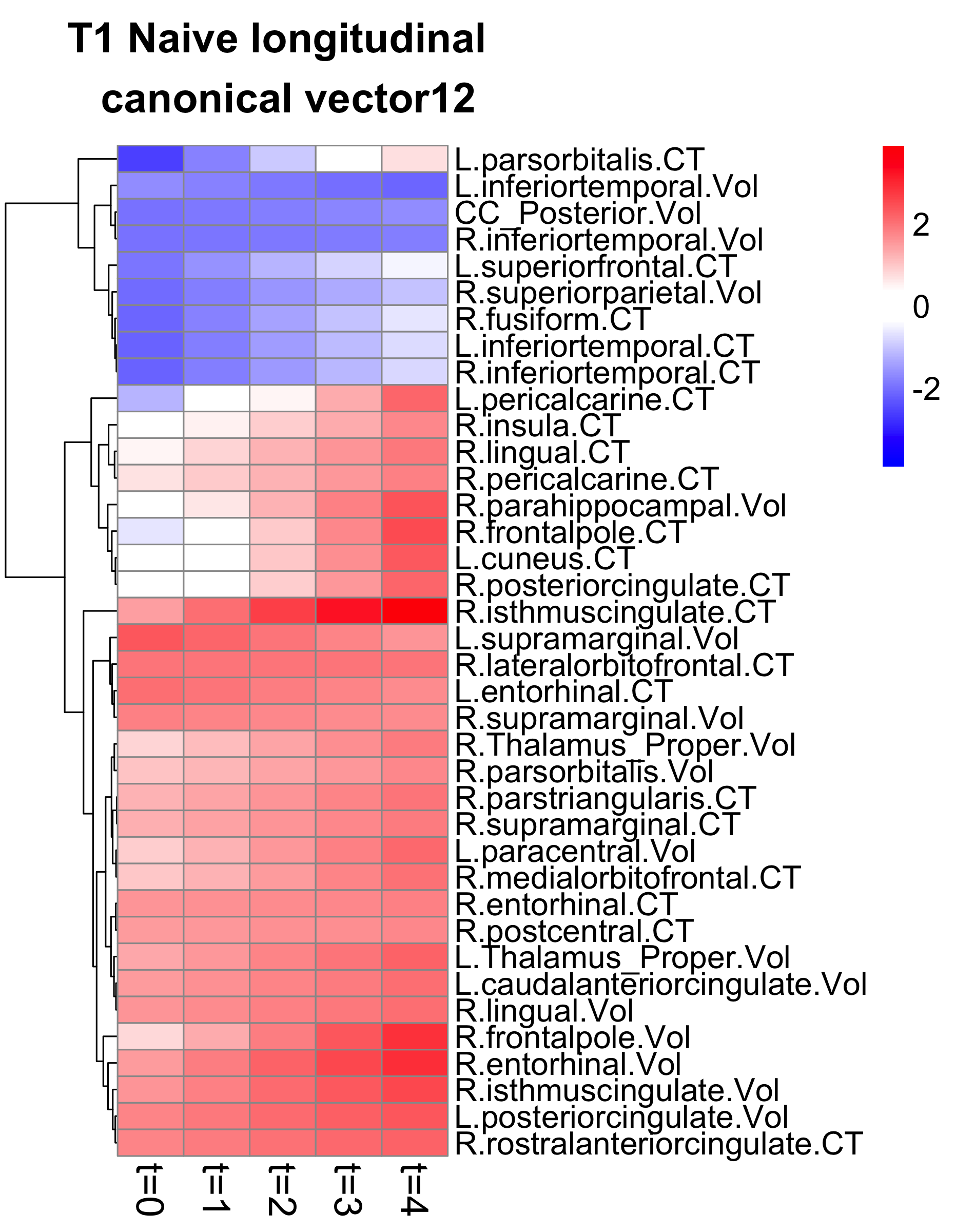}\\
\includegraphics[width=0.32\textwidth]{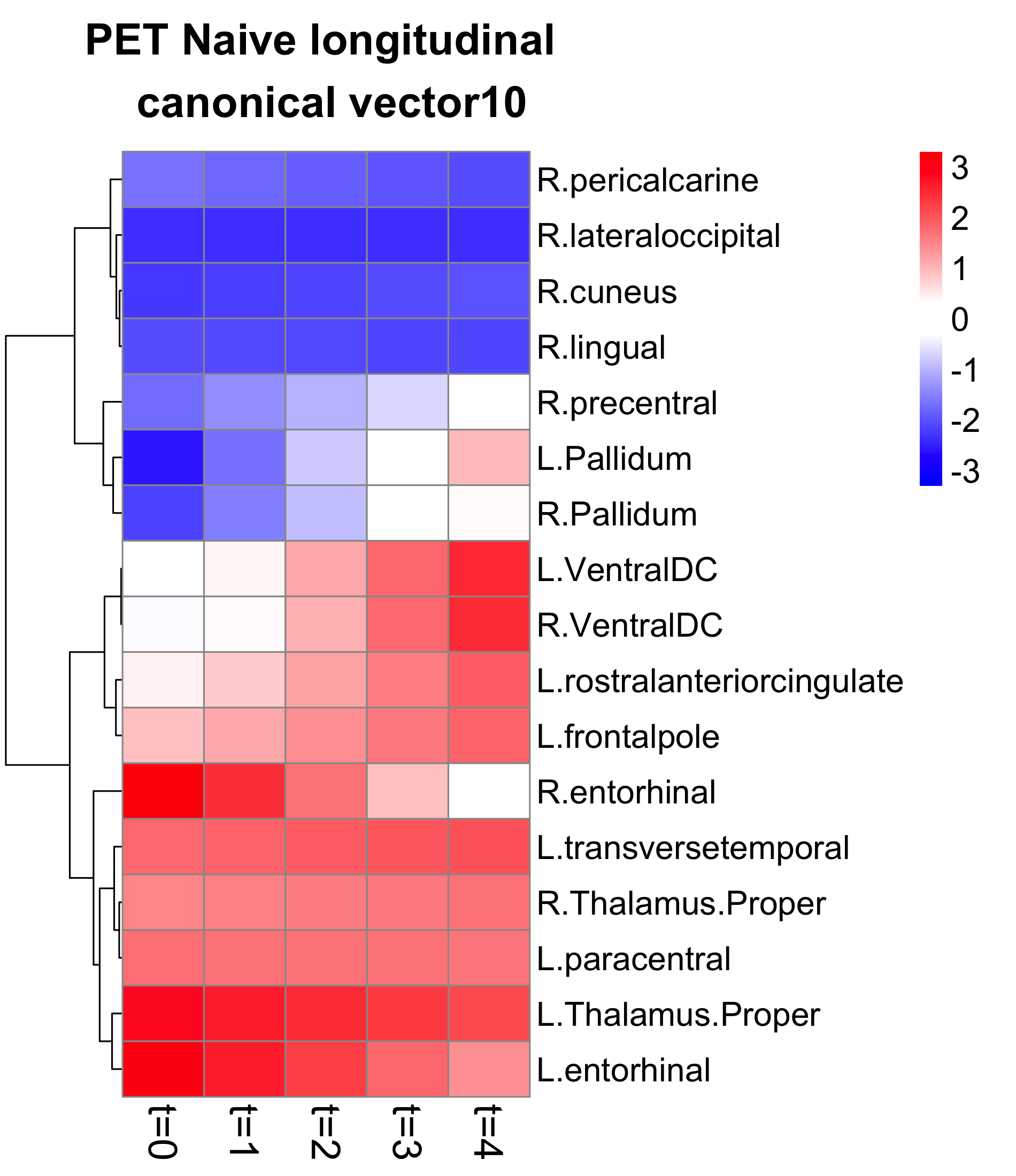}
\includegraphics[width=0.32\textwidth]{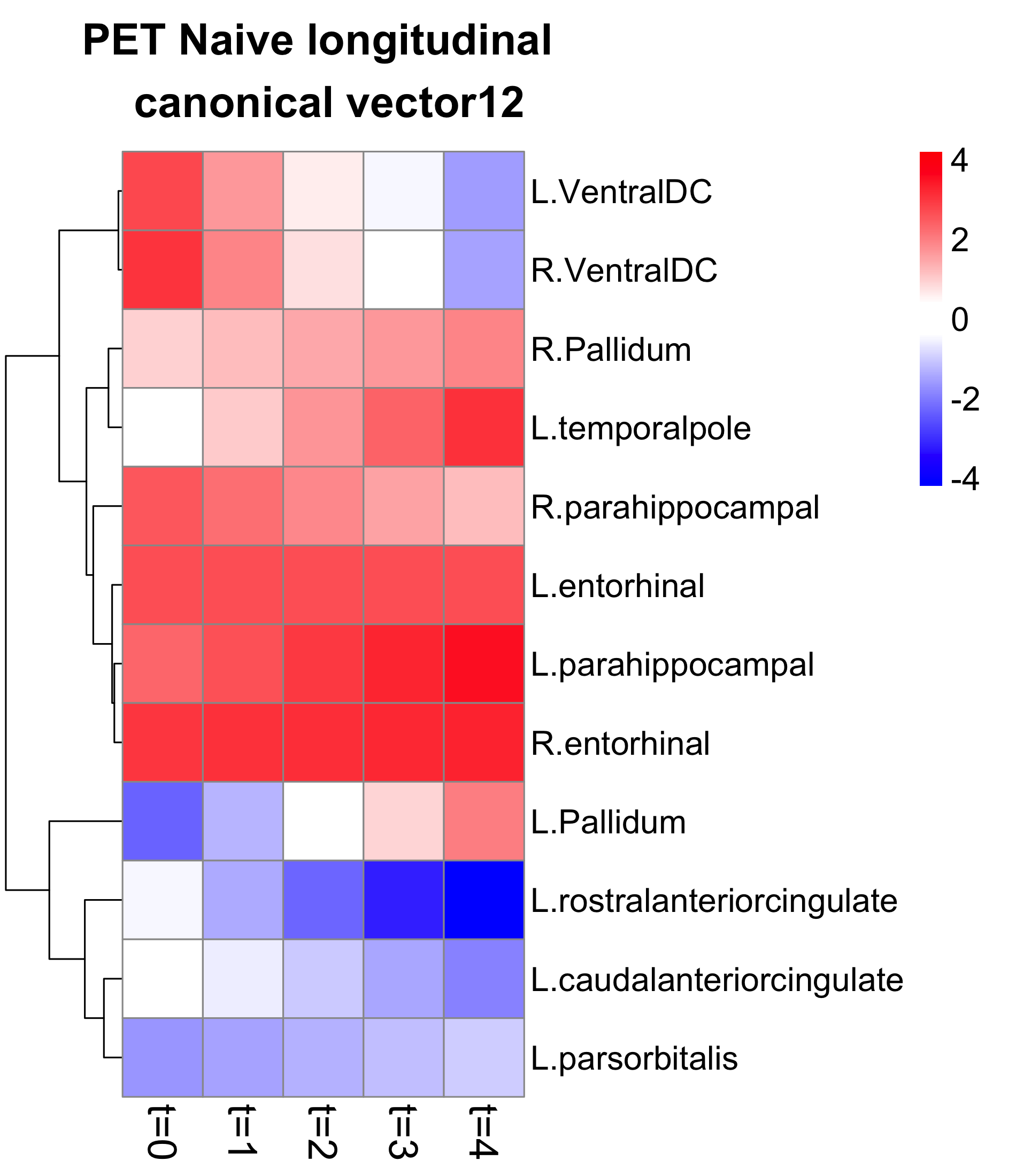}

\caption{Naive: Pairs of the longitudinal canonical vectors are displayed at time=0,1,2,3,4.}\label{fig:cw4}
\end{figure}

\begin{table}
\caption{\label{tab2} ANOVA of diagnosis group comparison and AD conversion with age, sex and education adjusted for LCCA.}
\begin{tabular}{*{10}{c}} \hline \hline
	&		&		&	T1		&		&		&		&	PET		&		&		\\ \hline
	&	LCV	&	F	&	p		&	$\eta^2_p$	&	outliers	&	F	&	p		&	$\eta^2_p$	&	outliers	\\ \hline
Group	&	1	&	3.47	&	0.032		&	0.010	&	0	&	0.57	&	0.566		&	0.002	&	1	\\
Comparison	&	2	&	19.83	&	$<$0.001	$^*$	&	0.056	&	0	&	38.73	&	$<$0.001	$^*$	&	0.103	&	0	\\
	&	3	&	2.13	&	0.120		&	0.006	&	1	&	1.08	&	0.342		&	0.003	&	2	\\
	&	4	&	4.86	&	0.008	$^*$	&	0.014	&	1	&	3.34	&	0.036		&	0.010	&	2	\\
	&	5	&	0.95	&	0.386		&	0.003	&	0	&	11.60	&	$<$0.001	$^*$	&	0.033	&	2	\\
	&	6	&	2.42	&	0.090		&	0.007	&	0	&	2.46	&	0.086		&	0.007	&	1	\\
	&	7	&	8.26	&	$<$0.001	$^*$	&	0.024	&	0	&	3.86	&	0.022		&	0.011	&	2	\\ \hline
AD	&	1	&	11.74	&	0.001	$^*$	&	0.018	&	0	&	5.76	&	0.017	$^*$	&	0.009	&	1	\\
Conversion	&	2	&	102.21	&	$<$0.001	$^*$	&	0.136	&	0	&	164.90	&	$<$0.001	$^*$	&	0.202	&	0	\\
	&	3	&	1.51	&	0.220		&	0.002	&	0	&	0.08	&	0.773		&	0.000	&	1	\\
	&	4	&	6.11	&	0.014	$^*$	&	0.009	&	0	&	7.70	&	0.006	$^*$	&	0.012	&	1	\\
	&	5	&	0.00	&	0.972		&	0.000	&	0	&	0.75	&	0.386		&	0.001	&	1	\\
	&	6	&	3.19	&	0.074		&	0.005	&	0	&	7.05	&	0.008	$^*$	&	0.011	&	0	\\
	&	7	&	2.80	&	0.095		&	0.004	&	0	&	3.44	&	0.064		&	0.005	&	2	\\ \hline \hline
		\end{tabular}
\end{table}

\begin{table}
\caption{\label{tab3} ANOVA of diagnosis group comparison and AD conversion with age, sex and education adjusted for the Naive approach. $^*$: adjusted p$<$0.05.}
\begin{tabular}{*{10}{c}} \hline \hline
	&		&		&	T1		&		&		&		&	PET		&		&		\\ \hline
	&	CV	&	F	&	p		&	$\eta^2_p$	&	outliers	&	F	&	p		&	$\eta^2_p$	&	outliers	\\ \hline
Group	&	1	&	0.02	&	0.984		&	0.000	&	3	&	3.23	&	0.040		&	0.010	&	4	\\
Comparison	&	2	&	4.24	&	0.015		&	0.012	&	3	&	1.18	&	0.308		&	0.004	&	4	\\
	&	3	&	47.83	&	$<$0.001	$^*$	&	0.124	&	1	&	51.68	&	$<$0.001	$^*$	&	0.133	&	0	\\
	&	4	&	0.75	&	0.472		&	0.002	&	0	&	1.92	&	0.148		&	0.006	&	1	\\
	&	5	&	8.74	&	$<$0.001	$^*$	&	0.025	&	1	&	2.17	&	0.115		&	0.006	&	1	\\
	&	6	&	2.03	&	0.132		&	0.006	&	2	&	8.47	&	$<$0.001	$^*$	&	0.025	&	1	\\
	&	7	&	0.24	&	0.787		&	0.001	&	1	&	1.19	&	0.304		&	0.004	&	3	\\
	&	8	&	0.78	&	0.458		&	0.002	&	0	&	1.60	&	0.202		&	0.005	&	1	\\
	&	9	&	0.53	&	0.589		&	0.002	&	0	&	1.71	&	0.183		&	0.005	&	2	\\
	&	10	&	0.25	&	0.780		&	0.001	&	0	&	1.47	&	0.230		&	0.004	&	0	\\
	&	11	&	1.00	&	0.367		&	0.003	&	0	&	1.71	&	0.182		&	0.005	&	2	\\
	&	12	&	1.17	&	0.310		&	0.003	&	0	&	4.96	&	0.007	$^*$	&	0.015	&	0	\\ \hline
AD	&	1	&	0.64	&	0.424		&	0.001	&	2	&	1.89	&	0.170		&	0.003	&	3	\\
Conversion	&	2	&	0.80	&	0.372		&	0.001	&	2	&	0.94	&	0.334		&	0.001	&	3	\\
	&	3	&	248.28	&	$<$0.001	$^*$	&	0.276	&	1	&	226.29	&	$<$0.001	$^*$	&	0.258	&	0	\\
	&	4	&	0.47	&	0.494		&	0.001	&	0	&	1.00	&	0.317		&	0.002	&	1	\\
	&	5	&	0.23	&	0.635		&	0.000	&	1	&	0.03	&	0.874		&	0.000	&	1	\\
	&	6	&	8.72	&	0.003	$^*$	&	0.013	&	2	&	13.11	&	$<$0.001	$^*$	&	0.020	&	1	\\
	&	7	&	0.38	&	0.541		&	0.001	&	1	&	0.81	&	0.368		&	0.001	&	3	\\
	&	8	&	0.02	&	0.886		&	0.000	&	0	&	3.43	&	0.064		&	0.005	&	1	\\
	&	9	&	0.09	&	0.769		&	0.000	&	0	&	1.16	&	0.282		&	0.002	&	2	\\
	&	10	&	5.84	&	0.016		&	0.009	&	0	&	11.10	&	0.001	$^*$	&	0.017	&	0	\\
	&	11	&	0.01	&	0.910		&	0.000	&	0	&	0.14	&	0.713		&	0.000	&	2	\\
	&	12	&	0.31	&	0.578		&	0.001	&	0	&	1.21	&	0.271		&	0.002	&	0	\\ \hline \hline
	\end{tabular}
\end{table}	

\begin{figure}
\includegraphics[width=0.5\textwidth]{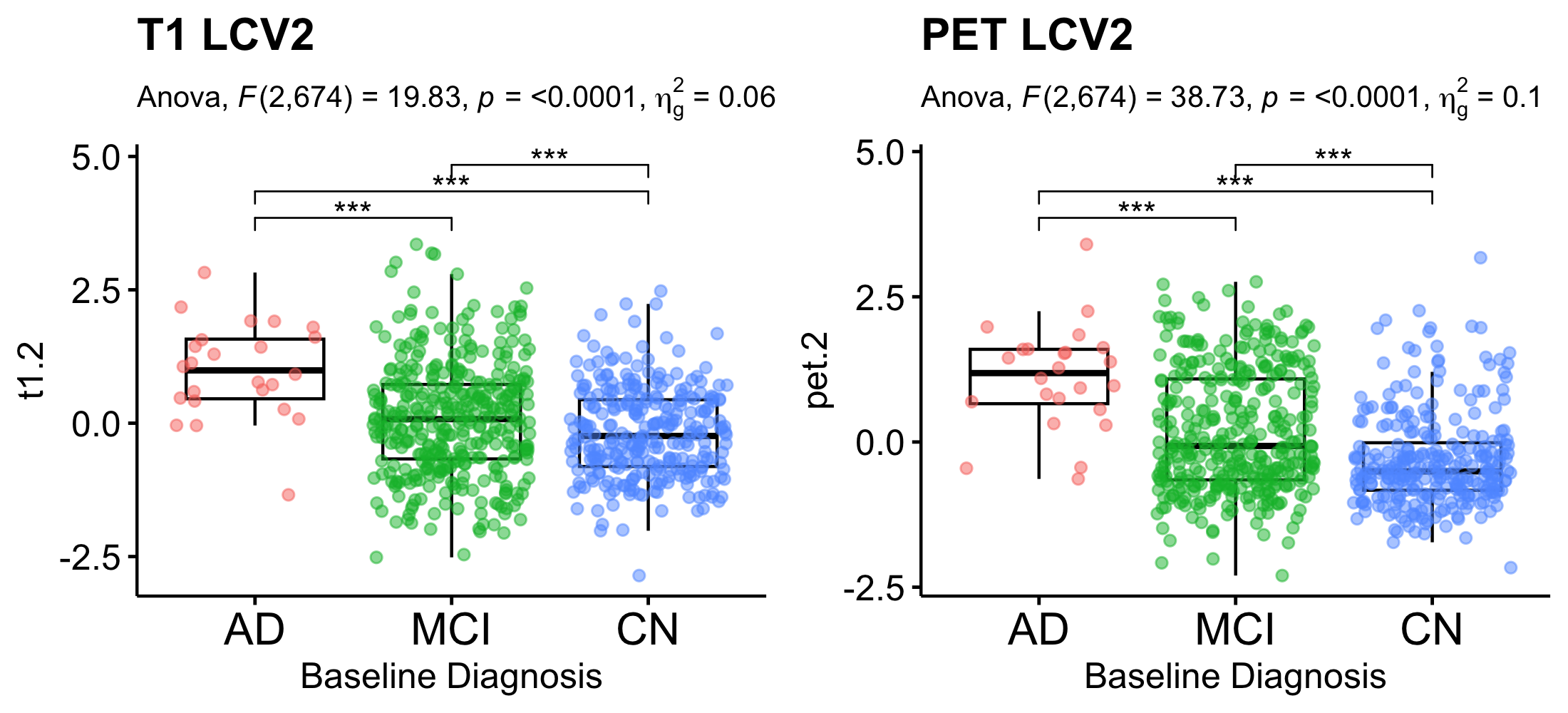}
\includegraphics[width=0.5\textwidth]{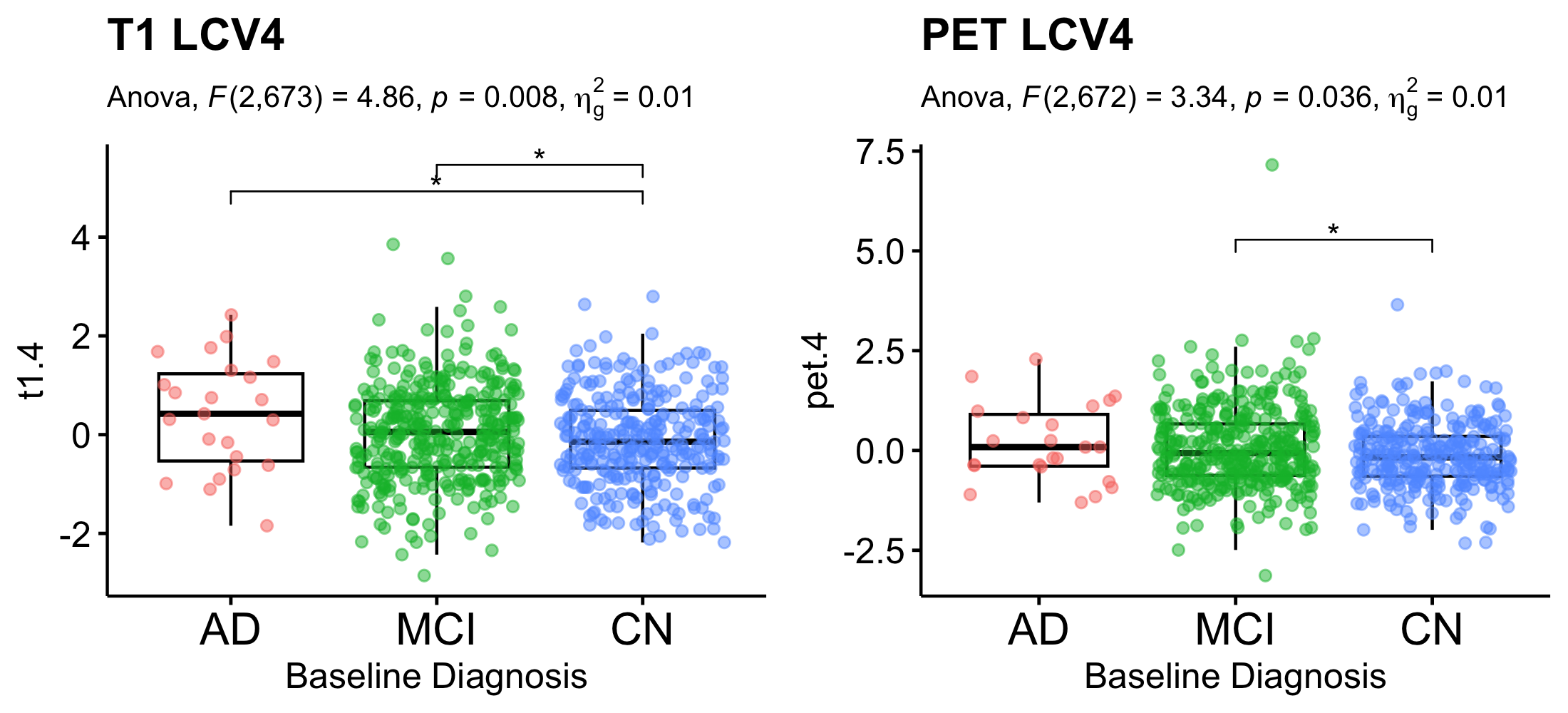}\\
\includegraphics[width=0.5\textwidth]{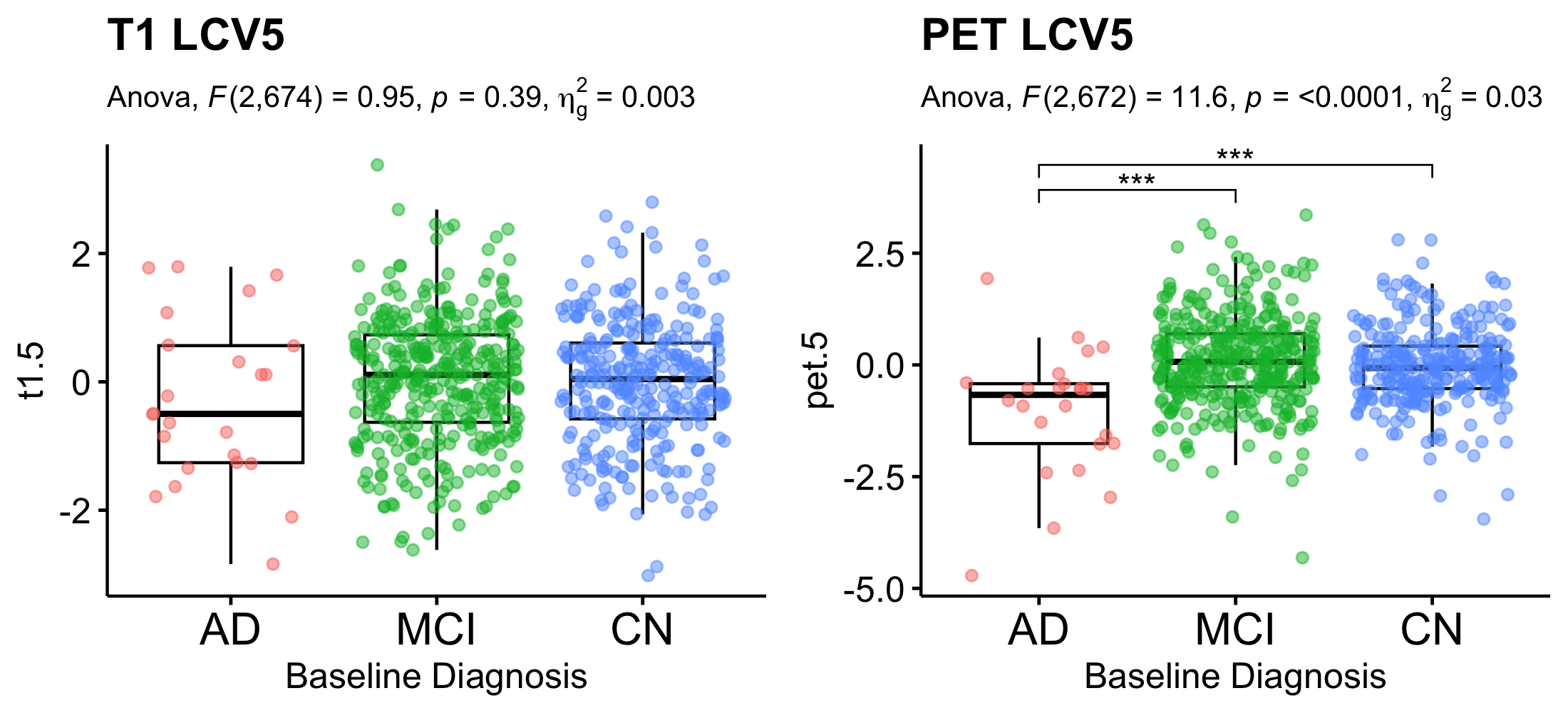}
\includegraphics[width=0.5\textwidth]{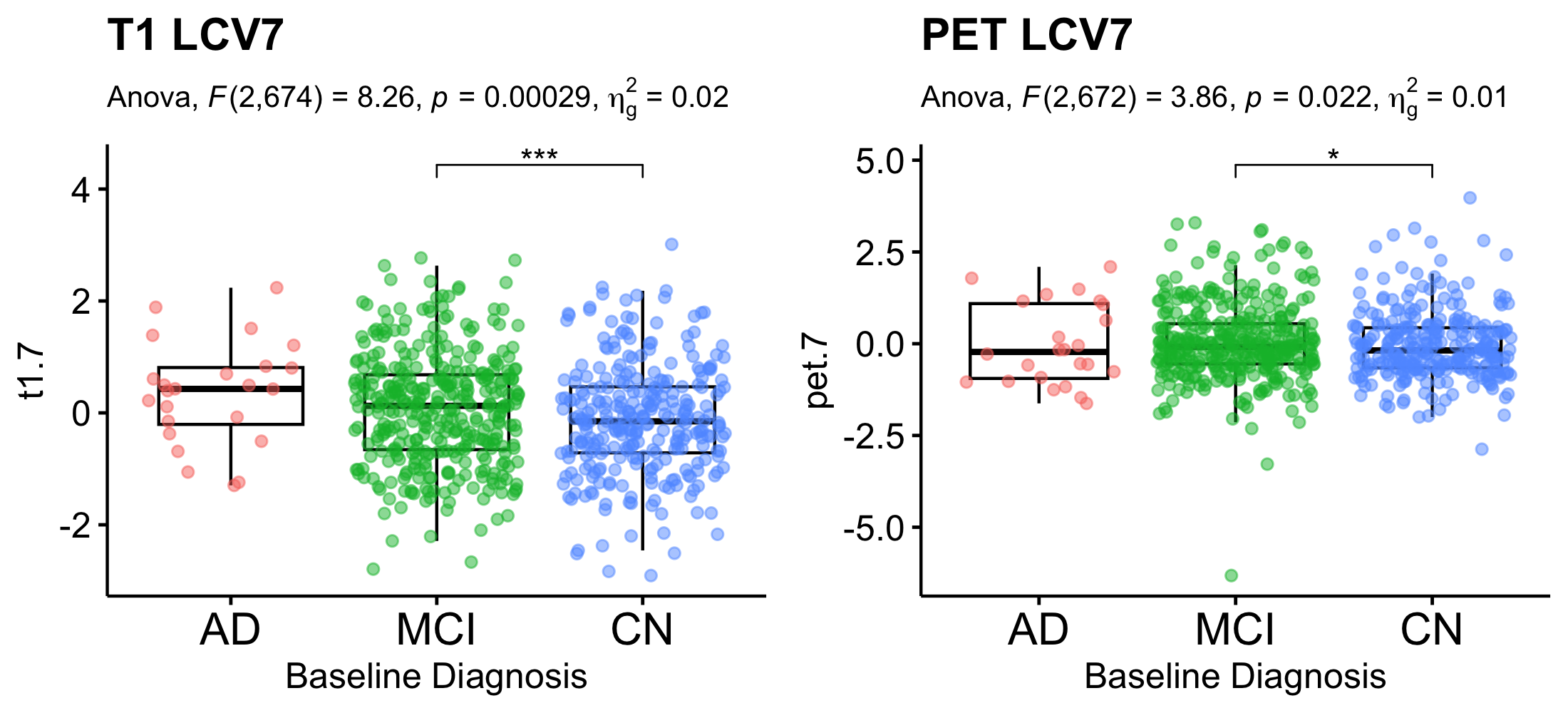}
\caption{LCCA: Baseline diagnosis status comparison. Significant CVs are reported. Age, sex and education were adjusted for statistical testing. For post-hoc pair-wise group comparison, least squared mean differences were computed, and unadjusted p-values are reported. 
}\label{fig:ccbox}
\end{figure}

Further, we investigated the longitudinal canonical variates (CV). 
First, we compared CVs by each subject's baseline clinical status. 
For group comparison, linear regression was conducted with age, sex and years of education as covariates, followed by 
multiple comparison correction controlling for false discovery rate \citep{benjamini1995controlling}. Table \ref{tab2} reports the 
F-statistics, unadjusted p-values and multiple comparison corrected p-values. 
For all analyses, we excluded extreme outliers beyond 3 IQR from the first and third quartiles. For post-hoc pair-wise group comparison, least squared mean differences were computed. Figure \ref{fig:ccbox} shows jittered boxplots and pair-wise comparison results with unadjusted p-values for the CVs with significant group difference after multiple comparison correction. 

The CV 2, 4, and 7 of T1 and CVs 2 and 5 of PET 
showed group differences after multiple comparison correction. The CV2 indicates that AD patients at baseline 
showed atrophy in bilateral hippocampus and amygdala. The amyloid deposition showed 
a decreased pattern in bilateral Pallidum, and increasing amyloid deposition in the cortical areas 
(right lingual, right pericalcarine, right fusiform, and inferior temporal cortices). The CV4 of T1 showed  the enlarged volume in right caudal middle frontal gyus and longitudinal atrophy in right pars orbitalis in AD and MCI comparing to CN, while PET did not showed any differences.
The CV5 of T1 did not show differences across baseline clinical status. 
The AD patients showed brain atrophy in the left Accumbens area volume, and 
right paracentral thickness, smaller volume in the left temporal transverse volume and 
right caudal middle frontal volume, while thicker thickness in the left isthmus cingulate 
and right lateral orbitofrontal gyrus. The CV5 of av45 showed a longitudinal increase in 
the left hippocampus and right Pallidum and left paracentral area in AD participants,
while MCI and CN participants did not show that pattern on average.  CV7 showed enlarged volumes in right lingual gyrus, but atrophy in left caudal ACC and pars orbitals, while no significant difference in amyloid deposition. The naive method identified CV 3 and 5 with significant T1 differences and CVs 9 and 12 for PET. Unlike LCCA, naive method did not identify CVs showed group differences between diagnosis status in both T1 and PET. For AD transition, only CV3 T1 and CV9 PET showed significant group difference. 

\begin{figure}
\includegraphics[width=0.5\textwidth]{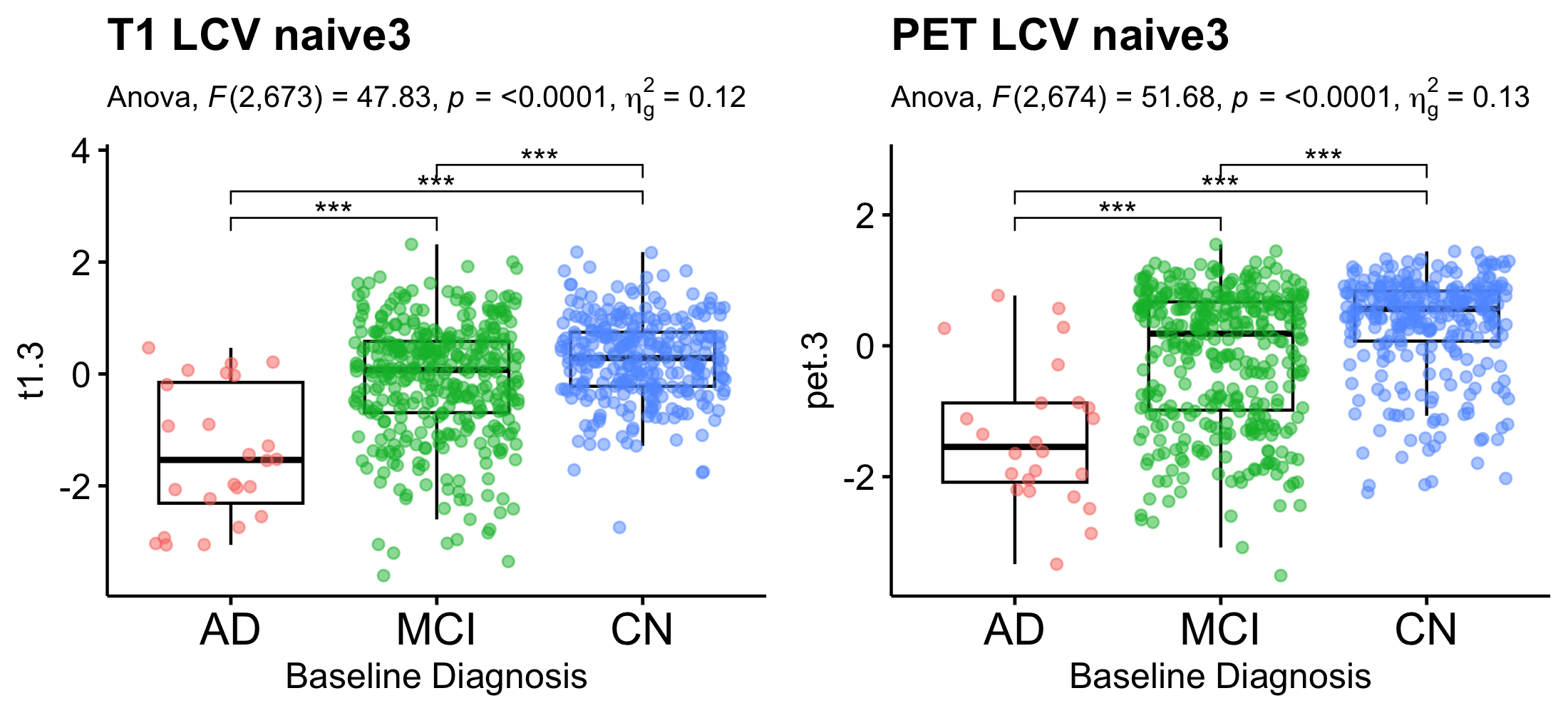}
\includegraphics[width=0.5\textwidth]{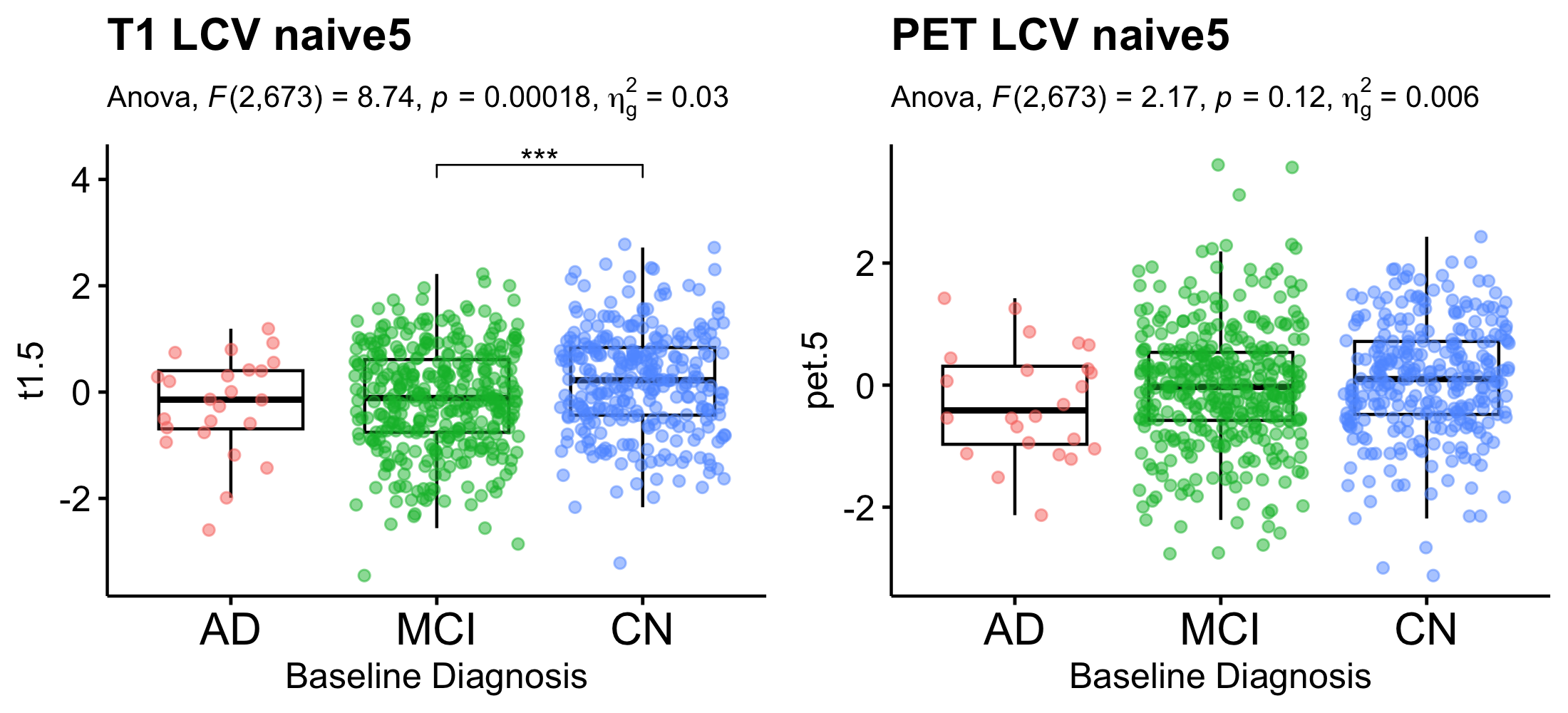}\\
\includegraphics[width=0.5\textwidth]{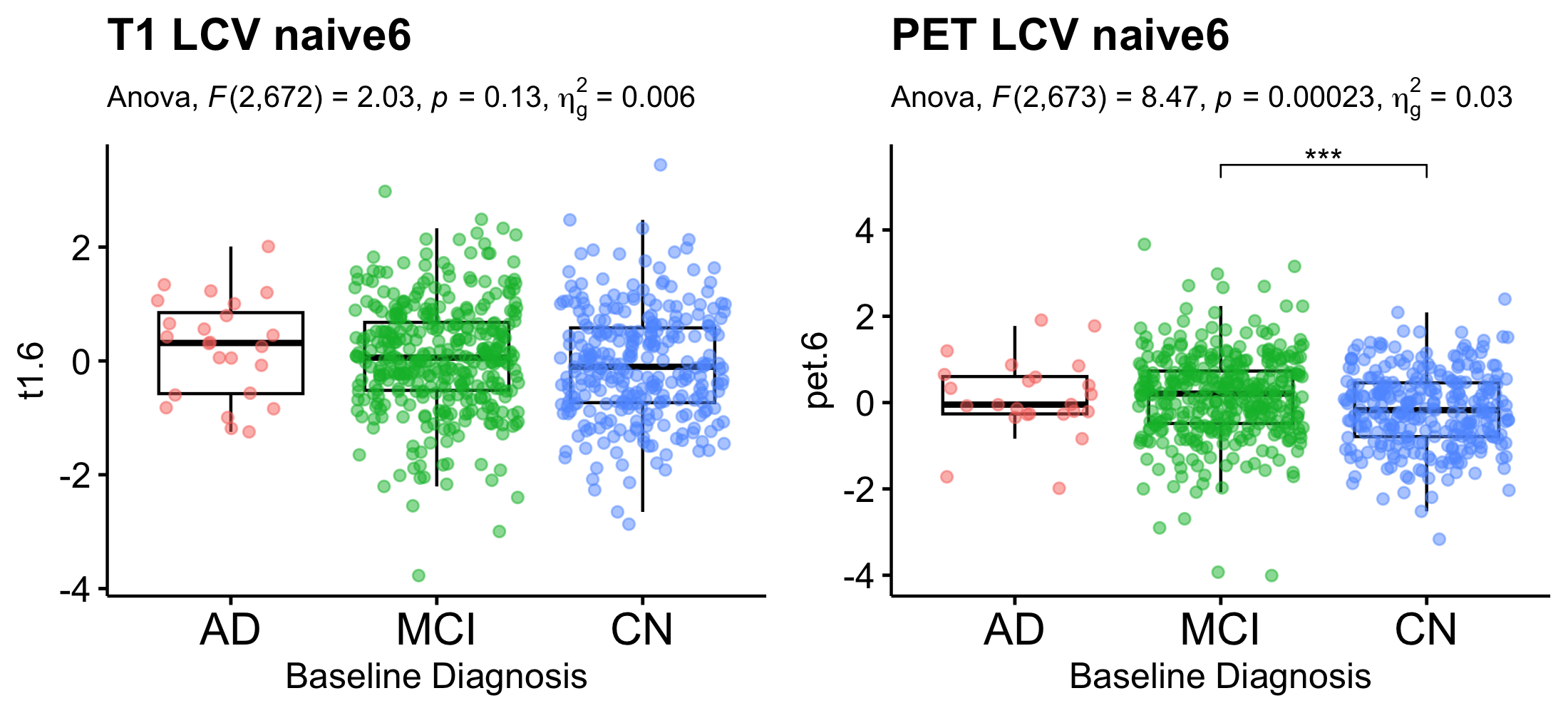}
\includegraphics[width=0.5\textwidth]{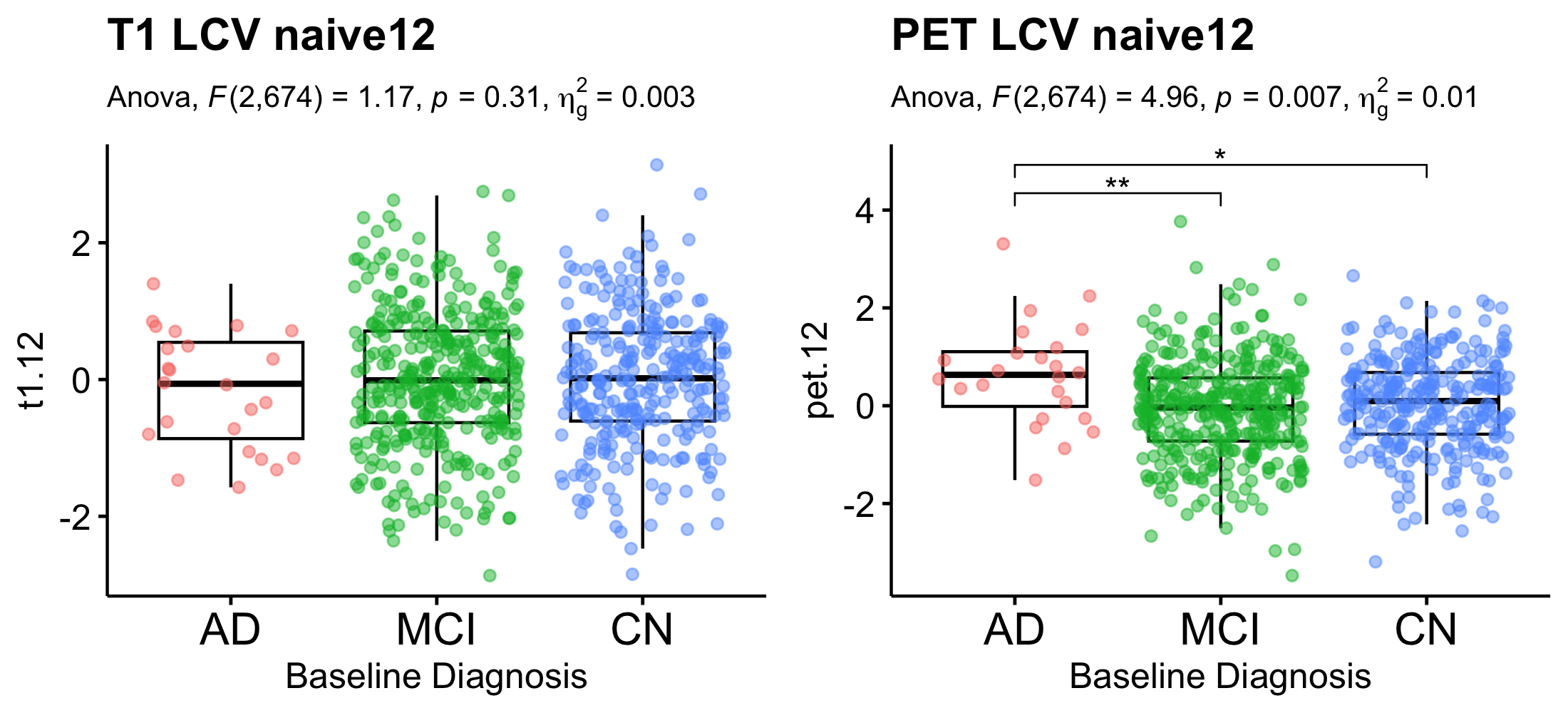}
\caption{Naive Approach: Baseline diagnosis status comparison. Significant CVs are reported. Age, sex and education were adjusted for statistical testing. For post-hoc pair-wise group comparison, least squared mean differences were computed, and unadjusted p-values are reported. 
}\label{fig:ccbox2}
\end{figure}

\begin{figure}
\includegraphics[width=0.5\textwidth]{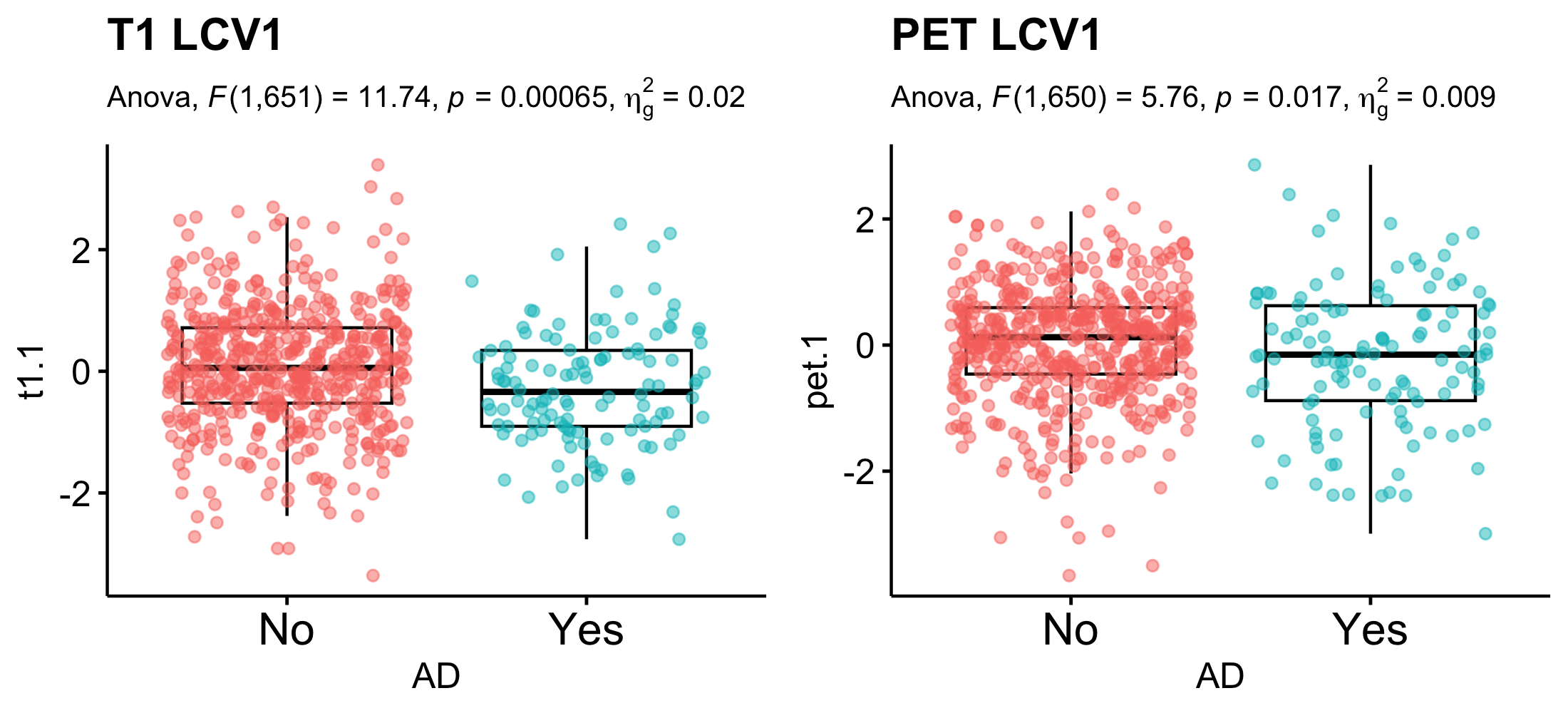}
\includegraphics[width=0.5\textwidth]{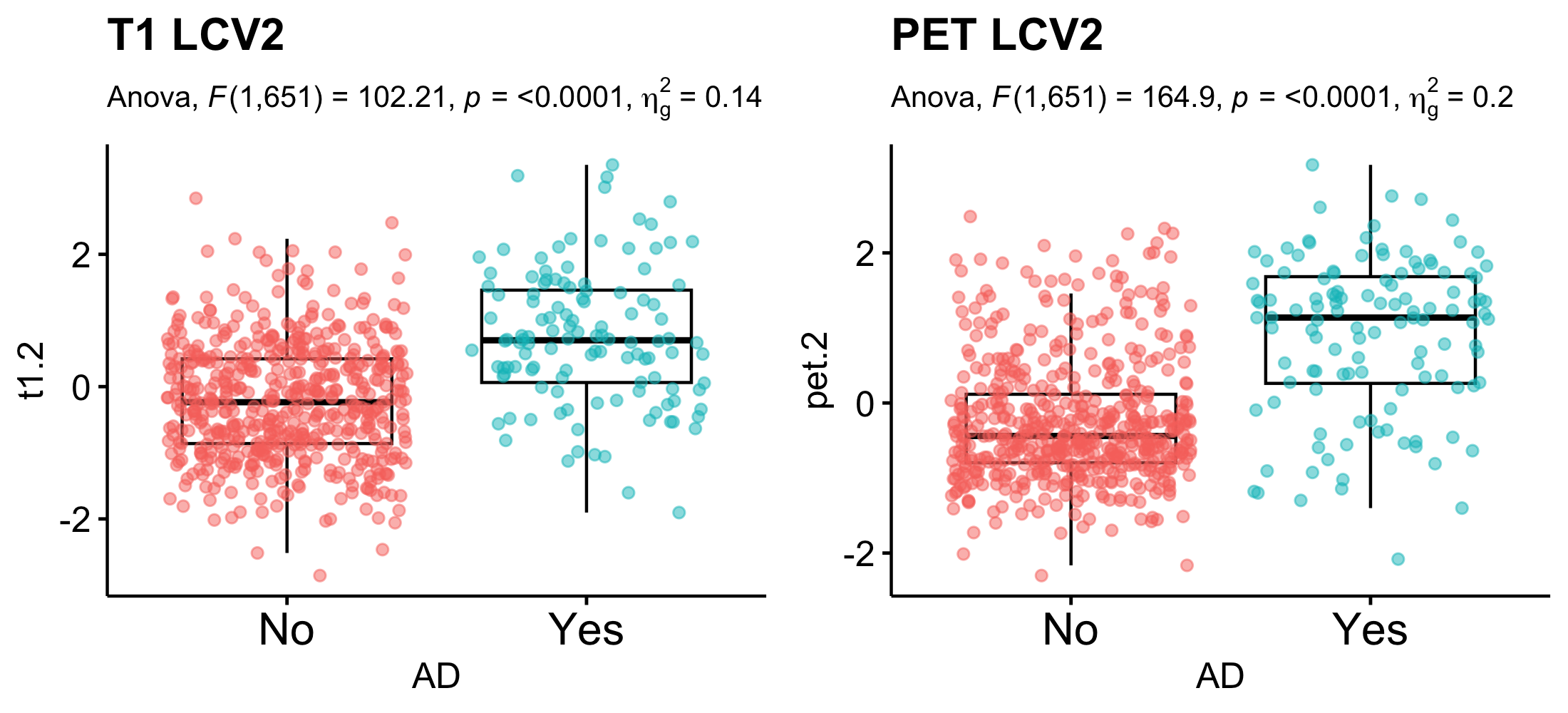}\\
\includegraphics[width=0.5\textwidth]{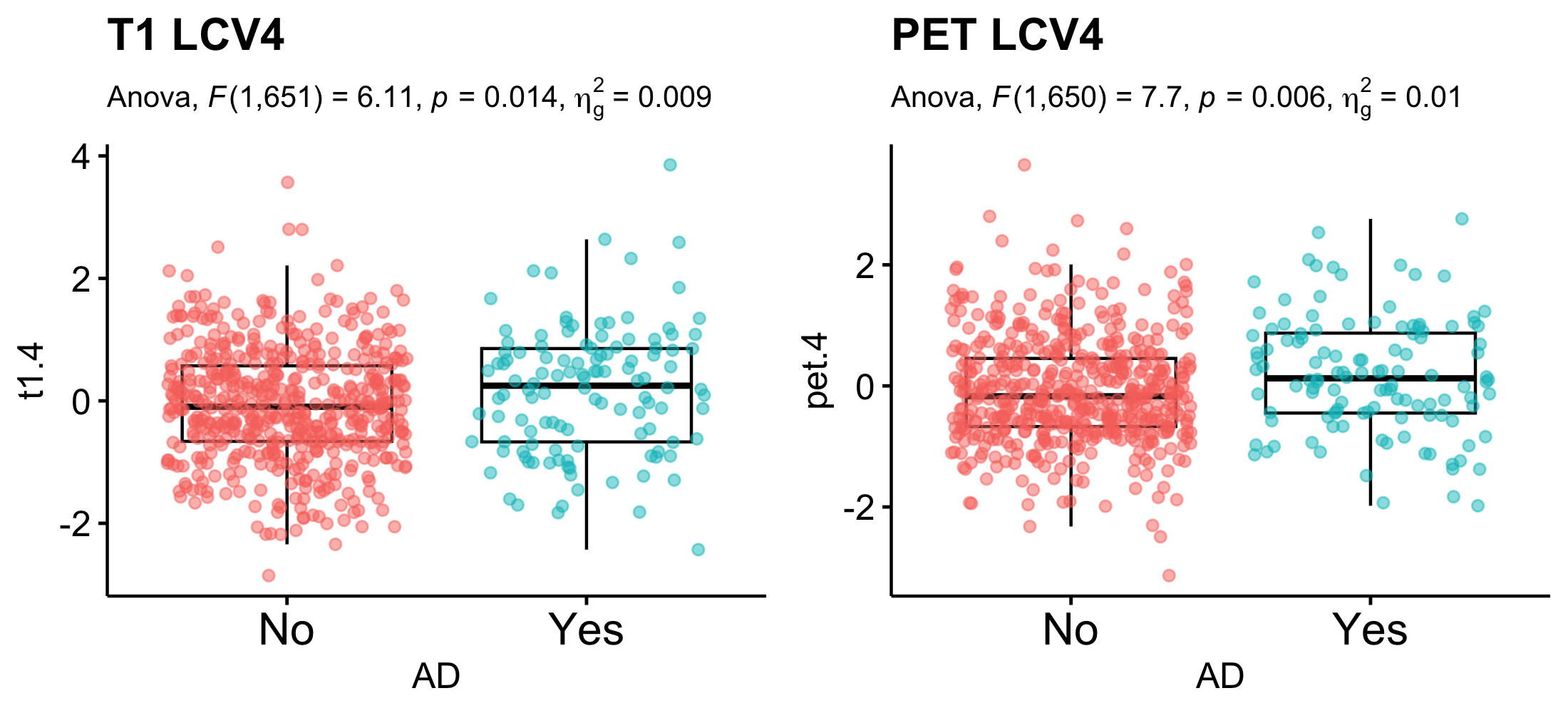}
\includegraphics[width=0.5\textwidth]{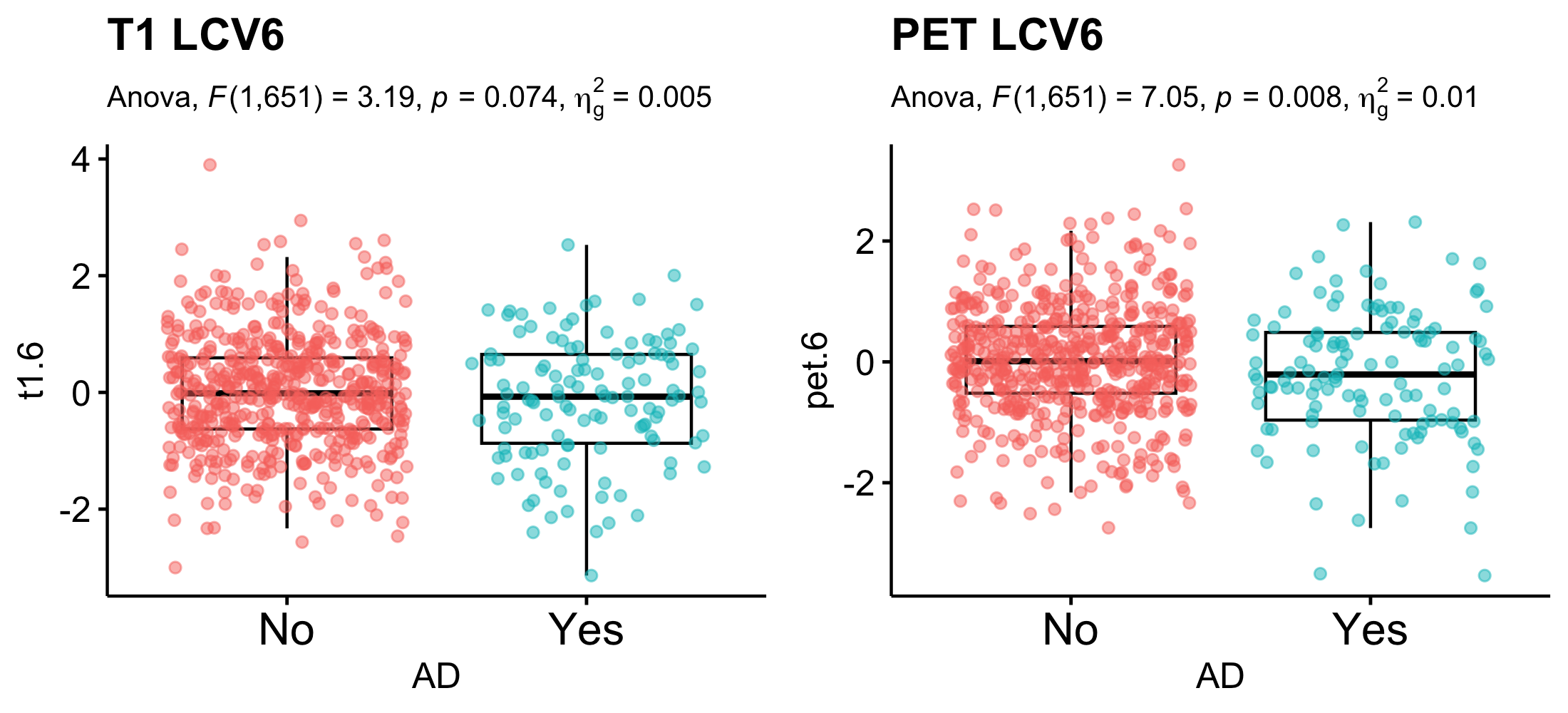}
\caption{LCCA: Group comparison of the canonical variates between participants who transition to dementia within 5 years compared to those who did not. Age, sex and education were adjusted. }\label{fig:ad}
\end{figure}

Secondly, among the non-demented participants at baseline, 
we compared CVs between the participants who translated to AD within 5 years
 and those who remained non-demented. Among 656 non-demented participants, 120 participants (18.3\%) converted to AD in 5 years. Table \ref{tab2} reports the F-statistics, unadjusted p-values and multiple comparison corrected p-values. 
For post-hoc pair-wise group comparison, least squared mean differences were computed. Figure \ref{fig:ad} shows jittered boxplots and pair-wise comparison results with unadjusted p-values for the CVs with significant group difference after multiple comparison correction. The CVs 1, 2 and 4 for T1 and CVs 1, 2, 4 and 6 of PET showed a significant difference between those who
 converted to AD and those who did not. 
 
\begin{figure}
\includegraphics[width=0.5\textwidth]{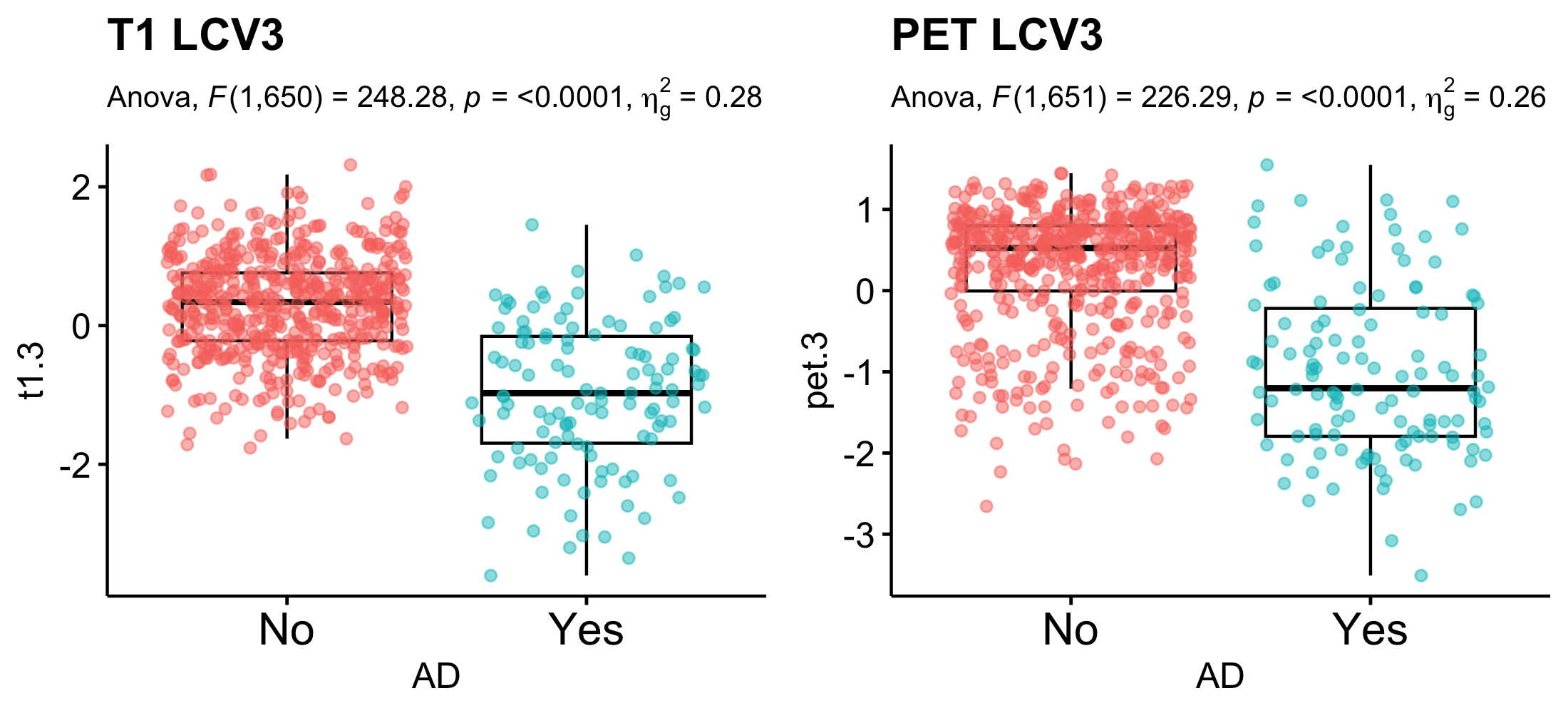}
\includegraphics[width=0.5\textwidth]{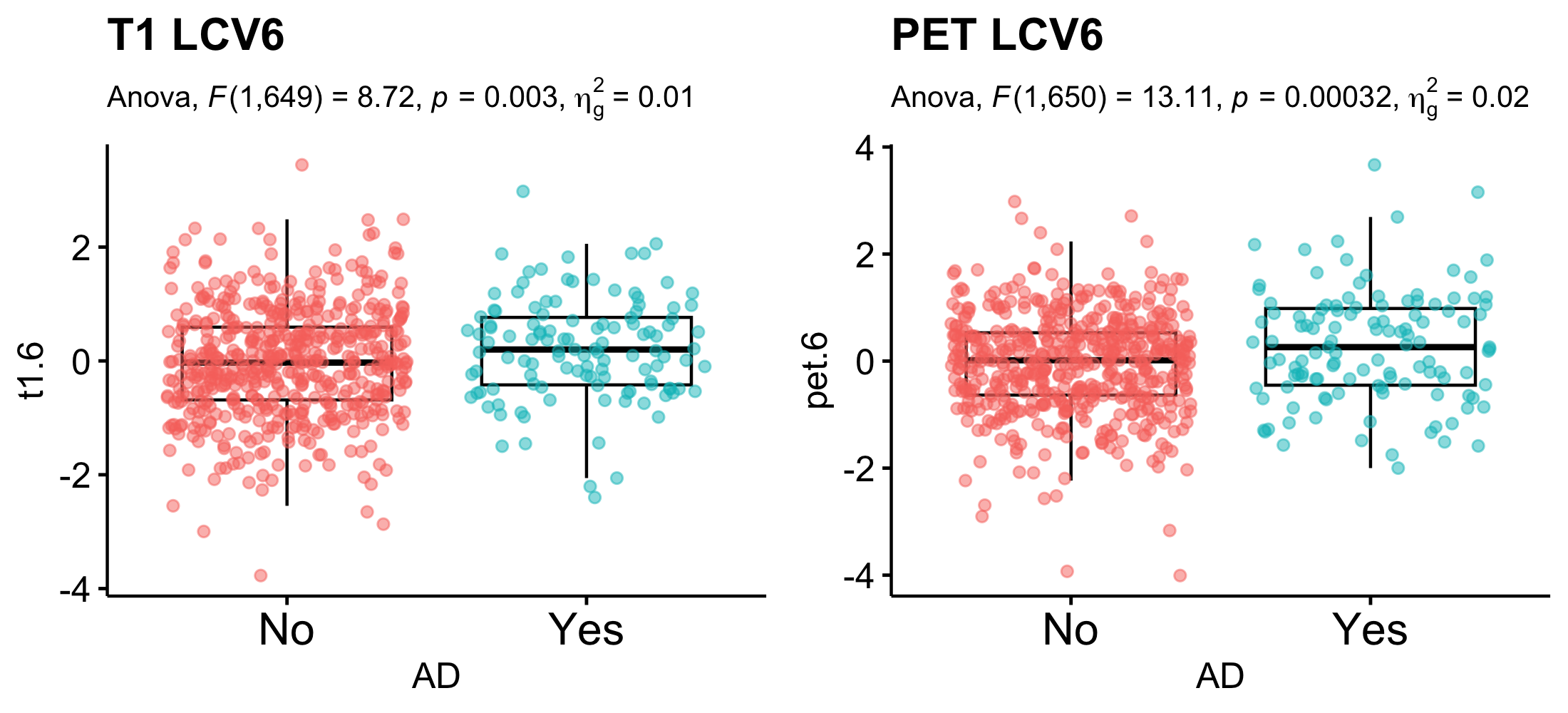}\\
\includegraphics[width=0.5\textwidth]{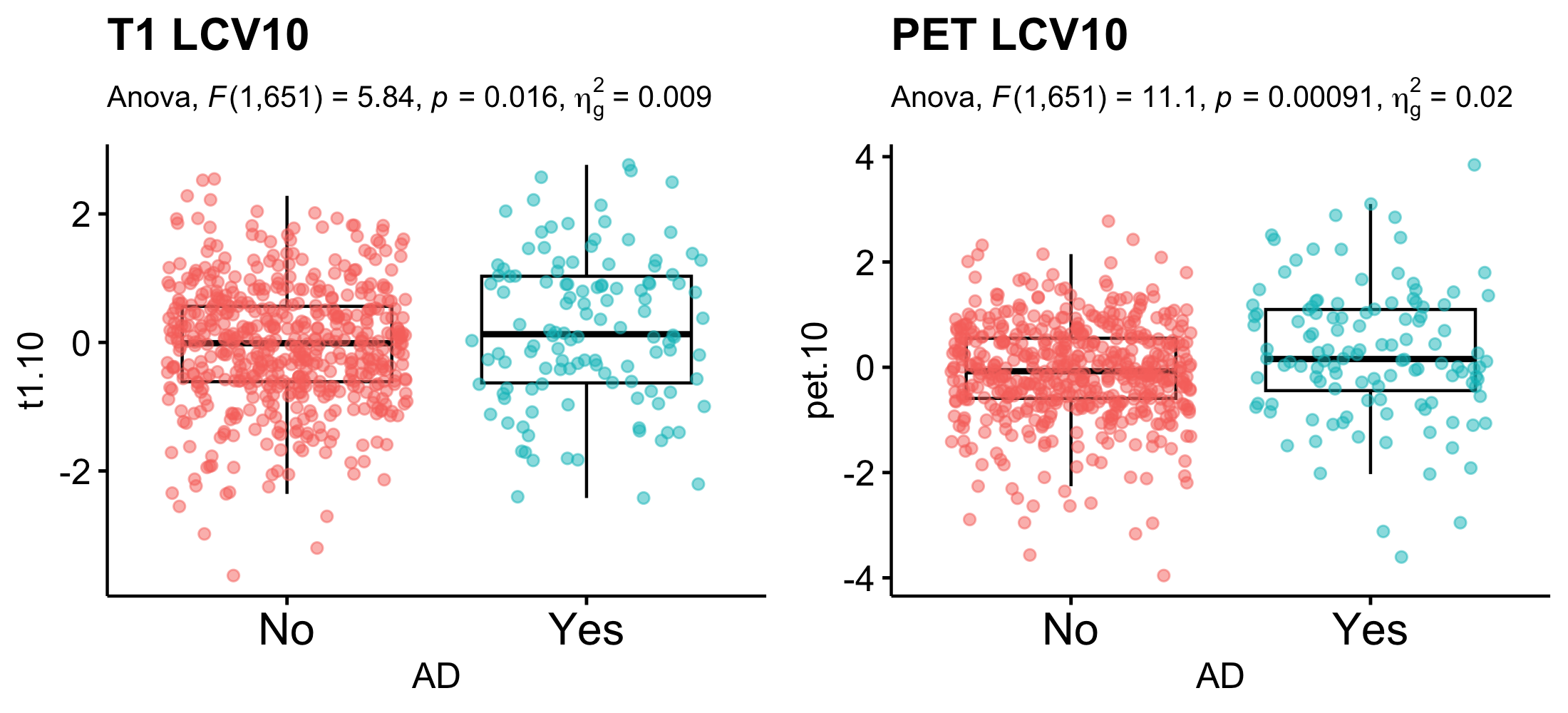}
\caption{Naive Approach: Group comparison of the canonical variates between participants who transition to dementia within 5 years compared to those who did not. Age, sex and education were adjusted. }\label{fig:ad2}
\end{figure}


\section{Simulations}\label{simulations}

We conducted extensive numerical experiments to
evaluate the performance of the LCCA comparing to the naive approach. The first simulation setting was
designed to evaluate general performance of LCCA when the data were
generated from the LCCA model. The second simulation was designed when the
data were generated with a low signal-to-noise ratio. We also conducted a third simulation to determine the ability of LCCA to recover subgroups across a range of sample sizes and a varied degree of subgroup imbalance.

For all simulation settings, we generated 100 independent datasets and
compared the performance of LCCA according to the following criteria: (1) number
of estimated canonical variates;(2) bias of the estimated canonical
correlation coefficients; (3) accuracy of the estimated canonical
loading using cosine similarity. For the third simulation, we report the correlation between the true and estimated canonical variates in the place of (2). 
Since the true number of canonical variates is one, we compared the first canonical variate's performance even if the LCCA or the naive approach selected different numbers of canonical variates. 

\subsection{Simulation Setting 1.}\label{simulation-setting-1}

\begin{figure}
\includegraphics[width=\textwidth]{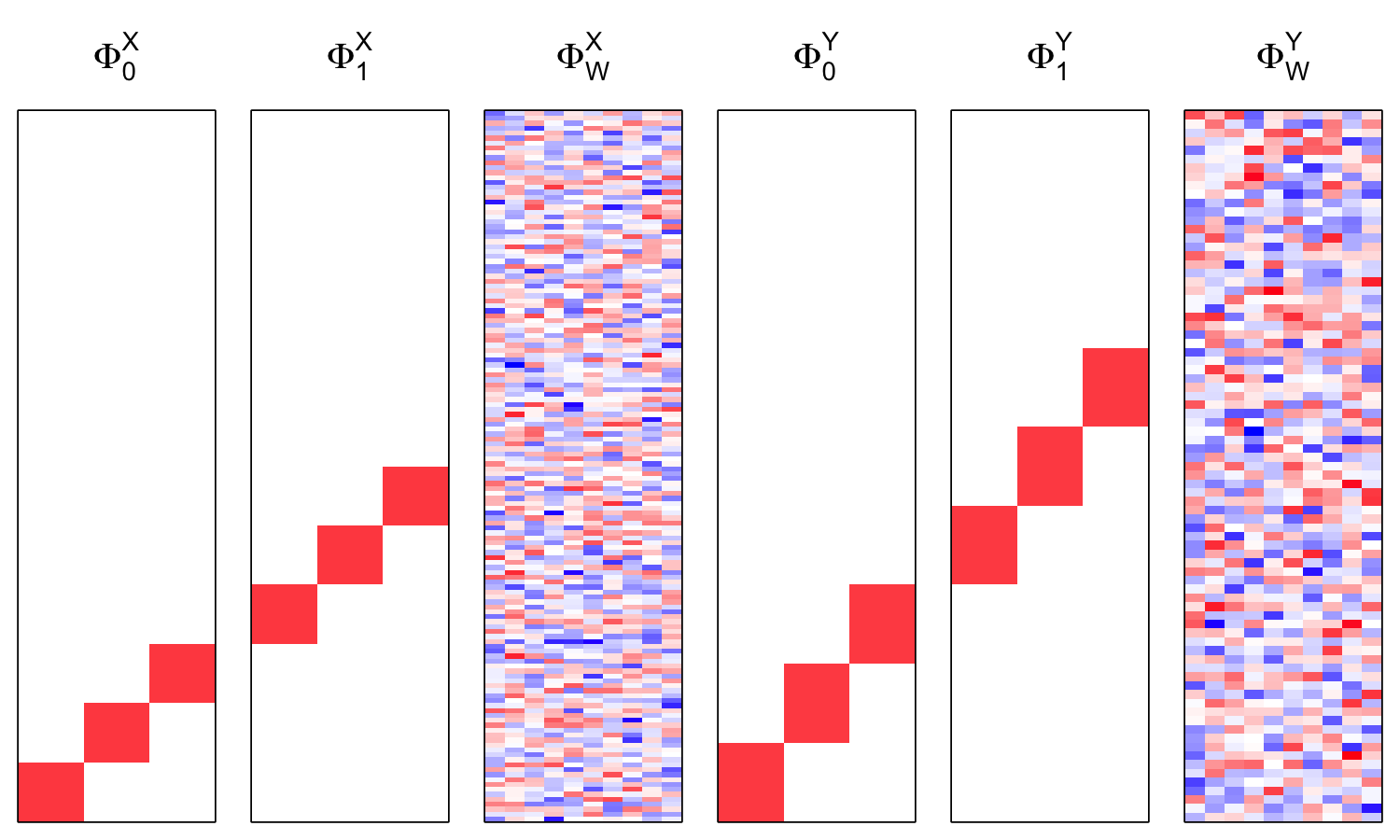}
\caption{Simulation 1. LFPCA models for simulation.}\label{sim1:phi}
\end{figure}

\begin{figure}
\centering
\includegraphics[width=0.6\textwidth]{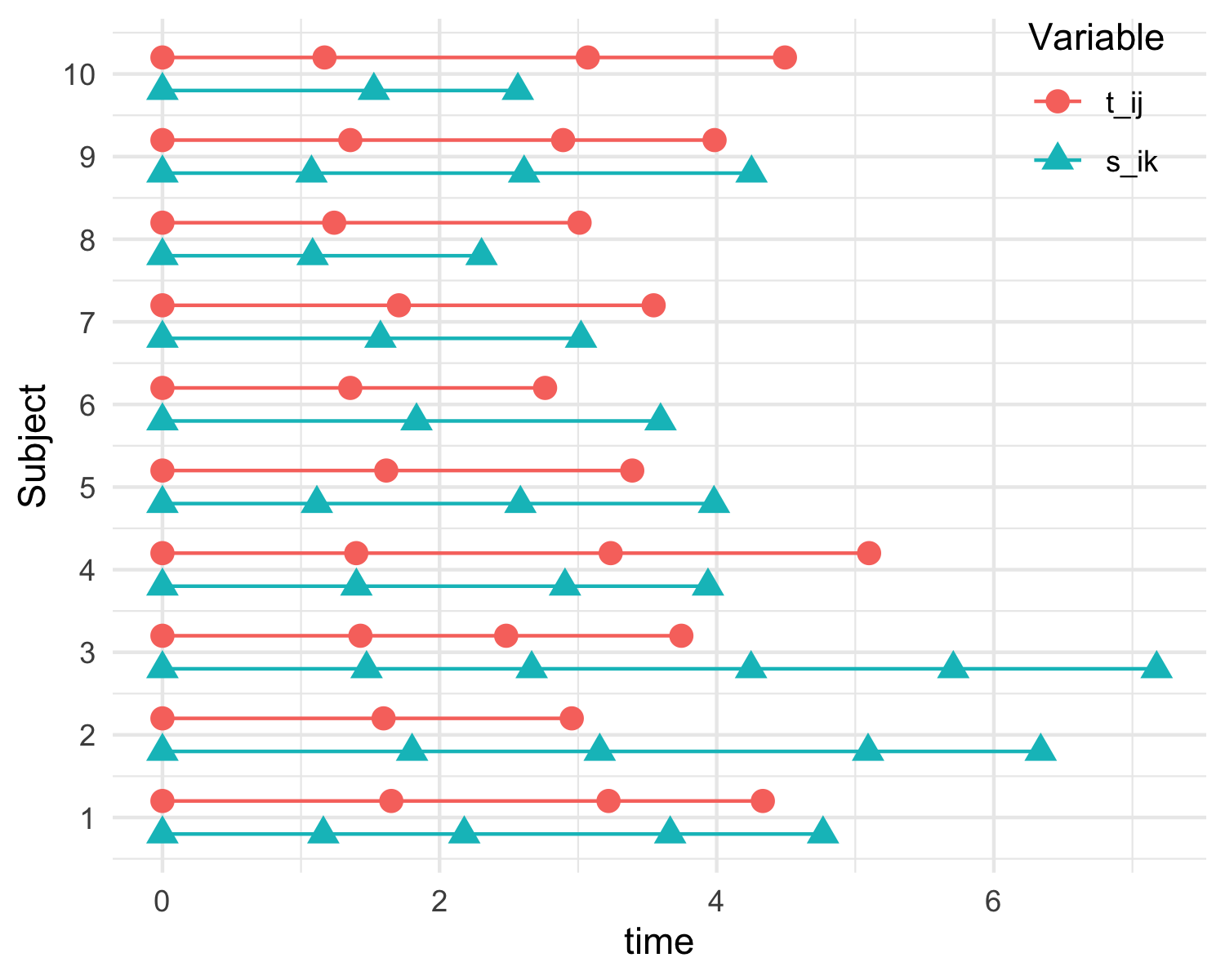}
\caption{Example of the time variables $t_{ij}$ and $s_{ik}$ from the first 10 subjects from a simulation setting. The time points between $X$ and $Y$ variables are not aligned, and the numbers of visits differ.}\label{sim1:time}
\end{figure}

Data were generated from the model
\ref{eqn:lpcax} setting \(N_{X}\) and \(N_{Y}\) to 3. The LPC loadings were 
generated as depicted in Figure \ref{sim1:phi}; we set
\(p=144\) and \(q=81\). The LPC scores \({\xi}_{ik}^{X}\)
and \(\xi_{ik}^{Y}\) were generated from the \(N(0,\lambda_k)\), \(\lambda_1=8,\lambda_2=4,\lambda_3=2\), where \(Cor({\xi}_{i2}^{X}, {\xi}_{i2}^{Y})=r\). We note that we correlated the second pairs of the LPCs to evaluate LCCA's performance when the correlated signals explain more minor variances. For each subject, the number of visits were generated from Poisson(1) followed by adding 3 such that each subject has on average 4 time points. The time intervals between visits were generated from \(U[1,2]\). Figure \ref{sim1:time} shows examples of the time variables $t_{ij}$ and $s_{ik}$ from the first 10 subjects from a simulation setting. The time points between $X$ and $Y$ variables are not aligned, and the numbers of visits differ. We conducted simulations for \(r=0.1,0.3,0.5\), at \(n=100,200,400\), and different threshold for LPC dimension selection (80\% or 90\%). For each scenario, 100 independent datasets were generated. 

\begin{figure}
\includegraphics[width=\textwidth]{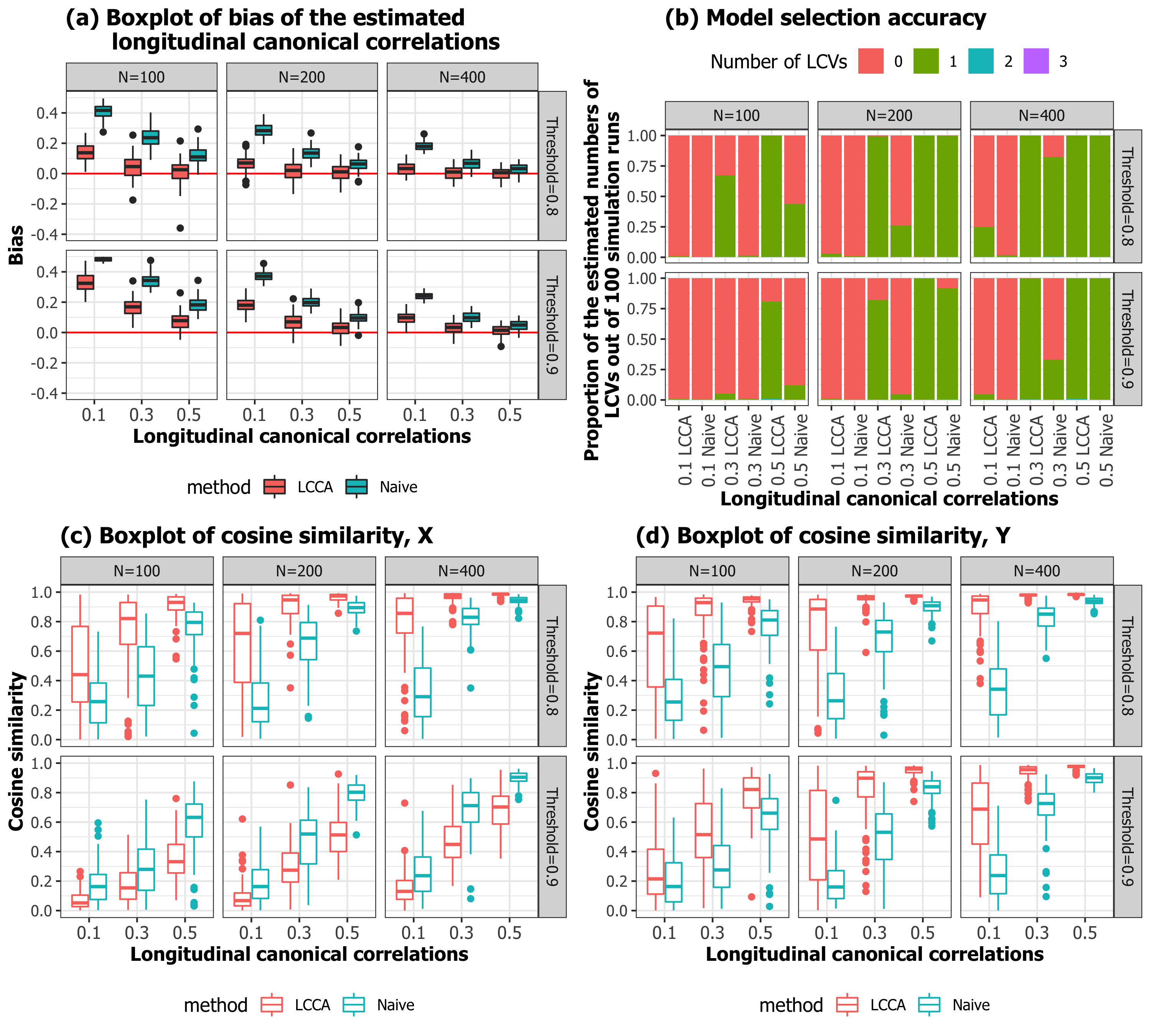}
\caption{Simulation 1. Performance evaluation.}\label{sim1:fig}
\end{figure} 

Overall, LCCA outperforms the naive approach in the three criteria. Figure \ref{sim1:fig} shows that both LCCA and the naive approach tend to overestimate canonical correlation when the sample size is small and the true canonical correlation is smaller, while LCCA performs better than the naive approach. The results show that the approximation-based CV dimension estimation performs better as the sample size increases, and the underlying true canonical correlation is larger. LCCA identifies the number of canonical variates more accurately than the naive approach. Similar patterns were found in the cosine similarity measures. The cosine similarity between estimated LCV loadings performs better as the sample size increases, and the underlying true canonical correlation is larger. 

\subsection{Simulation Setting 2.}\label{simulation-setting-2}
\begin{figure}
\includegraphics[width=\textwidth]{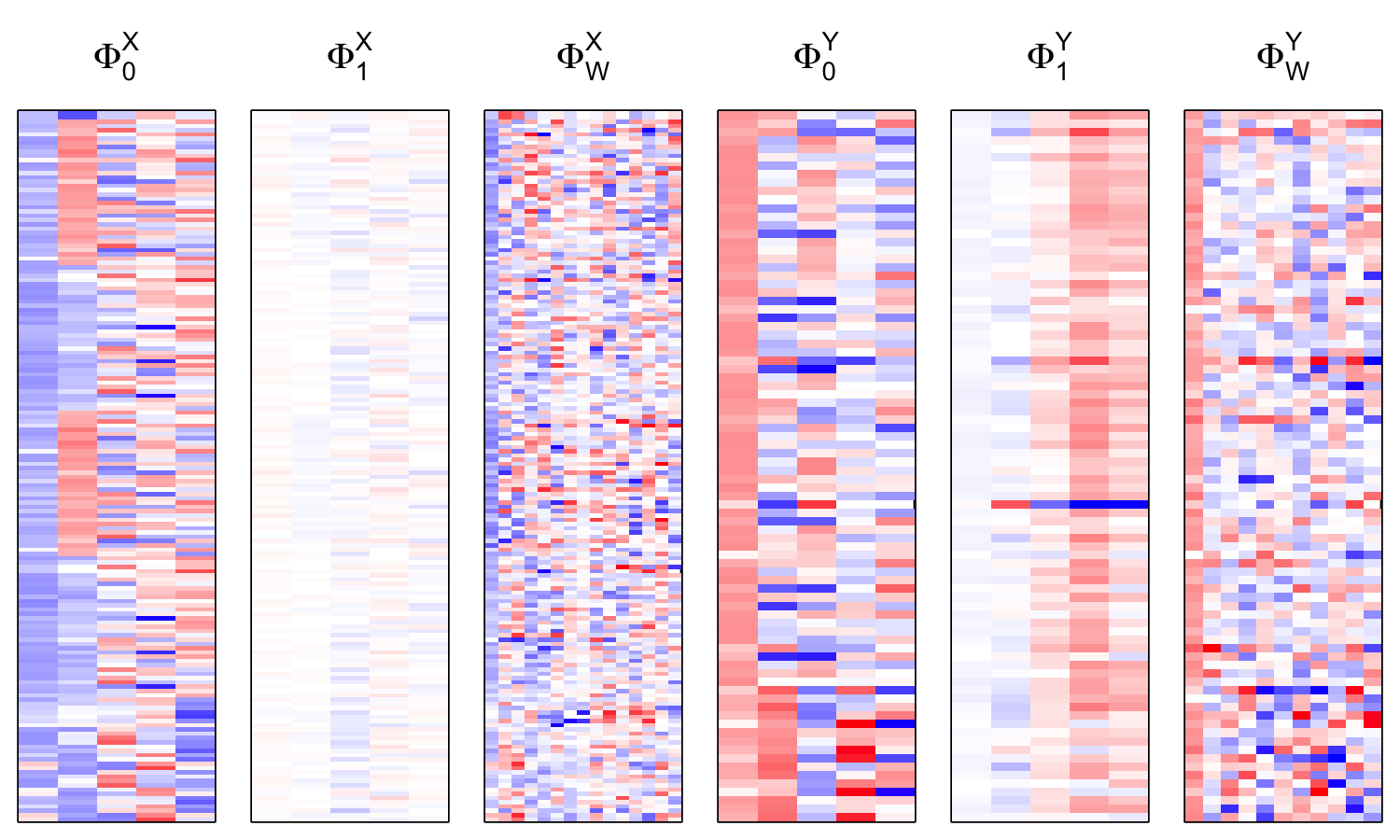}
\caption{Simulation 2. LFPCA models for simulation.}\label{sim2:phi}\end{figure} 

\begin{figure}
\includegraphics[width=\textwidth]{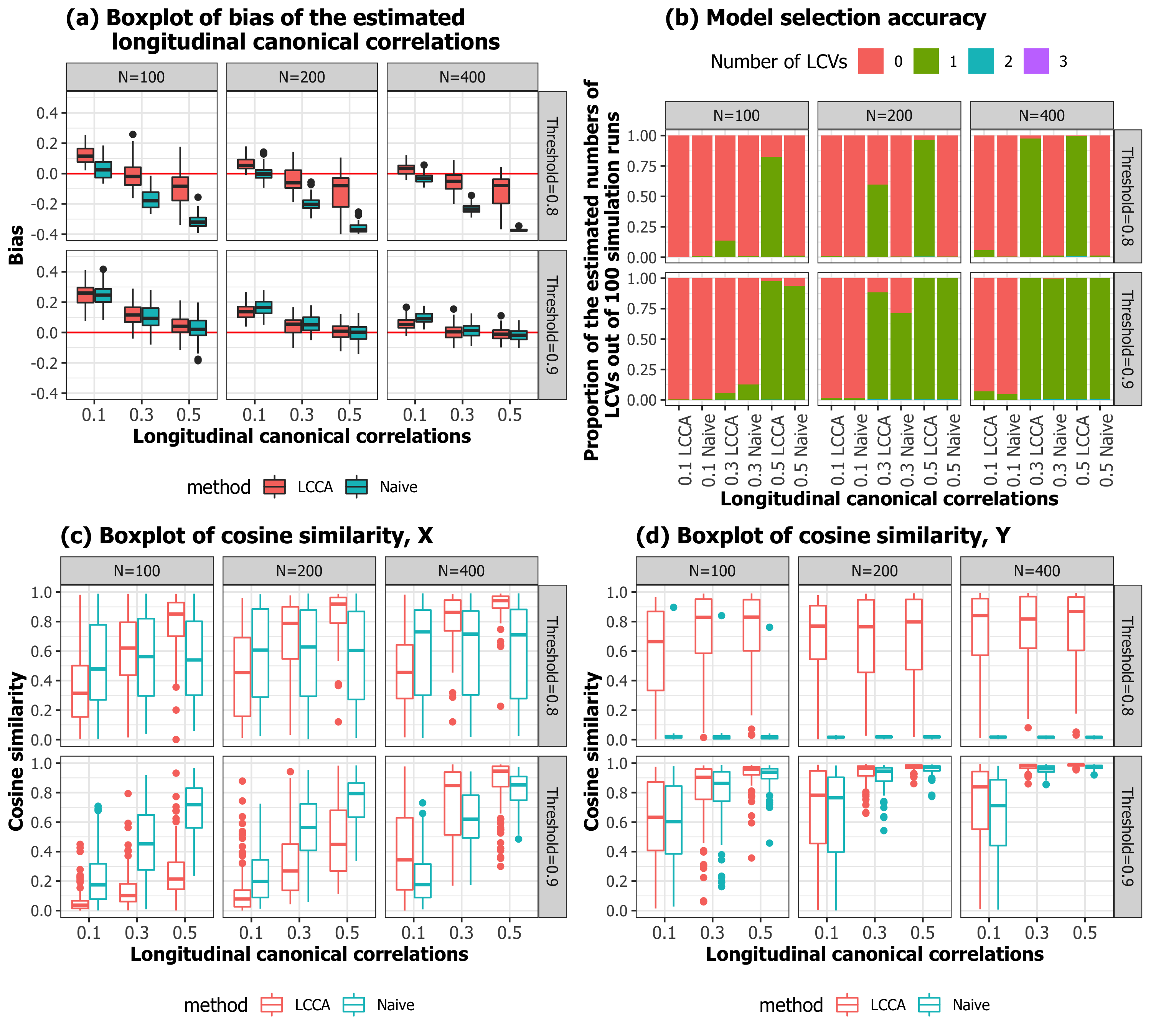}
\caption{Simulation 2. Performance evaluation.}\label{sim2:fig}
\end{figure}

Data were generated similarly to the first simulation, except the first five LPC loadings were taken from ADNI data from Section 3 (Figure \ref{sim2:phi}). The LPC scores \({\xi}_{ik}^{X}\)
and \(\xi_{ik}^{Y}\) were generated from the \(N(0,\lambda_k)\), \(\lambda_1=45.4,\lambda_2=17.2,\lambda_3=7.0,\lambda_4=4,1, \lambda_5=3.8\), where \(Cor({\xi}_{i2}^{X}, {\xi}_{i2}^{Y})=r\). For each subject, the number of visits were generated from Poisson(1) followed by adding 3 such that each subject has on average 4 time points. The time intervals between visits were generated from \(U[1,2]\). We conducted simulations for \(r=0.1,0.3,0.5\), at \(n=100,200,400\), and different threshold for LPC dimension selection (80\% or 90\%). For each scenario, 100 independent datasets were generated. We found very similar results as Simulation 1 (Figure \ref{sim2:fig}, assuring our data analysis results in Section 3 have less bias in estimating canonical weights and correlations. The both methods tend to underestimate canonical correlation when the threshold (=0.8) is smaller, while this issue does not persist when the threshold is more lenient. 

\subsection{Simulation Setting 3.}\label{simulation-setting-3}

\begin{figure}
\includegraphics[width=\textwidth]{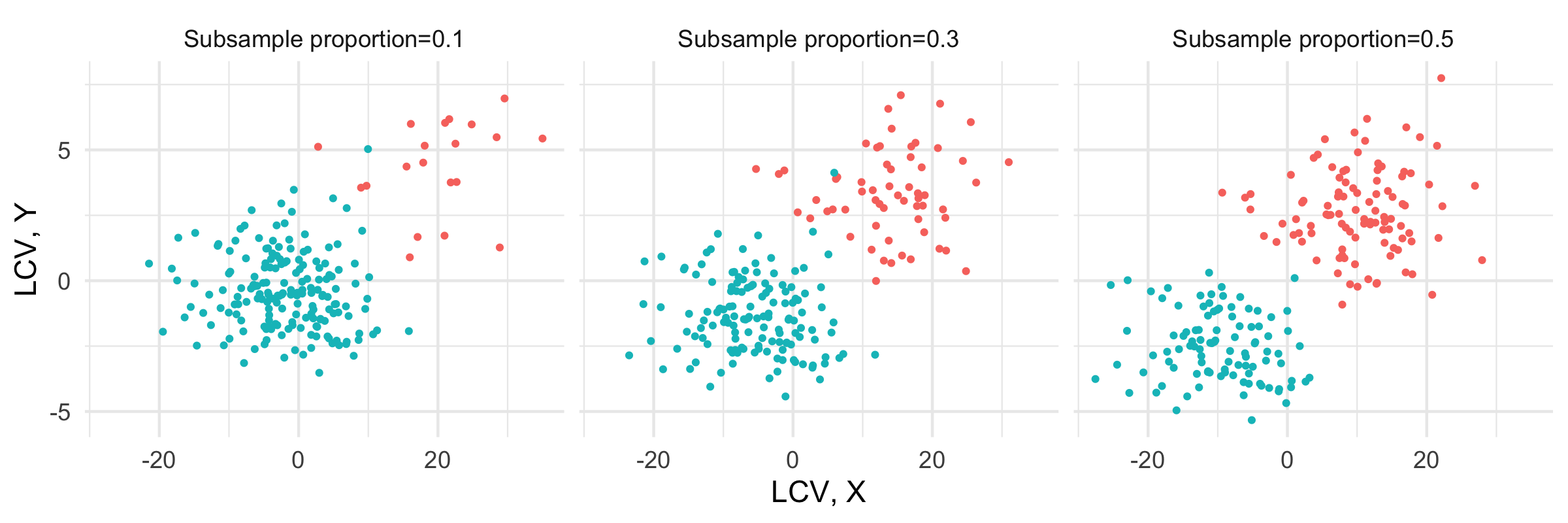}
\caption{Simulation 3. An example of the longitudinal canonical variates (LCVs) at the different ratios of subgroups.}\label{sim3:score}
\end{figure}

We also examined whether the LCCA can recover subgroups. We employed the same simulation settings from Simulation 2, except longitudinal canonical variates (LCVs). Instead of generating the LCVs from a multivariate normal distribution, we imposed two subgroups. One group was generated from $MVN(0,\lambda \mathbf{I}_2)$ and the other group was generated from $N(\mu \mathbf{1}_2, \mathbf{I}_2)$ as shown in Figure \ref{sim3:score}. We varied the proportion of subsamples from 0.1 to 0.5 and evaluated whether the LCCA can identify subgroups.

\begin{figure}
\includegraphics[width=\textwidth]{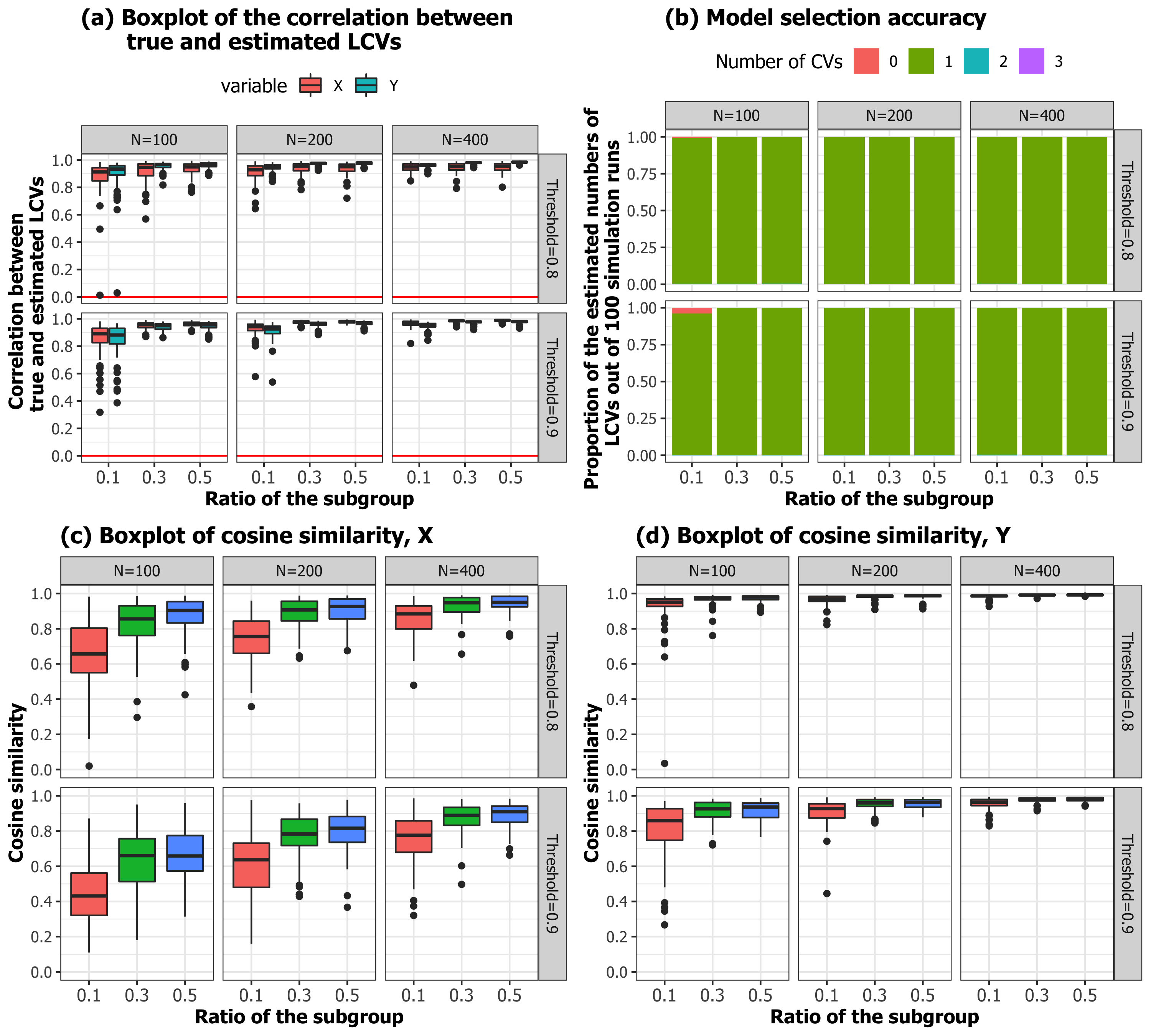}
\caption{Simulation 3. Performance evaluation.}\label{sim3:fig}
\end{figure}

To evaluate whether LCCA recovers the true LCVs with subgroups, the correlations between the true and estimated LCVs were calculated. Figure \ref{sim3:fig} shows that the LCCA identified LCVs, LC vectors, and the number of LCVs well, even with smaller sample sizes and a highly unbalanced subgroups (subsample proportion=0.1).

\section{Conclusion}\label{conclusion}

In this manuscript, we proposed a new longitudinal CCA method that can handle different temporal samplings and missing values that often occur in longitudinal data. The proposed method is very flexible in its ability to handle linear and nonlinear trajectories. The application to ADNI data, the longitudinal CCA revealed the most relevant patterns between Amyloid deposition associated neuronal loss measured as cortical thickness and subcortical volume. The canonical variates are associated with the baseline clinical status and predicted AD transition. The numerical experiments showed that LCCA outperformed the naive approach, also showing the performance of LCCA is not very sensitive to the selection of the threshold but requires adequate sample size or an effect size of the correlation. 

Our two-stage approach is computationally fast, and our numerical experiments showed good performance using a threshold of 80-90\% as a rule of thumb. However, the dimension reduction step using the LPCA possibly removes small but important biological signals in other neuroimaging modalities such as functional magnetic resonance imaging (fMRI). Furthermore, the two-step approach does not optimize both dimension reduction and dimension selection for CCA simultaneously, unlike the unified estimation approach for cross-sectional high-dimensional data \citep{song2016canonical}. Nonetheless, our novel LCCA approach offers a computationally efficient and well performing tool that could have important applications in neuroimaging and other settings producing high-dimensional, longitudinal, multivariate data.

\section*{Acknowledgement}\label{acknowledgement}

This work was supported by NIH R01AG062578 (PI: Lee).

Data collection and sharing for this project was funded by the Alzheimer's Disease Neuroimaging Initiative (ADNI) (National Institutes of Health Grant U01 AG024904) and DOD ADNI (Department of Defense award number W81XWH-12-2-0012). ADNI is funded by the National Institute on Aging, the National Institute of Biomedical Imaging and Bioengineering, and through generous contributions from the following: AbbVie, Alzheimer’s Association; Alzheimer’s Drug Discovery Foundation; Araclon Biotech; BioClinica, Inc.; Biogen; Bristol-Myers Squibb Company; CereSpir, Inc.; Cogstate; Eisai Inc.; Elan Pharmaceuticals, Inc.; Eli Lilly and Company; EuroImmun; F. Hoffmann-La Roche Ltd and its affiliated company Genentech, Inc.; Fujirebio; GE Healthcare; IXICO Ltd.; Janssen Alzheimer Immunotherapy Research \& Development, LLC.; Johnson \& Johnson Pharmaceutical Research \& Development LLC.; Lumosity; Lundbeck; Merck \& Co., Inc.; Meso Scale Diagnostics, LLC.; NeuroRx Research; Neurotrack Technologies; Novartis Pharmaceuticals Corporation; Pfizer Inc.; Piramal Imaging; Servier; Takeda Pharmaceutical Company; and Transition Therapeutics. The Canadian Institutes of Health Research is providing funds to support ADNI clinical sites in Canada. Private sector contributions are facilitated by the Foundation for the National Institutes of Health (www.fnih.org). The grantee organization is the Northern California Institute for Research and Education, and the study is coordinated by the Alzheimer’s Therapeutic Research Institute at the University of Southern California. ADNI data are disseminated by the Laboratory for Neuro Imaging at the University of Southern California.

\bibliographystyle{rss}
\bibliography{references}

\begin{thebibliography}{31}
\expandafter\ifx\csname natexlab\endcsname\relax\def\natexlab#1{#1}\fi
\expandafter\ifx\csname url\endcsname\relax
  \def\url#1{\texttt{#1}}\fi
\expandafter\ifx\csname urlprefix\endcsname\relax\def\urlprefix{URL: }\fi

\bibitem[{Adhikari et~al.(2019)Adhikari, Hong, Sampath, Chiappelli, Jahanshad,
  Thompson, Rowland, Calhoun, Du, Chen and
  Kochunov}]{adhikariFunctionalNetworkConnectivity2019}
Adhikari, B.~M., Hong, L.~E., Sampath, H., Chiappelli, J., Jahanshad, N.,
  Thompson, P.~M., Rowland, L.~M., Calhoun, V.~D., Du, X., Chen, S. and
  Kochunov, P. (2019) Functional network connectivity impairments and core
  cognitive deficits in schizophrenia.
\newblock \textit{Human Brain Mapping}.

\bibitem[{Avants et~al.(2010)Avants, Cook, Ungar, Gee and
  Grossman}]{avantsDementiaInducesCorrelated2010}
Avants, B.~B., Cook, P.~A., Ungar, L., Gee, J.~C. and Grossman, M. (2010)
  Dementia induces correlated reductions in white matter integrity and cortical
  thickness: A multivariate neuroimaging study with sparse canonical
  correlation analysis.
\newblock \textit{NeuroImage}, \textbf{50}, 1004--1016.

\bibitem[{Bao et~al.(2019)Bao, Hu, Pan and
  Zhou}]{baoCanonicalCorrelationCoefficients2019}
Bao, Z., Hu, J., Pan, G. and Zhou, W. (2019) Canonical correlation coefficients
  of high-dimensional {{Gaussian}} vectors: {{Finite}} rank case.
\newblock \textit{The Annals of Statistics}, \textbf{47}, 612--640.

\bibitem[{Bartlett(1947)}]{bartlett1947general}
Bartlett, M. (1947) The general canonical correlation distribution.
\newblock \textit{The Annals of Mathematical Statistics}, 1--17.

\bibitem[{Benjamini and Hochberg(1995)}]{benjamini1995controlling}
Benjamini, Y. and Hochberg, Y. (1995) Controlling the false discovery rate: a
  practical and powerful approach to multiple testing.
\newblock \textit{Journal of the Royal statistical society: series B
  (Methodological)}, \textbf{57}, 289--300.

\bibitem[{Deleus and {Van
  Hulle}(2011)}]{deleusFunctionalConnectivityAnalysis2011}
Deleus, F. and {Van Hulle}, M.~M. (2011) Functional connectivity analysis of
  {{fMRI}} data based on regularized multiset canonical correlation analysis.
\newblock \textit{Journal of Neuroscience Methods}, \textbf{197}, 143--157.

\bibitem[{Desikan et~al.(2006)Desikan, S{\'e}gonne, Fischl, Quinn, Dickerson,
  Blacker, Buckner, Dale, Maguire, Hyman et~al.}]{desikan2006automated}
Desikan, R.~S., S{\'e}gonne, F., Fischl, B., Quinn, B.~T., Dickerson, B.~C.,
  Blacker, D., Buckner, R.~L., Dale, A.~M., Maguire, R.~P., Hyman, B.~T. et~al.
  (2006) An automated labeling system for subdividing the human cerebral cortex
  on mri scans into gyral based regions of interest.
\newblock \textit{Neuroimage}, \textbf{31}, 968--980.

\bibitem[{Du et~al.(2019)Du, Liu, Zhu, Yao, Risacher, Guo, Saykin and
  Shen}]{duIdentifyingProgressiveImaging2019}
Du, L., Liu, K., Zhu, L., Yao, X., Risacher, S.~L., Guo, L., Saykin, A.~J. and
  Shen, L. (2019) Identifying progressive imaging genetic patterns via
  multi-task sparse canonical correlation analysis: A longitudinal study of the
  {{ADNI}} cohort.
\newblock \textit{Bioinformatics}, \textbf{35}, i474--i483.

\bibitem[{Fang et~al.(2016)Fang, Lin, Schulz, Xu, Calhoun and
  Wang}]{fangJointSparseCanonical2016}
Fang, J., Lin, D., Schulz, S.~C., Xu, Z., Calhoun, V.~D. and Wang, Y.-P. (2016)
  Joint sparse canonical correlation analysis for detecting differential
  imaging genetics modules.
\newblock \textit{Bioinformatics (Oxford, England)}, \textbf{32}, 3480--3488.

\bibitem[{Fischl et~al.(2002)Fischl, Salat, Busa, Albert, Dieterich,
  Haselgrove, {Van Der Kouwe}, Killiany, Kennedy, Klaveness and
  {others}}]{fischlWholeBrainSegmentation2002}
Fischl, B., Salat, D.~H., Busa, E., Albert, M., Dieterich, M., Haselgrove, C.,
  {Van Der Kouwe}, A., Killiany, R., Kennedy, D., Klaveness, S. and {others}
  (2002) Whole brain segmentation: Automated labeling of neuroanatomical
  structures in the human brain.
\newblock \textit{Neuron}, \textbf{33}, 341--355.

\bibitem[{Friederichs and Hense(2003)}]{friederichs2003statistical}
Friederichs, P. and Hense, A. (2003) Statistical inference in canonical
  correlation analyses exemplified by the influence of north atlantic sst on
  european climate.
\newblock \textit{Journal of Climate}, \textbf{16}, 522--534.

\bibitem[{Gossmann et~al.(2018)Gossmann, Zille, Calhoun and
  Wang}]{gossmannFDRCorrectedSparseCanonical2018}
Gossmann, A., Zille, P., Calhoun, V. and Wang, Y.-P. (2018) {{FDR}}-{{Corrected
  Sparse Canonical Correlation Analysis With Applications}} to {{Imaging
  Genomics}}.
\newblock \textit{IEEE transactions on medical imaging}, \textbf{37},
  1761--1774.

\bibitem[{Greve et~al.(2016)Greve, Salat, Bowen, Izquierdo-Garcia, Schultz,
  Catana, Becker, Svarer, Knudsen, Sperling and
  Johnson}]{greveDifferentPartialVolume2016}
Greve, D.~N., Salat, D.~H., Bowen, S.~L., Izquierdo-Garcia, D., Schultz, A.~P.,
  Catana, C., Becker, J.~A., Svarer, C., Knudsen, G.~M., Sperling, R.~A. and
  Johnson, K.~A. (2016) Different partial volume correction methods lead to
  different conclusions: {{An 18F}}-{{FDG}}-{{PET}} study of aging.
\newblock \textit{NeuroImage}, \textbf{132}, 334--343.

\bibitem[{Greve et~al.(2014)Greve, Svarer, Fisher, Feng, Hansen, Baare, Rosen,
  Fischl and Knudsen}]{greveCorticalSurfacebasedAnalysis2014}
Greve, D.~N., Svarer, C., Fisher, P.~M., Feng, L., Hansen, A.~E., Baare, W.,
  Rosen, B., Fischl, B. and Knudsen, G.~M. (2014) Cortical surface-based
  analysis reduces bias and variance in kinetic modeling of brain {{PET}} data.
\newblock \textit{NeuroImage}, \textbf{92}, 225--236.

\bibitem[{Greven et~al.(2011)Greven, Crainiceanu, Caffo and
  Reich}]{grevenLongitudinalFunctionalPrincipal2011}
Greven, S., Crainiceanu, C., Caffo, B. and Reich, D. (2011) Longitudinal
  functional principal component analysis.
\newblock In \textit{Recent {{Advances}} in {{Functional Data Analysis}} and
  {{Related Topics}}}, 149--154. {Springer}.

\bibitem[{Grosenick et~al.(2019)Grosenick, Shi, Gunning, Dubin, Downar and
  Liston}]{grosenickFunctionalOptogeneticApproaches2019}
Grosenick, L., Shi, T.~C., Gunning, F.~M., Dubin, M.~J., Downar, J. and Liston,
  C. (2019) Functional and {{Optogenetic Approaches}} to {{Discovering Stable
  Subtype}}-{{Specific Circuit Mechanisms}} in {{Depression}}.
\newblock \textit{Biological Psychiatry. Cognitive Neuroscience and
  Neuroimaging}, \textbf{4}, 554--566.

\bibitem[{Hao et~al.(2017)Hao, Li, Yan, Yao, Risacher, Saykin, Shen, Zhang and
  {Alzheimer's Disease Neuroimaging
  Initiative}}]{haoIdentificationAssociationsGenotypes2017}
Hao, X., Li, C., Yan, J., Yao, X., Risacher, S.~L., Saykin, A.~J., Shen, L.,
  Zhang, D. and {Alzheimer's Disease Neuroimaging Initiative} (2017)
  Identification of associations between genotypes and longitudinal phenotypes
  via temporally-constrained group sparse canonical correlation analysis.
\newblock \textit{Bioinformatics (Oxford, England)}, \textbf{33}, i341--i349.

\bibitem[{Kang et~al.(2016)Kang, Bowman, Mayberg and Liu}]{kang2016depression}
Kang, J., Bowman, F.~D., Mayberg, H. and Liu, H. (2016) A depression network of
  functionally connected regions discovered via multi-attribute canonical
  correlation graphs.
\newblock \textit{NeuroImage}, \textbf{141}, 431--441.

\bibitem[{Kim et~al.(2019)Kim, Won, Youn and
  Park}]{kimJointconnectivitybasedSparseCanonical2019}
Kim, M., Won, J.~H., Youn, J. and Park, H. (2019) Joint-connectivity-based
  sparse canonical correlation analysis of imaging genetics for detecting
  biomarkers of {{Parkinson}}'s disease.
\newblock \textit{IEEE transactions on medical imaging}.

\bibitem[{Lee et~al.(2015)Lee, Zipunnikov, Reich and
  Pham}]{leeStatisticalImageAnalysis2015}
Lee, S., Zipunnikov, V., Reich, D.~S. and Pham, D.~L. (2015) Statistical image
  analysis of longitudinal {{RAVENS}} images.
\newblock \textit{Frontiers in neuroscience}, \textbf{9}, 368.

\bibitem[{Lin et~al.(2014)Lin, Calhoun and
  Wang}]{linCorrespondenceFMRISNP2014a}
Lin, D., Calhoun, V.~D. and Wang, Y.-P. (2014) Correspondence between {{fMRI}}
  and {{SNP}} data by group sparse canonical correlation analysis.
\newblock \textit{Medical Image Analysis}, \textbf{18}, 891--902.

\bibitem[{Mihalik et~al.(2019)Mihalik, Ferreira, Rosa, Moutoussis, Ziegler,
  Monteiro, Portugal, Adams, Romero-Garcia, V{\a'e}rtes, Kitzbichler,
  V{\a'a}{{\v s}}a, Vaghi, Bullmore, Fonagy, Goodyer, Jones, Dolan and
  Mour{\~a}o-Miranda}]{mihalikBrainbehaviourModesCovariation2019}
Mihalik, A., Ferreira, F.~S., Rosa, M.~J., Moutoussis, M., Ziegler, G.,
  Monteiro, J.~M., Portugal, L., Adams, R.~A., Romero-Garcia, R., V{\a'e}rtes,
  P.~E., Kitzbichler, M.~G., V{\a'a}{{\v s}}a, F., Vaghi, M.~M., Bullmore,
  E.~T., Fonagy, P., Goodyer, I.~M., Jones, P.~B., Dolan, R. and
  Mour{\~a}o-Miranda, J. (2019) Brain-behaviour modes of covariation in healthy
  and clinically depressed young people.
\newblock \textit{Scientific Reports}, \textbf{9}.

\bibitem[{Mueller et~al.(2005)Mueller, Weiner, Thal, Petersen, Jack, Jagust,
  Trojanowski, Toga and Beckett}]{muellerAlzheimerDiseaseNeuroimaging2005}
Mueller, S.~G., Weiner, M.~W., Thal, L.~J., Petersen, R.~C., Jack, C., Jagust,
  W., Trojanowski, J.~Q., Toga, A.~W. and Beckett, L. (2005) The
  {{Alzheimer}}'s {{Disease Neuroimaging Initiative}}.
\newblock \textit{Neuroimaging Clinics of North America}, \textbf{15},
  869--877.

\bibitem[{Rao et~al.(1973)Rao, Rao, Statistiker, Rao and Rao}]{rao1973linear}
Rao, C.~R., Rao, C.~R., Statistiker, M., Rao, C.~R. and Rao, C.~R. (1973)
  \textit{Linear statistical inference and its applications}, vol.~2.
\newblock Wiley New York.

\bibitem[{Song et~al.(2016)Song, Schreier, Ram{\'\i}rez and
  Hasija}]{song2016canonical}
Song, Y., Schreier, P.~J., Ram{\'\i}rez, D. and Hasija, T. (2016) Canonical
  correlation analysis of high-dimensional data with very small sample support.
\newblock \textit{Signal Processing}, \textbf{128}, 449--458.

\bibitem[{Weiner et~al.(2012)Weiner, Veitch, Aisen, Beckett, Cairns, Green,
  Harvey, Jack, Jagust, Liu, Morris, Petersen, Saykin, Schmidt, Shaw, Siuciak,
  Soares, Toga, Trojanowski and {Alzheimer's Disease Neuroimaging
  Initiative}}]{weinerAlzheimerDiseaseNeuroimaging2012}
Weiner, M.~W., Veitch, D.~P., Aisen, P.~S., Beckett, L.~A., Cairns, N.~J.,
  Green, R.~C., Harvey, D., Jack, C.~R., Jagust, W., Liu, E., Morris, J.~C.,
  Petersen, R.~C., Saykin, A.~J., Schmidt, M.~E., Shaw, L., Siuciak, J.~A.,
  Soares, H., Toga, A.~W., Trojanowski, J.~Q. and {Alzheimer's Disease
  Neuroimaging Initiative} (2012) The {{Alzheimer}}'s {{Disease Neuroimaging
  Initiative}}: {{A}} review of papers published since its inception.
\newblock \textit{Alzheimer's \& Dementia}, \textbf{8}, S1--S68.

\bibitem[{Wilks(1935)}]{wilks1935independence}
Wilks, S. (1935) On the independence of k sets of normally distributed
  statistical variables.
\newblock \textit{Econometrica, Journal of the Econometric Society}, 309--326.

\bibitem[{Wilms and Croux(2016)}]{wilmsRobustSparseCanonical2016}
Wilms, I. and Croux, C. (2016) Robust sparse canonical correlation analysis.
\newblock \textit{BMC Systems Biology}, \textbf{10}, 72.

\bibitem[{Witten et~al.(2009)Witten, Tibshirani and
  Hastie}]{wittenPenalizedMatrixDecomposition2009}
Witten, D.~M., Tibshirani, R. and Hastie, T. (2009) A penalized matrix
  decomposition, with applications to sparse principal components and canonical
  correlation analysis.
\newblock \textit{Biostatistics}, \textbf{10}, 515--534.

\bibitem[{Zhang et~al.(2014)Zhang, Zhou, Jin, Wang and
  Cichocki}]{zhangFrequencyRecognitionSSVEPbased2014}
Zhang, Y., Zhou, G., Jin, J., Wang, X. and Cichocki, A. (2014) Frequency
  recognition in {{SSVEP}}-based {{BCI}} using multiset canonical correlation
  analysis.
\newblock \textit{International Journal of Neural Systems}, \textbf{24},
  1450013.

\bibitem[{Zipunnikov et~al.(2014)Zipunnikov, Greven, Shou, Caffo, Reich and
  Crainiceanu}]{zipunnikovLongitudinalHighdimensionalPrincipal2014}
Zipunnikov, V., Greven, S., Shou, H., Caffo, B., Reich, D.~S. and Crainiceanu,
  C. (2014) Longitudinal {{High}}-{{Dimensional Principal Components Analysis}}
  with {{Application}} to {{Diffusion Tensor Imaging}} of {{Multiple
  Sclerosis}}.
\newblock \textit{The annals of applied statistics}, \textbf{8}, 2175.

\end{thebibliography}

\end{document}